\documentclass[a4paper,11pt]{article}
\pdfoutput=1 

\usepackage{jheppub}

\usepackage{amsmath}
\usepackage{mathtools}
\usepackage{relsize}
\allowdisplaybreaks[2]

\usepackage{hhline}
\usepackage{bbm}
\usepackage[utf8]{inputenc}
\newcommand\id{\ensuremath{\mathbbm{1}}} 
\usepackage[T1]{fontenc} % if needed
\usepackage{bbm}
\usepackage{graphicx}
\usepackage{subfig}
\usepackage{booktabs} % For better table/tab
\usepackage{xcolor}

\title{\boldmath Dirac-Phase Thermal Leptogenesis in the extended Type-I Seesaw Model}
\usepackage{placeins}

\author[a]{Matthew J. Dolan}
\author[a,b]{ Tomasz P. Dutka}
\author[a]{ Raymond R. Volkas}

\affiliation[a]{ARC Centre of Excellence for Particle Physics at the Terascale, School of Physics,
The University of Melbourne, Victoria 3010, Australia}
\affiliation[b]{Corresponding Author}
\emailAdd{matthew.dolan@unimelb.edu.au}
\emailAdd{tdutka@student.unimelb.edu.au}
\emailAdd{raymondv@unimelb.edu.au}

\abstract{Motivated by the fact that $\delta_{CP}$, the Dirac phase in the PMNS matrix, is the only CP-violating parameter in the leptonic sector that can be measured in neutrino oscillation experiments, we examine the possibility that it is the dominant source of CP violation for leptogenesis caused by the out-of-equilibrium decays of heavy singlet fermions. We do so within a low-scale extended type-I seesaw model, featuring two Standard Model singlet fermions per family, in which lepton number is approximately conserved such that the heavy singlet neutrinos are pseudo-Dirac. We find that this produces a predictive model of leptogenesis. Our results show that for low-scale thermal leptogenesis, a pure inverse-seesaw scenario fails to produce the required asymmetry, even accounting for resonance effects, because wash-out processes are too efficient. Dirac-phase leptogenesis is, however, possible when the linear seesaw term is switched on, with the aid of the resonance contributions naturally present in the model. Degenerate and hierarchical spectra are considered -- both can achieve $\delta_{CP}$-leptogenesis, although the latter is more constrained. Finally, although unable to probe the parameter space of Dirac-phase leptogenesis, the contributions to unitarity violation of the PMNS matrix, collider constraints and charged-lepton flavour-violating processes are calculated and we further estimate the impact of the future experiments MEG-II and COMET for such models.}

\begin{document} 
\maketitle
\flushbottom
\parskip 5mm
\section{Introduction}
\label{sec:intro}

Two key puzzles of the Standard Model (SM) are the generation of the baryon asymmetry of the universe (BAU) and the origin of neutrino masses. These two problems can be linked to form one of the simplest realisations of Beyond the Standard Model (BSM) baryogenesis: thermal leptogenesis~\cite{Fukugita:1986hr} through the type-I seesaw mechanism~\cite{Minkowski:1977sc,Yanagida:1979as,Mohapatra:1979ia,Glashow:1979nm,GellMann:1980vs}. Here, heavy SM-singlet neutrinos (SN), which have masses far larger than the critical temperature $T_c$ of the electroweak phase transition, are added to the SM. They produce an initial excess in the family-lepton numbers $L_\alpha$ ($\alpha = e,\mu,\tau$) through the out-of-equilibrium, $L_\alpha$-violating decays of the SNs. The asymmetries in the $L_\alpha$ are then reprocessed into the observed asymmetry in baryon ($B$) number through the $(B+L)$ violating but $(B-L)$ conserving sphaleron interactions which are in thermal equilibrium in the early universe above $T_c$, but below that are highly suppressed with respect to the Hubble rate.

In the standard hierarchical, thermal, type-I leptogenesis scenario~\cite{Fukugita:1986hr}  (hereafter termed ``vanilla leptogenesis''), neutrino mass limits require experimentally unreachable mass scales -- typically above about $10^9$ GeV\footnote{We note that low-scale variations where the masses are close to (or below) the critical temperature do exist~\cite{Akhmedov:1998qx,Boubekeur:2004ez,Frigerio:2014ifa}.}  -- for the SNs. Furthermore, the required CP-violation in the theory can be completely decoupled from the low-energy leptonic sector of the SM, adding to the difficulty of experimentally probing vanilla leptogenesis. The scale of leptogenesis can be brought down to the TeV level if a quasi-degenerate spectrum is assumed for the SNs~\cite{Pilaftsis:2003gt,Pilaftsis:2005rv}, but this quasi-degeneracy should be motivated by a theory. Bringing the SNs to a low scale has the added consequence of suppressing the Yukawa couplings leading to highly suppressed discovery signals, such as in lepton flavour violation.

Extending the vanilla type-I seesaw scenario with extra sterile states, and promoting lepton number to being a ``good'' symmetry, leads to a class of neutrino mass models which have theoretically motivated low-scale SN states, with the potential for larger Yukawa couplings. The inclusion of the extra states can lead to a double suppression of the light neutrino masses allowing much larger Yukawa couplings for the SNs at TeV scale masses. The degeneracy in masses amongst the heavy SNs is naturally realised in this model since, in the limit that the Majorana masses are turned off, lepton number conservation is restored; those Majorana mass parameters, when small, thus obey technical naturalness. In this scenario, the heavy SNs mostly gain their masses from explicit Dirac mass terms, with the small Majorana masses lifting the mass degeneracy between the states leading to a natural possibility of resonant enhancement. In limiting cases these models are known as the ``inverse seesaw (ISS)''~\cite{Wyler:1982dd,Mohapatra:1986aw,Mohapatra:1986bd,Ma:1987zm,GonzalezGarcia:1988rw,GonzalezGarcia:1990fb,Deppisch:2004fa} and the ``linear seesaw (LSS)''~\cite{Akhmedov:1995vm,Barr:2003nn,Malinsky:2005bi,Gu:2010xc,Dib:2014fua} models. They are linked by a rotation~\cite{Ma:2009du}, although it proves convenient to distinguish between the two if other BSM symmetries exist. Of particular relevance is the left-right symmetric model~\cite{Pati:1974yy,Mohapatra:1974hk,Senjanovic:1975rk,Mohapatra:1979ia,Mohapatra:1980yp} (LRSM) where an underlying  $SU(2)_L \times SU(2)_R \times U(1)_{B-L}$ gauge symmetry is adopted that distinguishes between the two SM-singlet fermions in each family: one is in a right-handed doublet, while the other is a LRSM gauge singlet. In the context of such a model, the additional symmetries preclude the freedom in rotating the basis, justifying our choice to consider these contributions as independent.\footnote{Although motivated by such extended models, we operate under the assumption that any additional gauge and matter content beyond the additional SNs is sufficiently decoupled in order to not affect the phenomenological calculations below. See~\cite{Garayoa:2006xs,Blanchet:2010kw} for examples where additional field content is not decoupled.} See e.g.~\cite{Wang:2015saa,Das:2017ski} for alternative models.  

These models can be naturally low-scale and therefore more experimentally testable. Due to the low scale of the SNs, a fully-flavoured calculation of leptogenesis must be undertaken, since the out-of-equilibrium condition is satisfied in a regime where the lepton flavours $(e, \mu, \tau)$ are completely distinguishable in the thermal bath. In the fully-flavoured regime, it has been shown that the low-energy CP-violating observables of the active neutrinos (in this work we will be considering the PMNS parametrisation) are linked with the high energy CP-violation \textit{required} for the generation of BAU, and in fact can be solely responsible for leptogenesis. The Dirac phase, $\delta_{CP}$ (and Majorana phases, $\alpha_{ij}$) of the PMNS matrix can solely account for asymmetry generation in vanilla leptogenesis~\cite{Pascoli:2006ie,Pascoli:2006ci,Anisimov:2007mw}, but only for very high-mass new physics as a result of the Davidson-Ibarra bound~\cite{Davidson:2002qv,Giudice:2003jh}.\footnote{Note that the Davidson-Ibarra bound can be somewhat lowered by exploiting flavour effects, but the resulting scale is still very high~\cite{Racker:2012vw}.}

Motivated by current hints of $\delta_{CP}$~\cite{Esteban:2016qun} being maximally violating, we aim to consider the scenario of \textit{Dirac-phase leptogenesis} in which the BAU is driven dominantly by $\delta_{CP}$.\footnote{For the sake of simplicity, all other potential CP-violating phases are set to zero. In reality, it is expected that several of these phases will contribute substantially to the BAU. The simplifying assumption is used to establish that $\delta_{CP}$ can be a substantial contributor to the BAU is some regions of parameter space.} The low-scale nature of the extended type-I seesaw leads to a flavoured regime of leptogenesis making PMNS phases important whilst having large Yukawa couplings and avoiding the Davidson-Ibarra bound.

This paper is organised as follows. Section~\ref{sec:Model} explains how neutrino masses arise in this low-scale model and demonstrates how to guarantee the correct light, active neutrino physics. Section~\ref{sec:Leptogenesis} describes the leptogenesis calculations. Section~\ref{sec:Constraints} details the constraints used and estimates the possibility of future experiments in constraining such models. Finally, Sec.~\ref{sec:Numerical} presents numerical results obtained in the analysis and discusses implications of these results. In two appendices we present the decay and scattering used, and discuss the $\delta_{CP}$ dependence of the asymmetry in more detail. 

\section{Model}
\label{sec:Model}

The extended seesaw model is produced through the addition of three extra SM gauge singlet fermions alongside the three right-handed neutrinos that are added to the minimal SM to obtain the type-I seesaw model. The extra singlets lead to an extension of the SM Lagrangian of the form,
\begin{equation}
\label{lagrangian}
\begin{split}
-\mathcal{L}_{BSM} = \lambda_{D}  \overline{l_{L}} \,\widetilde{\Phi} N_{R} &+ \lambda_{L}\overline{l_{L}} \,\widetilde{\Phi} \left( S_L \right)^{c}
 + M_{R}\overline{\left( N_{R}  \right)^c } \left( S_{L} \right)^{c} \\
 &+ \frac{1}{2}\mu \,\overline{S_{L}} \left( S_{L}\right)^{c}+ \frac{1}{2} \mu^{\prime}\, \overline{(N_{R})^{c}}N_{R} +h.c,
 \end{split}
\end{equation}
and $\widetilde{\Phi}$ denotes charge conjugation. After the SM Higgs doublet $\Phi$ gains a vacuum expectation value, $v$, and breaks $SU(2)_{L}\otimes U(1)_{Y} \rightarrow U(1)_{\rm{EM}}$, 
the neutral lepton mass matrix arising from Eq.~\eqref{lagrangian} has the general form\footnote{For the remainder of this paper $\mu^{\prime}$ will be set to zero as unlike $\mu$ it does not have a first order contribution to the light neutrino masses~\cite{Bazzocchi:2010dt,Dev:2012sg}.}, in the $[ (\nu_{L})^c,\,N_{R},\, (S_{L})^c ]^{T}$ basis,
\begin{equation}
\label{fullmatrix}
M_{\nu}=\begin{pmatrix}
0 & m_{D} & m_{L} \\
m_{D}^{T} & 0 & M_{R} \\
m_{L}^{T} & M_{R}^{T} & \mu \end{pmatrix} ,
\end{equation}
where $\mu$ is in general a complex, symmetric $3 \times 3$ matrix and $m_{D}=\lambda_D v$, $m_{L}=\lambda_L v$ and $M_{R}$ are in general complex $3 \times 3$ mass matrices.
The full mass matrix of Eq.~\eqref{fullmatrix} has a number of limiting cases, according to assumptions on the allowed couplings and their strengths.

\subsubsection{Inverse seesaw (ISS)}

The ISS mechanism arises from Eq.~\eqref{fullmatrix} in the parameter regime $\mu \ll m_{D} \ll M_{R}$ and $m_{L}\rightarrow0$, leading to the neutrino mass matrix
\begin{equation}
\label{inversematrix}
M_{\nu}=\begin{pmatrix}
0 & m_{D} & 0 \\
m_{D}^{T} & 0 & M_{R} \\
0 & M_{R}^{T} & \mu \end{pmatrix} .
\end{equation}
Upon block diagonalisation of Eq.~\eqref{inversematrix}, one can calculate the light neutrino mass matrix to first order,
\begin{equation}
\label{inversemasslight}
m_{\nu} = m_{D}\left(M_{R}^{T}\right)^{-1} \mu\, M_{R}^{-1}\, m_{D}^{T}.
\end{equation}
The heavy Majorana neutrinos will be primarily admixtures of the $N_{R}$ and $(S_{L})^c$ states with mass matrices of the form (above the critical temperature and again to first order)~\cite{Baldes:2013eva},
\begin{equation}
\label{inversemassheavy}
M_{N} = M_{R} \pm \frac{1}{2}\mu.
\end{equation}
By contrast with the type-I seesaw model, there is now a double suppression of the light neutrino states in Eq.~\eqref{inversemasslight} due to the smallness of $\left( m_{D} M_{R}^{-1} \right)^{2}$ as well as the smallness of the parameter $\mu$ as discussed above. In the limit of $\mu\rightarrow0$ the light neutrinos become massless and the heavy sterile states form heavy Dirac singlets in a way that is independent of any other parameter in the theory restoring lepton number conservation.

The scale required in order to achieve the necessary active neutrino masses can be estimated\footnote{In the one generation approximation.}~\cite{Deppisch:2004fa}
\begin{equation}
\left(\frac{m_\nu}{0.1~\mbox{eV}}\right) = \left(\frac{m_{D}}{100~\mbox{GeV}}\right)^{2} \left(\frac{\mu}{1~\mbox{keV}}\right) \left(\frac{M_{R}}{10^{4}~\mbox{GeV}}\right)^{-2},
\end{equation}
where now the mass scale of the heavy SNs has been brought down to a relatively low, potentially experimentally testable level and with electroweak scale couplings.

\subsubsection{Linear Seesaw (LSS)}

The limit $\mu \rightarrow 0$ of Eq.~\eqref{fullmatrix} is known as the linear see-saw (LSS), with a mass matrix of the form
\begin{equation}
\label{linearmatrix}
M_{\nu}=\begin{pmatrix}
0 & m_{D} & m_{L} \\
m_{D}^{T} & 0 & M_{R} \\
m_{L}^{T} & M_{R}^{T} & 0 \end{pmatrix} .
\end{equation}
Block-diagonalising the upper-left $3 \times 3$ block in the limit that the term $m_{L} M_{R}^{-1}$ is small, and assuming the usual hierarchy $m_{D} \ll M_{R}$ exists, the effective light neutrino mass matrix is now given by~\cite{Forero:2011pc},
\begin{equation}
\label{linearmasslight}
m_{\nu} = - \left[ m_{D} \left( m_{L} M_{R}^{-1} \right)^{T}+\left( m_{L} M_{R}^{-1} \right) m_{D}^{T} \right],
\end{equation}
and the heavy SNs have mass matrix (above the critical temperature) 
\begin{equation}
\label{lineramassheavy}
M_{N}=M_{R}.
\end{equation}
Similar to above the correct active neutrino masses are possible with low-scale heavy SNs,
\begin{equation}
\left(\frac{m_\nu}{0.1~\mbox{eV}}\right) = \left(\frac{m_{D}}{100~\mbox{GeV}}\right) \left(\frac{m_{L}}{10~\mbox{eV}}\right) \left(\frac{M_{R}}{10^{4}~\mbox{GeV}}\right)^{-1}.
\end{equation}
Massless light neutrinos are recovered in the limit of vanishing $m_{L}$, and, in complete analogy with the ISS parameter $\mu$, this is the parameter in the LSS that explicitly breaks lepton number and therefore can be set small from the point of view of technical naturalness.

\subsection{Casas-Ibarra parametrisation }
\label{sec:type1}

The Casas-Ibarra parametrisation of the Dirac mass matrix $m_{D}$~\cite{Casas:2001sr} can be generalised in order to satisfy the light neutrino constraints for this extended type-I scenario.
In the conventional type-I scenario, the effective light neutrino mass matrix is given by~\cite{Lindner:2001hr}
\begin{equation}
\label{type1lightmass}
m_{\nu}=-m_{D}M_{R}^{-1}m_{D}^{T} 
\end{equation}
in the limit $m_{D} \ll M_{R}$, and with the effective mass term being $\bar{\nu}_L m_\nu (\nu_L)^c + h.c.$.
This $3 \times 3$ block is approximately diagonalised by the unitary PMNS matrix $U_\text{PMNS}$,
\begin{equation}
\label{type1}
U_\text{PMNS}^{\dag}\,  m_{\nu}\,  U_\text{PMNS}^* = m_{n} \equiv \text{diag}(m_1, m_2, m_3),
\end{equation}
where $m_{1,2,3}$ are the light neutrino mass eigenvalues, and
\begin{equation}
\label{PMNS}
\begin{split}
U_\text{PMNS} = & \left(\begin{array}{ccc}
c_{12}c_{13} & s_{12}c_{13} & s_{13}e^{-i\delta_{CP}} \\ 
-s_{12}c_{23}-c_{12}s_{23}s_{13}e^{i\delta_{CP}} & c_{12}c_{23}-s_{12}s_{23}s_{13}e^{i\delta_{CP}} & s_{23}c_{13} \\ 
s_{12}s_{23}-c_{12}c_{23}s_{13}e^{i\delta_{CP}} & -c_{12}s_{23}-s_{12}c_{23}s_{13}e^{i\delta_{CP}} & c_{23}c_{13}
\end{array}  \right) \times \\
& \times \left(\begin{array}{ccc}
1 & 0 & 0 \\ 
0 & e^{i\alpha_{21}/2} & 0 \\ 
0 & 0 & e^{i\alpha_{31}/2}
\end{array}\right).
\end{split}
\end{equation}
For the remainder of this paper, the low energy Majorana CP-violating phases $\alpha_{21}\, \rm{and} \,\alpha_{31}$ will be set to zero 
as our aim is to see if leptogenesis may be successfully driven solely by $\delta_{CP}$. We also suppress the $\text{PMNS}$ subscript on $U_\text{PMNS}$.

Equations \eqref{type1lightmass} and \eqref{type1} can be solved for the Dirac mass matrix $m_D$ to obtain the Casas-Ibarra parametrisation,
\begin{equation}
\label{type1casas1}
m_{D} =  \textit{i} \, U \, m_{n}^{1/2} \, R \, M_{R}^{1/2} ,
\end{equation}
where the complex matrix $R$ must be orthogonal: $R R^{T} = \id$. Thus, inputting $U$ and two out of the three masses in $m_n$ from experiment, this equation specifies the required neutrino Dirac
mass matrix for given choices of the unknown absolute neutrino mass scale for the light neutrino sector, and the matrices $M_R$ and $R$.\footnote{Note that the generated Dirac mass matrix will only be guaranteed to recover 
the correct mass differences among the light neutrinos if it continues to satisfy the approximation used for the 
block diagonalisation of the original mass matrix, namely $m_{D} \ll M_{R}$ in the case of the type-I seesaw, and any other approximation used in extended or different seesaw models.}

\subsection{Extended Casas-Ibarra parametrisations}
We now derive equivalent Casas-Ibarra-type parametrisations in different extended type-I seesaw models for $m_D$ in order to ensure agreement with neutrino oscillation experimental results.

%%%%%%%%%%%%%%%%%%%%%%%%%%%%%%%%%%%%%%%%%%%
\subsubsection{Inverse Seesaw}
%%%%%%%%%%%%%%%%%%%%%%%%%%%%%%%%%%%%%%%%%%%%%%

In the case of the ISS, combining Eqs.~\eqref{inversemasslight} and \eqref{type1} gives
\begin{equation}
\left( m_{n}^{-1/2}\, U^{\dag}\, m_{D} \,\left( M_{R}^{T}\right)^{-1}\, \mu^{1/2} \right) \left( \mu^{1/2} \,M_{R}^{-1} \,m_{D}^{T}\, U^* \,m_{n}^{-1/2} \right) = \id .
\end{equation}
Since $\mu$ is symmetric, we define an $R$ matrix and rearrange for $m_{D}$ to obtain 
\begin{equation}
m_{D} = U \, m_{n}^{1/2}\, R\, \mu^{-1/2}\, M_{R}^{T} \,.
\end{equation}
In general $R$ is a complex orthogonal matrix. However, in order to maintain $\delta_{CP}$ as the only source of CP-violation, we restrict $R$ to being real orthogonal, parametrised by
\begin{equation}
\label{inverseR}
R = \left(
\begin{array}{ccc}
 c_{y} c_{z} & -s_{x} c_{z} s_{y}-c_{x} s_{z} & s_{x} s_{z}-c_{x} s_{y} c_{z} \\
 c_{y} s_{z} & c_{x} c_{z}-s_{x} s_{y} s_{z} & -c_{z} s_{x}-c_{x} s_{y} s_{z} \\
 s_{y} & s_{x} c_{y} & c_{x} c_{y} \\
\end{array}
\right)
\end{equation}
where $c_{x} = \cos x$ and $s_{x} = \sin x$ and so on for $x,y,z \in \mathbb{R}$.

%%%%%%%%%%%%%%%%%%%%%%%%%%%%%%%%%%%%%%%%%%%%%%
\subsubsection{Linear Seesaw}
%%%%%%%%%%%%%%%%%%%%%%%%%%%%%%%%%%%%%%%%%%%%%%%

For completeness the parametrisation required for the linear seesaw is also presented.\footnote{The pure linear seesaw will not be considered further as above the 
critical temperature, relevant for thermal leptogenesis, the SNs will have degenerate masses and an asymmetry will not be generated~\cite{Pilaftsis:1998pd} 
This mass degeneracy can potentially be lifted if the thermal mass splitting of the neutrinos in the bath was considered~\cite{Aoki:2015owa}.}
Combining \eqref{linearmasslight} and \eqref{type1} leads to
\begin{equation}
- m_{n}^{-1/2} \,U^{\dag}\, \left[ m_{D} \left(m_L M_{R}^{-1}\right)^{T}\right]\, U^*\, m_{n}^{-1/2}\, + m_{n}^{-1/2}\, U^{\dag} \left[ \left(m_{L} M_{R}^{-1} \right) m_{D}^{T}\right] \,U^*\, m_{n}^{-1/2} = \id.
\end{equation}
Defining
\begin{equation}
\label{linearcasas}
R = - m_{n}^{-1/2}\, U^{\dag}\, \left[ m_{D} \left(m_L M_{R}^{-1}\right)^{T}\right]\, U^*\, m_{n}^{-1/2}
\end{equation}
implies that the $R$ matrix must now satisfy $R + R^{T} = \id$.
Rearranging \eqref{linearcasas} gives the parametrisation for $m_{D}$,
\begin{equation}
m_{D} = - U\, m_{n}^{1/2}\, R\, m_{n}^{1/2}\, U^{T}\, \left(m_{L}^{T}\right)^{-1} M_{R}^{T} \, .
\end{equation}

%%%%%%%%%%%%%%%%%%%%%%%%%%%%%%%%%%%%%%%%%%%%%%%%%%%%%%%%%%%%
\subsection{Inverse + Linear Seesaws (ISS + LSS)}
%%%%%%%%%%%%%%%%%%%%%%%%%%%%%%%%%%%%%%%%%%%%%%%%%%%%%%%%%%%%%%

In the case where terms from both the linear and inverse seesaws are present, the effective light neutrino mass matrix at first order is a linear combination of \eqref{inversemasslight} and \eqref{linearmasslight},
\begin{equation}
\label{linearandinverselightmass}
m_{\nu} = m_{D}\,\left( M_{R}^{T}\right)^{-1}\mu\, M_{R}^{-1}\, m_{D}^{T} - \left[( m_{D} \left( m_{L} M_{R}^{-1} \right)^{T} + \left( m_{L} M_{R}^{-1}\right) m_{D}^{T}\right].
\end{equation}
Combining with \eqref{type1} leads to
\begin{equation}
m_{n}^{-1/2}\,U^{\dag}\left[m_{D} \,\left(M_{R}^{T}\right)^{-1} \mu\, M_{R}^{-1}\, m_{D}^{T} - m_{D}\left(m_L M_{R}^{-1}\right)^{T} - \left(m_{L} M_{R}^{-1}\right)m_{D}^{T} \right]\, U^*\, m_{n}^{-1/2} = \id.
\label{full4}
\end{equation}
It is convenient to define a new matrix
\begin{align}
A \;=&\;\frac{1}{2}\, m_{D} \left(M_{R}^{T}\right)^{-1} \mu \,M_{R}^{-1} \,m_{D}^{T} - m_D \left(m_L M_R^{-1}\right)^{T},\nonumber
\\
A^{T} \;=& \;\frac{1}{2}\, m_{D} \left(M_{R}^{T}\right)^{-1} \mu \,M_{R}^{-1} \,m_{D}^{T} - \left( m_L M_R^{-1}\right) m_D^{T}.
\label{full1}
\end{align}
Rewriting \eqref{full4} in terms of this new A matrix gives
\begin{equation}
m_{n}^{-1/2} \,U^{\dag} \left[ A + A^{T} \right] U^*\, m_{n}^{-1/2} = \id.
\label{full2}
\end{equation}
We thus define
\begin{equation}
R = m_{n}^{-1/2}\, U^{\dag}\, A\, U^*\, m_{n}^{-1/2} \, ,
\label{full3}
\end{equation}
where $R$ now satisfies $R + R^{T} = \id$.
Combining \eqref{full4} to \eqref{full3} leads to the non-linear matrix equation
\begin{equation}
\label{linearandinversemd}
U\, m_{n}^{1/2}\, R \,m_{n}^{1/2}\, U^{T} = \frac{1}{2}\, m_{D} \left(M_{R}^{T}\right)^{-1} \mu \,M_{R}^{-1} \,m_{D}^{T} - m_D \left(m_L M_R^{-1}\right)^{T} 
\end{equation}
which will satisfy the light neutrino bounds when solved for $m_D$.

The choice of $R$ above is not a unique parametrisation of $m_D$ for the neutrino mass matrix. It is, however, convenient. Due to the non-linearity of Eq.~\eqref{linearandinversemd} 
it is difficult to find an analytic solution for $m_D$. We therefore solve this equation numerically to obtain our results in Sec.~\ref{sec:Numerical}, using an $R$-matrix of the form
\begin{equation}
\label{Rmatrix}
R = \begin{pmatrix}
\frac{1}{2} & r_{1} & r_{2} \\
-r_{1} & \frac{1}{2} & r_{3} \\
-r_{2} & -r_{3} & \frac{1}{2} \end{pmatrix} 
\end{equation}
where $r_i\in\mathbb{R}$ in order to make $\delta_{CP}$ the sole source of CP-violation.

%%%%%%%%%%%%%%%%%%%%%%%%%%%%%%%%%%%%%
\section{Leptogenesis}
\label{sec:Leptogenesis}
%%%%%%%%%%%%%%%%%%%%%%%%%%%%%%%%%%%%%%%%%%

In order to assess the viability of low-scale, Dirac-phase leptogenesis it is necessary to change basis such that the SN mass sub-matrix is 
diagonal and real in order to consider the decays of physical SNs to the standard leptons and Higgs boson. At this time in the early universe, above the electroweak phase transition critical temperature $(T_c)$,
the SM leptons are massless and the SNs decay either to a charged lepton and a charged scalar of the opposite sign, or a neutral lepton and a neutral scalar.  Taking Eq.~\eqref{fullmatrix}, 
performing a block diagonalisation of the lower right $2 \times 2$ block such that the SNs are in their mass basis, and 
rotating the Yukawa couplings to the SM leptons, transforms the mass matrix to the form \cite{Aoki:2015owa}
\begin{equation}
\label{blockrotation}
M_{\nu} \rightarrow M_{\nu}^{\prime} \simeq \begin{pmatrix}
0 & \frac{v}{\sqrt{2}} y^{\prime}_{D} & \frac{v}{\sqrt{2}} y^{\prime}_{L} \\
\frac{v}{\sqrt{2}}\left(y^{\prime}_{D}\right)^{T} & M_{R} - \frac{1}{2}\mu & 0 \\
\frac{v}{\sqrt{2}}\left(y^{\prime}_{L}\right)^{T} & 0 & M_{R}+\frac{1}{2}\mu \end{pmatrix} ,
\end{equation}
where $y^{\prime}_{D}$ and $y^{\prime}_{L}$ correspond to the rotated $3 \times 3$ Yukawa matrices coupling the heavy, sterile fermions to the SM leptons. For this $3 + 6$ generation case, the rotation will be performed numerically. In the one generation, it case can be expressed analytically (to lowest order) as~\cite{Blanchet:2009kk}
\begin{eqnarray}
y'_{D} & \simeq & \frac{i}{v}\left[\left( 1 + \frac{\mu}{4M_{R}} \right) m_{D} - m_{L}\right]\nonumber\\
y'_{L} & \simeq & \frac{1}{v}\left[\left( 1 - \frac{\mu}{4M_{R}} \right) m_{D} + m_{L}\right].
\end{eqnarray}
The rotated couplings are a linear combination of the unrotated couplings. Note that the coupling $m_L$ does not influence the heavy SN masses at this temperature.

Thermal leptogenesis proceeds because, once the temperature of the universe drops below the mass regime of the heavy SNs [in our case $\mathcal{O}(\text{TeV})]$, the inverse decays producing SNs will fall out of 
thermal equilibrium and a net lepton asymmetry will be generated by forward decays due to the non-zero Dirac phase, $\delta_{CP}$, in the PMNS matrix.

Because of the low-temperature regime in which the lepton asymmetry will be generated, where the individual lepton flavours are distinguishable, it is necessary to adopt flavour-dependent Boltzmann equations, 
which is nicely consistent with our requirement that the low-energy CP-violating phases contribute to leptogenesis. 
As can be seen in Eq.~\eqref{blockrotation}, and given that the parameter $\mu$ can be small in a technically-natural way, 
a small mass splitting exists amongst the SNs and therefore the lepton asymmetry generated may be enhanced due to a 
resonance effect \cite{Pilaftsis:2003gt,Pilaftsis:2005rv}. It is therefore necessary to account for this potential resonant enhancement in the calculation of the generated lepton asymmetry.

The CP asymmetry generated due to the decays of an SN, $N_{iR}$, to a specific lepton flavour $\alpha$ is defined as
\begin{equation}
\label{cpasymmetry1}
	\varepsilon^{\alpha}_{i} = \frac{\Gamma\left(N_{iR} \rightarrow l_{\alpha} \Phi \right) - \Gamma\left( N_{iR} \rightarrow \left(l_{\alpha}\right)^{c} \Phi^{\dag}\right)}{\sum\limits_{\alpha}^{ } \Big[\, \Gamma\left(N_{iR} \rightarrow l_{\alpha} \Phi \right) + \Gamma\left( N_{iR} \rightarrow \left(l_{\alpha}\right)^{c} \Phi^{\dag}\right) \Big]}.
\end{equation}
For later use we also define the individual branching ratios of each SN into a specific lepton flavour as
\begin{equation}
\label{branchingfraction}
B^{\alpha}_{i}=\frac{\Gamma\left(N_{iR} \rightarrow l_{\alpha} \Phi \right) + \Gamma\left( N_{iR} \rightarrow l^{c}_{\alpha} \Phi^{\dag}\right)}{\sum\limits_{\alpha}^{ } \Big[\, \Gamma\left(N_{iR} \rightarrow \left(l_{\alpha}\right)^{c} \Phi \right) + \Gamma\left( N_{iR} \rightarrow \left(l_{\alpha}\right)^{c} \Phi^{\dag}\right) \Big]}.
\end{equation}

The CP asymmetry parameter $\varepsilon^{\alpha}_{i}$ has been calculated previously for the general case, incorporating both the hierarchical as well as degenerate scenarios as limiting cases, and including potential resonance effects \cite{Adhikary:2014qba}:
\begin{eqnarray}
\label{cpasymmetry2}
\varepsilon^{\alpha}_{i} = &\mathlarger{\frac{1}{8\pi \left(h_{\nu}^{\dag} h_{\nu}\right)_{ii}}} \,\mathlarger{\mathlarger{\sum\limits_{j\ne i}}}\, \text{Im}\left( \left(h_{\nu}^{\dag} h_{\nu}\right)_{ij} \left(h_{\nu}^{\dag}\right)_{i\alpha}\left(h_{\nu}\right)_{\alpha j} \right)\left[ f(x_{ij}) + \frac{\sqrt{x_{ij}} \left(1 - x_{ij} \right)}{\left(1-x_{ij}\right)^{2} + \frac{1}{64\pi^{2}}\left(h_{\nu}^{\dag} h_{\nu}\right)^{2}_{jj}}\right]\nonumber\\
&+\mathlarger{\frac{1}{8\pi \left(h_{\nu}^{\dag} h_{\nu}\right)_{ii}}} \,\mathlarger{\mathlarger{\sum\limits_{j\ne i}}}\, \frac{\left(1-x_{ij}\right) \text{Im}\left( \left(h_{\nu}^{\dag} h_{\nu}\right)_{ji} \left(h_{\nu}^{\dag}\right)_{i\alpha}\left(h_{\nu}\right)_{\alpha j} \right)}{\left(1-x_{ij}\right)^{2} + \frac{1}{64\pi^{2}}\left(h_{\nu}^{\dag} h_{\nu}\right)^{2}_{jj}} + \mathcal{O}(h_\nu^6)\dots ,
\end{eqnarray}
where $h_{\nu} = \frac{\sqrt{2}}{v} \left( m^{\prime}_{D}, m^{\prime}_{L} \right)$, $x_{ij} = \left(\frac{m_{N_j}}{ m_{N_i}} \right)^{2}$ and the loop function of the vertex diagram contribution $f(x_{ij})$, is given by~\cite{Fukugita:1986hr}
\begin{equation}
\label{loopfunction}
f(x_{ij})=\sqrt{x_{ij}}\left[1-(1+x_{ij})\ln\left(\frac{1+x_{ij}}{x_{ij}}\right)\right].
\end{equation}
Eq.~\eqref{cpasymmetry2} includes a resonant enhancement in the CP asymmetry of a SN decay if and only if 
\begin{equation}
\label{resonantcondition}
1-x_{ij} \simeq \frac{1}{8\pi} \left(h_{\nu}^{\dag} h_{\nu}\right)_{jj}
\end{equation}
 is satisfied.

%%%%%%%%%%%%%%%%%%%%%%%%%%%%%%%%%%%%%%%%%%
\subsection{Boltzmann equations}
%%%%%%%%%%%%%%%%%%%%%%%%%%%%%%%%%%%%%%%%%%

The flavour-dependent Boltzmann equations valid for the low-scale regime of interest for the extended type-I seesaw considered are given by \cite{Pilaftsis:2005rv}:
\begin{eqnarray}
  \label{boltzn} 
\frac{d \eta_{N_{i}}}{dz} &=& \frac{z}{H(z=1)}\
\Bigg[\,\Bigg( 1 \: -\: \frac{\eta_{N_{i}}}{\eta^{eq}_{N_{i}}}\,
\Bigg)\, \sum_{\beta\,=\,e,\mu,\tau} \bigg(\,
\Gamma^{D\; (i \beta)} \: +\: \Gamma^{S\; (i \beta)}_\text{Yuk}\: +\:
\Gamma^{S\; (i \beta)}_\text{Gauge}\, \bigg) \nonumber\\ &&-\,
\frac{2}{3}\, \sum_{\beta\,=\,e,\mu,\tau} \eta^{\beta}_{l}\,
\varepsilon^{\beta}_{i}\, \bigg(\,
\widehat{\Gamma}^{D\; (i \beta)} \: +\: 
\widehat{\Gamma}^{S\; (i \beta)}_\text{Yuk} \: +\:
\widehat{\Gamma}^{S\; (i \beta)}_\text{Gauge}\, \bigg)\,\Bigg]\,,\\[5mm]
  \label{boltzl} 
\frac{d \eta_\alpha}{dz} &=& 
\frac{z}{H(z=1)}\, \Bigg\{\, \sum\limits_{i=1}^6\,
\varepsilon^{\alpha}_{i}\ \Bigg(
\frac{\eta_{N_{i}}}{\eta^{eq}_{N_{i}}} \: -\: 1\,\Bigg)\, 
\sum_{\beta\,=\,e,\mu,\tau} \bigg(\,
\Gamma^{D\; (i \beta)} \: +\: \Gamma^{S\; (i \beta)}_\text{Yuk}\:
+\: \Gamma^{S\; (i \beta)}_\text{Gauge}\, \bigg) \nonumber\\
&&-\,\frac{2}{3}\, \eta_\alpha\, \Bigg[\, \sum\limits_{i=1}^6\,
B^{\alpha}_{i}\,
\bigg(\, \widetilde{\Gamma}^{D\; (i \alpha)} \: +\: 
\widetilde{\Gamma}^{S\;(i \alpha)}_\text{Yuk}\: +\: 
\widetilde{\Gamma}^{S\; (i \alpha)}_\text{Gauge}\: +\: 
\Gamma^{W\; (i \alpha)}_\text{Yuk} + \Gamma^{W\;(i \alpha)}_\text{Gauge}
\,\bigg)\nonumber\\
&& \qquad\qquad\,\,\;
\: +\: \sum_{\beta\,=\,e,\mu,\tau}\,\bigg(\,\Gamma^{\Delta L =2\:(\alpha
\beta)}_\text{Yuk} \: +\:\Gamma^{\Delta L =0\:(\alpha \beta)}_\text{Yuk}\,\bigg) \Bigg] \nonumber\\
&&-\, \frac{2}{3}\,\sum_{\beta\,=\,e,\mu,\tau} \eta_{\beta}\,
\Bigg[\,\sum\limits_{i=1}^6\,
\varepsilon^{\alpha}_{i}\,\varepsilon^{\beta}_{i}
\bigg(\,\Gamma^{W\;(i \beta)}_\text{Yuk}\:
+\: \Gamma^{W\; (i \beta)}_\text{Gauge}\,\bigg)\nonumber\\ 
&& \qquad\qquad\qquad\qquad\qquad\quad\,\, \: +\: \Gamma^{\,\Delta L =2\:
(\beta \alpha)}_\text{Yuk} \: -\: \Gamma^{\,\Delta L =0\: (\beta \alpha)}_\text{Yuk}
\Bigg]\,\Bigg\}\,,
\end{eqnarray}
where the functions $\Gamma^{X}_{\dots}$ represent various decay and scattering cross sections normalised to photon density. They are explicitly rewritten in Appendix~\ref{sec:Appendix A}. An alternative approach is to use flavour-covariant rate equations relevant for resonant scenarios which incorporate quantum decoherence effects~\cite{Dev:2014laa,Dev:2014wsa}. 

The Boltzmann equations \eqref{boltzn} and \eqref{boltzl} involve the rescaled variable,
\begin{equation}
\label{rescaledvar}
z = \frac{m_{N_1}}{T}
\end{equation}
and the number density of each particle species has been normalised to the photon density
\begin{equation}
\label{normalizedeta}
\eta_{a}(z) \ =\ \frac{n_{a}(z)}{n_{\gamma}(z)}
\end{equation}
with
\begin{equation}
\label{photondens}
n_\gamma (z)\ =\ \frac{2\,T^3}{\pi^2}\ =\ 
               \frac{2\, m^3_{N_1}}{\pi^2}\,\frac{1}{z^3} .
\end{equation}
The term $H(T)$ corresponds to the Hubble parameter
\begin{equation}
\label{hubble}
H(T) = \sqrt{\frac{4\pi^3}{45}} \, g_{\star}^{1/2} \frac{T^{2}}{M_{\rm{Planck}}} \implies H(z) =\sqrt{\frac{4\pi^3}{45}} \, g_{\star}^{1/2} \frac{m_{N_1}^{2}}{z^{2} M_{\rm{Planck}}} ,
\end{equation}
where $M_{\rm{Planck}} = 1.2 \times 10^{19} \rm{\,GeV}$ and $g_{\star} = g_{SM} = 106.75$ is the number of relativistic degrees of freedom of the SM at the temperature relevant to leptogenesis at $z \gtrsim 1$.

The Boltzmann equations above utilise all $\Delta L = 0,1,2$ scattering terms.\footnote{In flavoured leptogenesis it is necessary to include $\Delta L = 0$ lepton-flavour-violating but lepton-number-conserving interactions.} As shown in Ref.~\cite{Blanchet:2009kk}, the $\Delta L = 2$ contributions can be of particular importance in this low-scale realisation, where lepton number is approximately conserved, for reducing the total effective washout.

%%%%%%%%%%%%%%%%%%%%%%%%%%%%%%%%%%%%%%%%%%
\subsection{Baryon asymmetry}
%%%%%%%%%%%%%%%%%%%%%%%%%%%%%%%%%%%%%%%%%%

As the SN decays generate the majority of the asymmetry at around the TeV scale, we adopt the approximation that the critical temperature is a hard cutoff for reprocessing the lepton asymmetry into a baryon asymmetry. In other words, we will not be considering the temperature dependence of the sphaleron rate as the temperature passes across the electroweak phase transition, with~\cite{Cline:1993bd}
\begin{equation}
\label{criticaltemp}
T_{c} = \left( 246\ \text{GeV} \right) \left( \frac{1}{2} + \frac{3 g^{2}}{16 \lambda} + \frac{g^{'2}}{16 \lambda} + \frac{h_{t}}{4 \lambda} \right)^{-1/2} \, ,
\end{equation}
where $h_t$ is the top-quark Yukawa coupling, $\lambda$ is the Higgs quartic coupling, and $g$ and $g'$ are the $U(1)_Y$ and $SU(2)$ gauge couplings.
Therefore the lepton asymmetry will simply be read off at the critical temperature, which is when $z = z_c$ with
\begin{equation}
\label{zc}
z_{c} = \frac{m_{N_1}}{T_{c}} .
\end{equation}
This asymmetry is reprocessed by sphaleron effects into a baryon asymmetry, and in the case of three families and one Higgs doublet we have
\begin{equation}
\label{leptotobaryo}
\eta_{B} = -\frac{28}{51} \frac{1}{27}  \sum_{\alpha = e,\mu,\tau} \eta_{\alpha} .
\end{equation}
The factor of $28/51$ arises from the fraction of lepton asymmetry reprocessed into a baryon asymmetry by the electroweak sphalerons \cite{Harvey:1990qw}, while $1/27$ is the dilution factor 
from photon production until the recombination epoch \cite{Deppisch:2010fr}.

As a consistency check, the asymmetries calculated through numerical solution of the Boltzmann equations were compared with an approximate analytic 
solution which is valid in the large washout regime, $K_\alpha^{\rm{eff}} \geq 5$, where
\begin{equation}
\label{washout}
K^{\alpha}_{\rm{eff}} = \kappa^{\alpha} \sum_i K_i B_{i}^{\alpha},
\end{equation}
and
\begin{equation}
\label{kappa}
\kappa^{\alpha} = 2 \sum_{i,j(j\neq i)}\frac{\rm{Re}\left[ \left(h_\nu\right)_{i \alpha} \left(h_\nu\right)_{j \alpha}^{*} \left( h_\nu h_\nu^{\dag} \right)_{ij} \right] + \rm{Im}\left[ \left(\left(h_\nu\right)_{i \alpha} \left(h_\nu\right)_{j \alpha}^{*}\right)^{2}\right]}{\rm{Re}\left[\left(h_\nu^{\dag}h_\nu\right)_{\alpha \alpha} \left\lbrace \left(h_\nu h_\nu^{\dag}\right)_{ii}+\left(h_\nu h_\nu^{\dag}\right)_{jj}\right\rbrace\right]} \left( 1 - 2i\frac{m_i-m_j}{\Gamma_i+\Gamma_j}\right)^{-1} .
\end{equation}
These low-scale models easily satisfy the $K^\alpha_{\rm{eff}} \geq 5$ condition~\cite{Deppisch:2010fr,Borah:2017qdu}. The approximate asymmetry is then estimated as
\begin{equation}
\label{anal}
\eta_{B}^{\rm{approx.}} = -\frac{28}{51}\frac{1}{27}\frac{3}{2}\sum_{\alpha,i} \frac{\varepsilon_{i}^{ \alpha}}{K^{\alpha}_{\rm{eff}} \rm{Min}[z_c,1.25\ln({25 K_{\rm{eff}}^{\alpha}})]}.
\end{equation}
A relatively close agreement was found between the numerical results and the analytic approximations in both the pure ISS and the ISS+LSS scenario demonstrated in Table~\ref{table2}.

\begin{table}[t]
\begin{center}
{%\renewcommand{\arraystretch}{1.2}
\scalebox{1.0}{
\begin{tabular}{ccc}
\toprule
Example parameters (ISS+LSS) & $\left|\eta_{B}\right|$ & $\left|\eta_{B}^{\rm{approx.}}\right|$ \\ 
\midrule 
\midrule
$m_L = 7.42 \times 10^{-5}\, [\rm{GeV}]$ & & \\
$\quad\,\mu = 7.27 \times 10^{-14}\, [\rm{GeV}]$ & $3.03 \times 10^{-12}$ & $5.11 \times 10^{-12}$ \\ 
$r_1 = 0.092,\quad r_2 = 0.163,\quad r_3 = 0.741\quad$ & & \\
% & & \\
\midrule
$m_L = 2.69 \times 10^{-3}\, [\rm{GeV}]$ & & \\
$\quad\,\mu = 6.59 \times 10^{-10}\, [\rm{GeV}]$ & $2.39 \times 10^{-14}$ & $6.70 \times 10^{-14}$ \\ 
$r_1 = 0.013,\,\quad r_2 = 0.014,\,\quad r_3 = 0.800\quad$ & & \\
% & & \\
\midrule
$m_L = 9.72 \times 10^{-5}\, [\rm{GeV}]$ & & \\
$\,\,\,\,\,\mu = 3.44 \times 10^{-6}\, [\rm{GeV}]$ & $1.29 \times 10^{-10}$ & $2.94 \times 10^{-10}$ \\ 
$r_1 = 0.129,\,\quad r_2 = 0.134,\,\quad r_3 = 0.202\quad$ & & \\
% & & \\
\midrule
$m_L = 7.71 \times 10^{-4}\, [\rm{GeV}]$ & & \\
$\quad\,\mu = 1.49 \times 10^{-17}\, [\rm{GeV}]$ & $7.17 \times 10^{-17}$ & $1.98 \times 10^{-16}$ \\ 
$r_1 = 0.084,\,\quad r_2 = 0.359,\,\quad r_3 = 0.026\quad$ & & \\
\bottomrule
\end{tabular}
}
}
\end{center}
\caption{Comparison between the numerically computed asymmetry $\left| \eta_B \right|$ and the analytic approximation $\left| \eta_B^{\rm{approx.}}\right|$ from Eq.~\eqref{anal} for example points of parameter space in the ISS+LSS regime. } 
\label{table2}
\end{table}
%%%%%%%%%%%%%%%%%%%%%%%%%%%%%%%%%%%%%%%%%%%%
\section{Constraints}
\label{sec:Constraints}
%%%%%%%%%%%%%%%%%%%%%%%%%%%%%%%%%%%%%%%%%%%%%%%%%%

In this section we detail the constraints we impose upon our model. Firstly, we place a perturbativity constraint on the Yukawa couplings generated using the Casas-Ibarra formalism, $\left| h_{ij} \right|^{2} \leq 4\pi$. An alternative perturbative unitarity constraint often used is $\frac{\Gamma_i}{m_i} < \frac{1}{2}$ which results in a similar effect on the parameter space. We also require that the light, active neutrino masses satisfy current oscillation limits taken from Ref.~\cite{Esteban:2016qun} and listed in Table \ref{table}. We consider only the normal hierarchy for the light neutrinos as the inverted hierarchy is disfavoured in the inverse seesaw model~\cite{Abada:2014vea}. We rejected points that did not satisfy the current mass difference squared best fit values to within 1$\sigma$. For definiteness, the lightest neutrino mass was fixed to $0.1~\text{eV}$ and the Dirac phase $\delta_{CP}$ fixed to the current approximate best fit value of $3\pi/2$. The values of the three mixing angles were fixed to their best fit points.

\begin{table}[h]
\begin{center}
{%\renewcommand{\arraystretch}{1.2}
\scalebox{1.0}{
\begin{tabular}{ccc}
\toprule
Parameter & Value \\ 
\midrule
\midrule
$\sin^{2} \theta_{12}$ & 0.306 \\ 
%\midrule 
$\sin^{2} \theta_{23}$ & 0.441 \\ 
%\midrule 
$\sin^{2} \theta_{13}$ & 0.02166 \\ 
\midrule 
$\Delta m_{21}^{2}/(10^{-5}\; \rm{eV}^{2})$ & $7.5^{+ 0.19}_{-0.17}$\\ 
%\midrule 
$\Delta m_{31}^{2}/(10^{-3} \;\rm{eV}^{2})$ & $2.524^{+0.039}_{-0.040}$ \\ 
\midrule 
$m_{1}/\,\rm{eV}$ & 0.1 \\ 
%\midrule 
$\delta_{CP}$ & $3\pi/2$ \\ 	
\bottomrule 
\end{tabular}
}
}
\end{center}
\caption{List of experimental results relevant to the PMNS matrix and the light neutrino mass splittings fixed by active neutrino oscillation experiments. The light neutrino mass differences were allowed to vary within 1$\sigma$ of their current best fit values~\cite{Esteban:2016qun}.} 
\label{table}
\end{table}

%%%%%%%%%%%%%%%%%%%%%%%%%%%%%%%%%%%%%%%%%%
\subsection{Unitarity}
%%%%%%%%%%%%%%%%%%%%%%%%%%%%%%%%%%%%%%%%%%

The inclusion of heavy SN states in the theory results in a violation of unitarity of the PMNS matrix due to the mixing between the active and sterile states. Unlike for the vanilla type-I case, the unitarity violation required in the leptonic sector can be potentially used as a discovery signal. To estimate the level of unitarity violation present we define the matrix
\begin{equation}
\label{umatrix}
V = \begin{pmatrix}
V_{3 \times 3} & V_{3 \times 6} \\
V_{6 \times 3} & V_{6 \times 6} \end{pmatrix} 
\end{equation}
which diagonalises the \textit{full} $9 \times 9$ neutrino matrix of Eq.~\eqref{fullmatrix} through
\begin{equation}
V^{T} M_{\nu} V = M^\text{diag}_{\nu} \, .
\end{equation}
The matrix $V$ has an exact representation derived using~\cite{Korner:1992zk,Dev:2012sg,Das:2014jxa},
\begin{equation}
\label{umatrix1}
V = \begin{pmatrix}
\left(\id_{3 \times 3} + \zeta^{*}\zeta^{T} \right)^{-1/2} & \zeta^{*} \left( \id_{6 \times 6} + \zeta^{T} \zeta ^{*} \right)^{-1/2} \\
-\zeta^{T} \left( \id_{3 \times 3} + \zeta^{*} \zeta^{T} \right)^{-1/2} & \left( \id_{6\times 6} + \zeta^{T} \zeta^{*} \right)^{-1/2} \end{pmatrix} \begin{pmatrix}
U & 0 \\
0 & V^{\prime} \end{pmatrix} 
\end{equation}
where $U$ is the usual PMNS matrix which would exactly diagonalise the light neutrino block in the absence of the heavy SNs. $V^{\prime}$ is the unitary matrix diagonalising the heavy neutrino mass sub-matrix,
\begin{equation}
\label{Vheavy}
\left(V^{\prime}\right)^{T}\begin{pmatrix}
0 & M_R \\
M^{T}_R & \mu \end{pmatrix} V^{\prime} = \begin{pmatrix}
M^{-} & 0 \\
0 & M^{+} \end{pmatrix}
\end{equation}
and $M^{\pm} = M_R \pm \mu$. While $V^{\prime}$ will be numerically computed, it can be approximated in the case where $M_{R}$ and $\mu$ are diagonal to lowest order by
\begin{equation}
\label{u6x6}
V^{\prime} = \frac{\sqrt{2}}{2}\begin{pmatrix}
\mathbbm{1}_3 & -i\mathbbm{1}_3 \\
\mathbbm{1}_3 & i\mathbbm{1}_3 \end{pmatrix} + \mathcal{O}(\mu M_R^{-1}) ,
\end{equation}
where $\zeta$ is a $3 \times 6$ matrix 
\begin{equation}
\label{zeta}
\zeta = \left( 0_{3 \times 3} , M_D M_R^{-1} \right) \equiv \left( 0_{3 \times 3}, F\right),
\end{equation}
which satisfies\footnote{In the inverse and linear approximations.} $||\zeta||<1 $. The matrix $F$ is often used instead of $\zeta$ in the literature~\cite{Dev:2009aw,Abdallah:2011ew}.

Each component of $V$ in Eq.~\eqref{umatrix1} can be expressed in terms of $\zeta$, or $F$ if expanding at lowest order,
\begin{equation}
\label{u3x3}
V_{3 \times 3} = \left( 1_{3 \times 3} + \zeta^{*} \zeta^{T} \right)^{-1/2} U \simeq \left( 1_{3 \times 3} - \frac{1}{2} F^{*} F^{T} \right) U = \left( 1_{3 \times 3} - \eta \right) U
\end{equation}
where  $\eta$ expresses the ``deviation from unitarity'', 
\begin{equation}
\eta \equiv \id_{3 \times 3} - \left(\id_{3 \times 3} + \zeta^{*} \zeta^{T} \right)^{-1/2} \simeq \frac{1}{2}F^{*} F^{T}.
\end{equation}
The remaining components of $V$ can be similarly expressed,
\begin{equation}
\label{u3x6}
V_{3 \times 6} = \zeta^{*} \left( \id_{6\times 6} + \zeta^{T} \zeta^{*} \right)^{-1/2}  V^{\prime} \simeq \left(0_{3 \times 3}, F^{*}\right) V^{\prime},
\end{equation}
\begin{equation}
\label{u6x3}
V_{6 \times 3} = -\zeta^{T} \left( \id_{3\times 3} + \zeta^{*} \zeta^{T} \right)^{-1/2}  U \simeq - \begin{pmatrix}
0_{3\times 3}  \\
F^{T} \end{pmatrix}  U,
\end{equation}
\begin{equation}
\label{u6x6}
V_{6 \times 6} = \left( \id + \zeta^{T} \zeta^{*} \right)^{-1/2} V^{\prime} \simeq \left(\id_{6 \times 6} - \frac{1}{2} \begin{pmatrix}
0_{3 \times 3} & 0_{3 \times 3} \\
0_{3 \times 3}& F^{T}F \end{pmatrix} \right) V^{\prime}.
\end{equation}

It is clear from the above that requiring $F = m_D m_R^{-1} \ll \id_{3 \times 3}$ is naturally realised in the ISS and LSS framework (to achieve low-scale neutrino masses) and therefore naturally can suppress the unitarity violation effects.

The deviation from unitarity allowed by experimental observation from universality tests of weak interactions, rare leptonic decays,
the invisible width of Z-boson and neutrino oscillation data is expressed through maximum allowed values for the entries in the matrix $\eta$~\cite{Fernandez-Martinez:2016lgt},
\begin{equation}
\label{unitarity}
\lvert \eta_{ij} \rvert < \begin{pmatrix} 
 1.3 \times 10^{-3}\ & 1.2 \times 10^{-5}\ & 1.4 \times 10^{-3}\\
 1.2\times 10^{-5}\  & 2.2 \times 10^{-4}\ & 6.0 \times 10^{-4}\\
 1.4\times 10^{-3}\ & 6.0 \times 10^{-4}\ & 2.8 \times 10^{-3}
\end{pmatrix} .
\end{equation}

%%%%%%%%%%%%%%%%%%%%%%%%%%%%%%%%%%%%%%%%%%%%%%%%%%%%
\subsection{Charged Lepton Flavour Violation (cLFV)}
%%%%%%%%%%%%%%%%%%%%%%%%%%%%%%%%%%%%%%%%%%%%%%%%%%%%

\subsubsection{ $\mu \rightarrow e \gamma$ }

The current 90\% C.L. upper bound on the branching ratio of $\mu \rightarrow e \gamma$ from the MEG collaboration is~\cite{TheMEG:2016wtm}
\begin{equation}
\label{meg1}
\text{BR}(\mu \rightarrow e\gamma) < 4.2 \times 10^{-13} \,.
\end{equation}
The branching ratio for this process in the presence of additional sterile neutrinos is given by~\cite{Abada:2014vea,Ilakovac:1994kj}
\begin{equation}
\label{branchingmu}
\textrm{BR}(\mu \rightarrow e\gamma) = \frac{\alpha_{w}^{3} s_{w}^{2}}{256 \pi^{2}}\frac{m_\mu^5}{M_{W}^{4}}\frac{1}{\Gamma_{\mu}} \left| \sum_{i=1}^{9} V_{\mu i}^{\star}V_{e i}\, G(y_{i})\right|^{2} \, ,
\end{equation}
where the loop function is given by 
\begin{equation}
G(x) = -\frac{2x^3+5x^2-x}{4(1-x)^3}-\frac{3x^3}{2(1-x)^4}\ln{x}
\end{equation}
and $y_i=m_{i}^{2}/M_{W}^{2}$,\, $\alpha_w = g_w^2/4\pi,\, s_w^2=1-(M_W/M_Z)^2$, $m_i$ refers to both the active and sterile neutrinos and the matrix $V$ is defined above \eqref{umatrix}.

The design sensitivity of the proposed MEGII detector \cite{Cattaneo:2017psr} is expected to increase the  90\% C.L. bound of Eq.~\eqref{meg1} to
\begin{equation}
\label{meg2}
\textrm{BR}(\mu \rightarrow e\gamma) < 5 \times 10^{-14} \,.
\end{equation}

%%%%%%%%%%%%%%%%%%%%%%%%%%%%%%%%%%%%%%%%%%%%%
\subsubsection{ $\mu\to e $ conversion}
%%%%%%%%%%%%%%%%%%%%%%%%%%%%%%%%%%%%%%%%%%%%%

The current strongest limit on the branching ratio for $\mu\to e$ coherent conversion on nuclei is from experiments involving gold~\cite{1674-1137-40-10-100001},
\begin{equation}
\label{currentconversionlimit}
\text{BR}(\mu \,\text{Au} \rightarrow e \,\text{Au}) < 7.0 \times 10^{-13}.
\end{equation}
The expected sensitivity of the future experiments COMET and Mu2e is $\rm{BR}(\mu^- N \to e^- N) \lesssim 10^{-17}$~\cite{Krikler:2015msn}, providing a potentially competitive cLFV limit to $\rm{BR}(\mu \rightarrow e \gamma)$.

It was shown in Ref.~\cite{Deppisch:2005zm} that in ISS models the long range, photonic contribution dominates in coherent conversion\footnote{For singlet fermion masses in these ISS models much lower than 1 TeV the photonic contribution can become sub-dominant to other diagrams making this approximation invalid~\cite{Abada:2014kba}.} and so there is a tight relationship between the expected branching ratios of $\mu \to e$ conversion and $\mu \rightarrow e \gamma$.
When the photonic contribution dominates, $\mu \to e $ conversion can expressed in terms of the $\textrm{BR}(\mu \to e \gamma) $ as~\cite{Kuno:1999jp}
\begin{equation}
\rm{BR}(\mu^-\, N \rightarrow e^-\, N) \sim \frac{B(A,Z)}{428} \, \rm{BR}(\mu \rightarrow e \gamma),
\end{equation}
where B(A,Z) corresponds to a rate dependence on the mass number $A$ and atomic number $Z$ for nucleus $N$. In this limit it is thus possible to use  $\mu \to e$ conversion to constrain $\mu \rightarrow e \gamma$ (or vice versa). Both Mu2E and COMET will initially make use of an aluminium target for which $B(A,Z)\sim 1.1$, implying
\begin{equation}
\label{approxconversion}
\rm{BR}(\mu^-\, N \rightarrow e^-\, N) \sim 2.6 \times 10^{-3}\;\rm{BR}(\mu \rightarrow e \gamma).
\end{equation}
The first run of COMET in 2018 has an expected sensitivity of 
\begin{equation}
\rm{BR}(\mu^-\, N \rightarrow e^-\, N) \sim 3 \times \; 10^{-15} .
\end{equation}
Comparing this to Eq.~\eqref{approxconversion} constrains $\textrm{BR}(\mu \rightarrow e\gamma) \lesssim  1.15 \times 10^{-12}$, a less competitive limit to the current limit directly measured [Eq.~\eqref{meg1}]. However, the lifetime sensitivity of the COMET experiment is expected to be $\mathcal{O}(10^{-17})$, implying a constraint of $\textrm{BR}(\mu \rightarrow e\gamma) \lesssim 3.84 \times 10^{-15}$. This would be an order of magnitude more sensitive in constraining models of heavy sterile neutrinos compared to the expectation in Eq.~\eqref{meg2}.

%%%%%%%%%%%%%%%%%%%%%%%%%%%%%%%%%%%%%%%%%%%%%%%%%%%%%%
\subsection{Neutrinoless Double Beta Decay}
%%%%%%%%%%%%%%%%%%%%%%%%%%%%%%%%%%%%%%%%%%%%%%%%%%%%%%%%%

A Majorana mass contribution for neutrinos implies the possibility of neutrinoless double beta decay. In our model, all the SN states are much heavier than the momentum transfer involved in double beta decay $m_{N_i} \gg 100 \,\rm{MeV}$. Accordingly the light neutrino propagator can be approximated by~\cite{Blennow:2010th}
\begin{equation}
\frac{1}{p^2-m_{N_i}^2} = - \frac{1}{m_{N_i}^2}+\mathcal{O}\left(\frac{p^2}{m_{N_i}^4}\right).
\end{equation} 
The dominant contribution will come from  the active, light neutrinos, and the allowed region of the lightest neutrino masses will remain the same as in the SM.\footnote{A possible caveat to this is when one-loop corrections and higher order terms become non-negligible. SNs as heavy as 5 GeV can dominate in certain limits~\cite{LopezPavon:2012zg} or if very large complex entries in the R-matrix are considered~\cite{Lopez-Pavon:2015cga}.}. Reference~\cite{Abada:2016awd} also demonstrates that, in our mass-range of interest, contributions to the electron electric dipole moment are not significant enough to have measurable effects.

%%%%%%%%%%%%%%%%%%%%%%%%%%%%%%%%%%%%%%%%%%%%
\section{Numerical Results}
\label{sec:Numerical}
%%%%%%%%%%%%%%%%%%%%%%%%%%%%%%%%%%%%%%%%%%

We consider the $3+3$ scenario in which three right-handed neutrinos $\nu_R$ and three singlets $S_L$ are added to the SM. The minimal scenario allowed by experiment is actually a $2+2$ theory, since one of the active neutrinos is allowed to be massless. However as we are considering SNs with similar TeV-scale masses, a degenerate spectrum of active neutrino masses is assumed due to the degenerate spectrum of sterile neutrinos considered. 

\begin{figure}[t]
\subfloat{
  \includegraphics[width=0.46\linewidth]{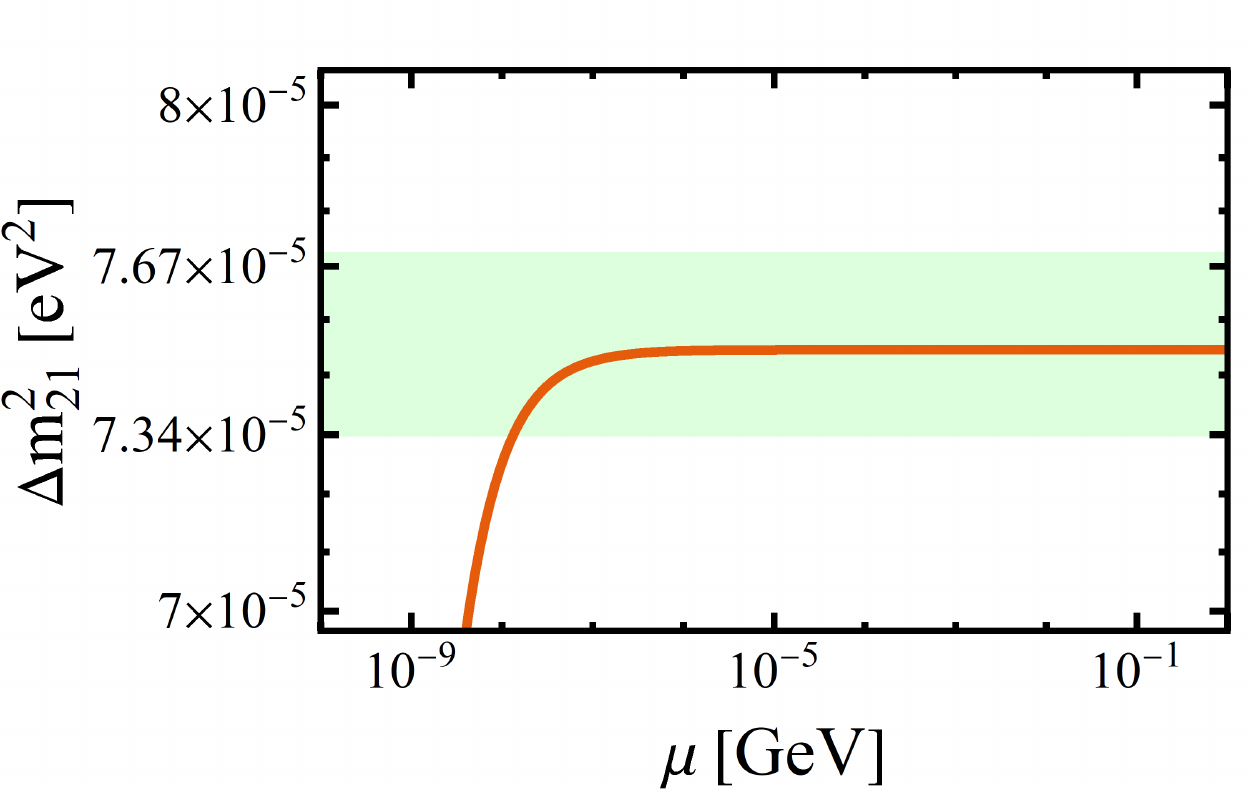}
} \hspace{0.5cm}
\subfloat{
  \includegraphics[width=0.46\linewidth]{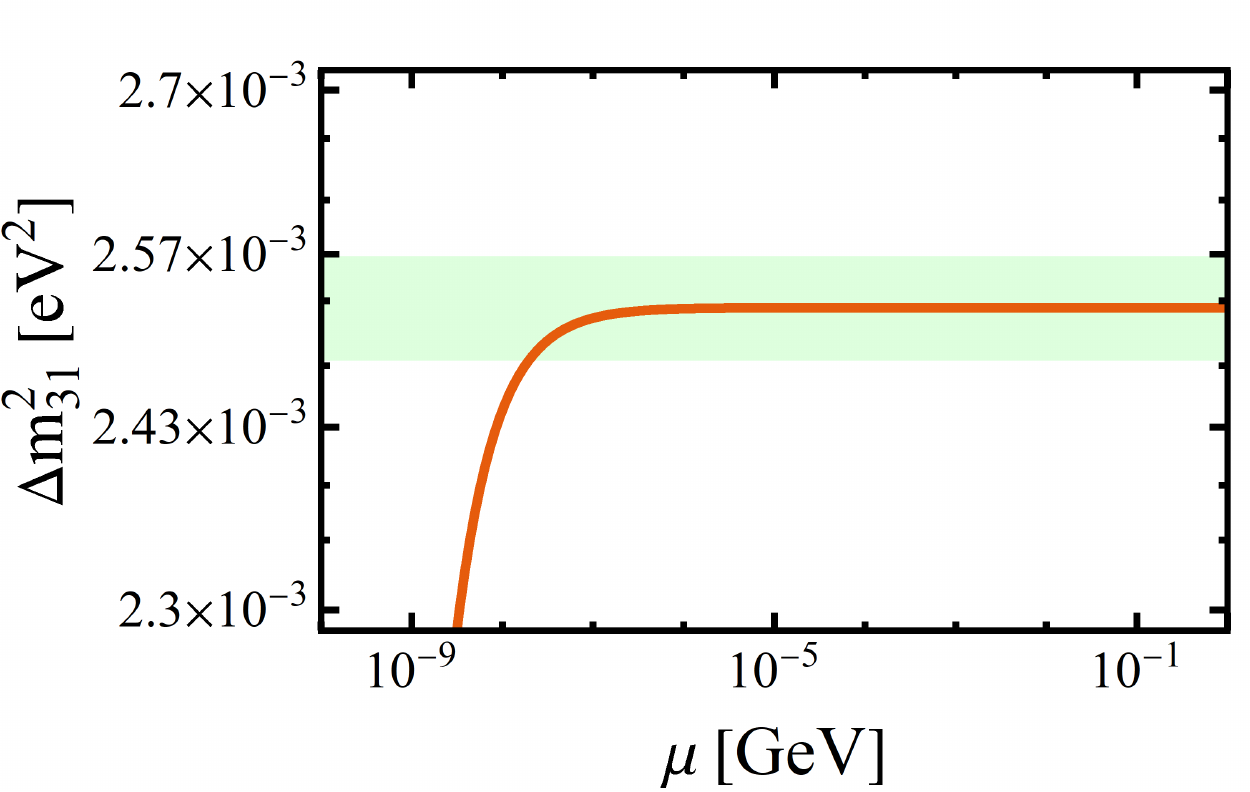}
}
\caption{Log-Log scale plots of the light neutrino mass splittings (orange curves) as a function of the Majorana parameter $\mu$ in the ISS, for normal ordering and parameter choices as described in the text. The coloured bands correspond to the current $1\sigma$ experimental limits on the allowed mass splittings between the light neutrino mass eigenstates. For regions below $\mu \lesssim 10^{-8}$ GeV, the generated Dirac matrix $m_{D}$ grows too large and the block diagonalisation assumptions in Eq.~\ref{inversemasslight} break down.}
\label{inverselimits}
\end{figure}

\begin{figure}[t]
\subfloat{
  \includegraphics[width=0.46\linewidth]{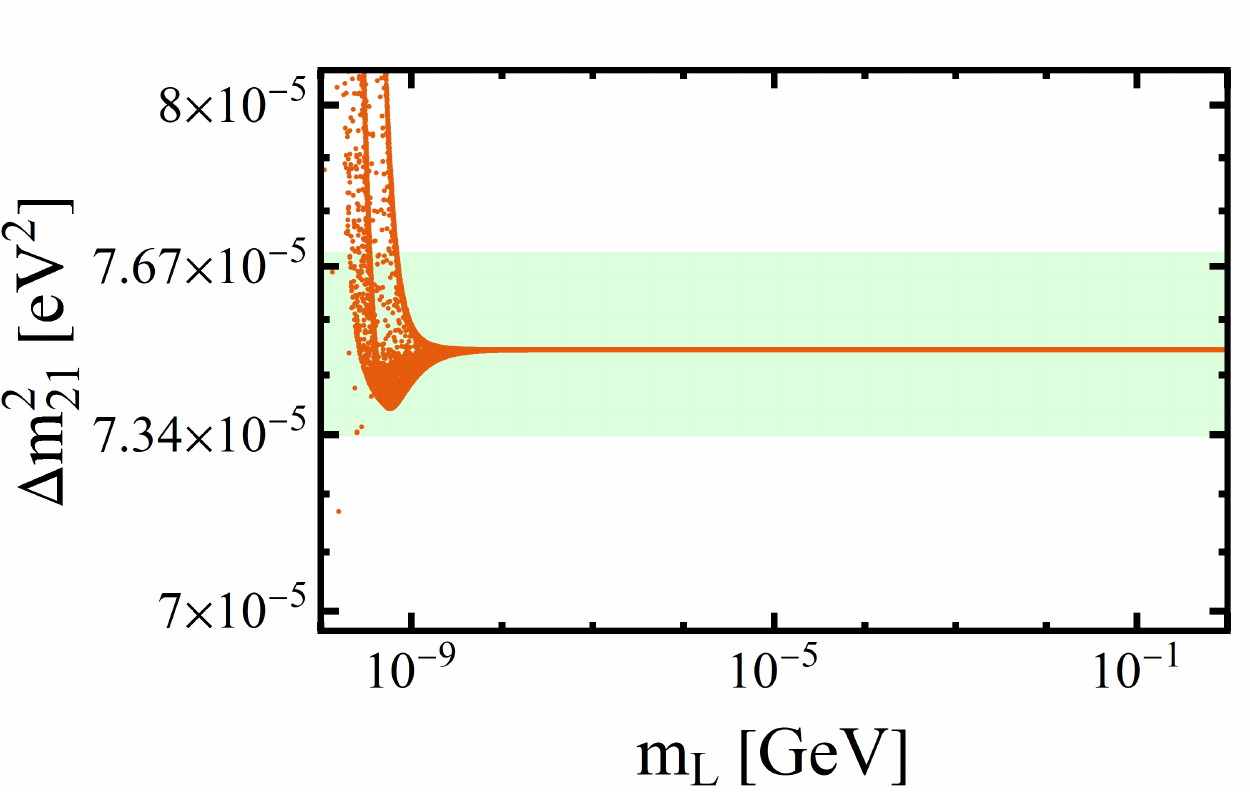}
}\hspace{0.5cm}
\subfloat{
  \includegraphics[width=0.46\linewidth]{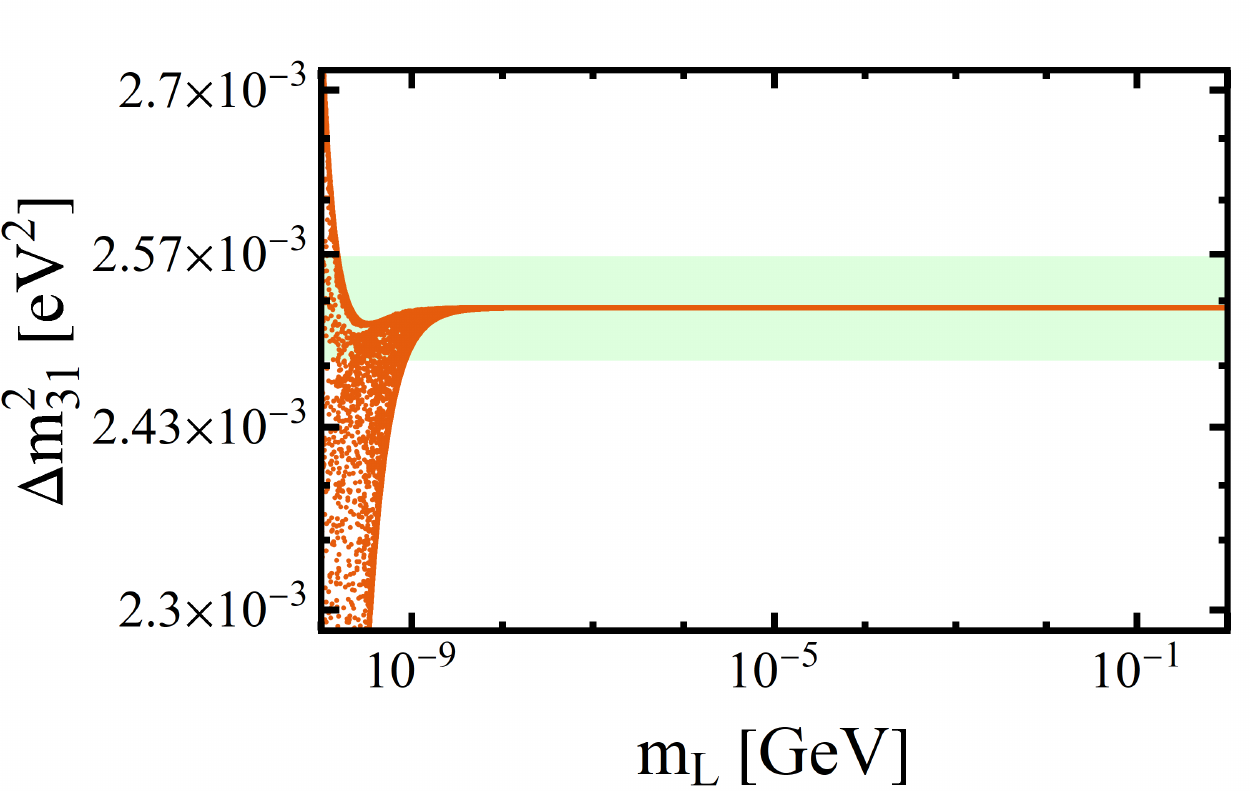}
}
\newline
\subfloat{
  \includegraphics[width=0.46\linewidth]{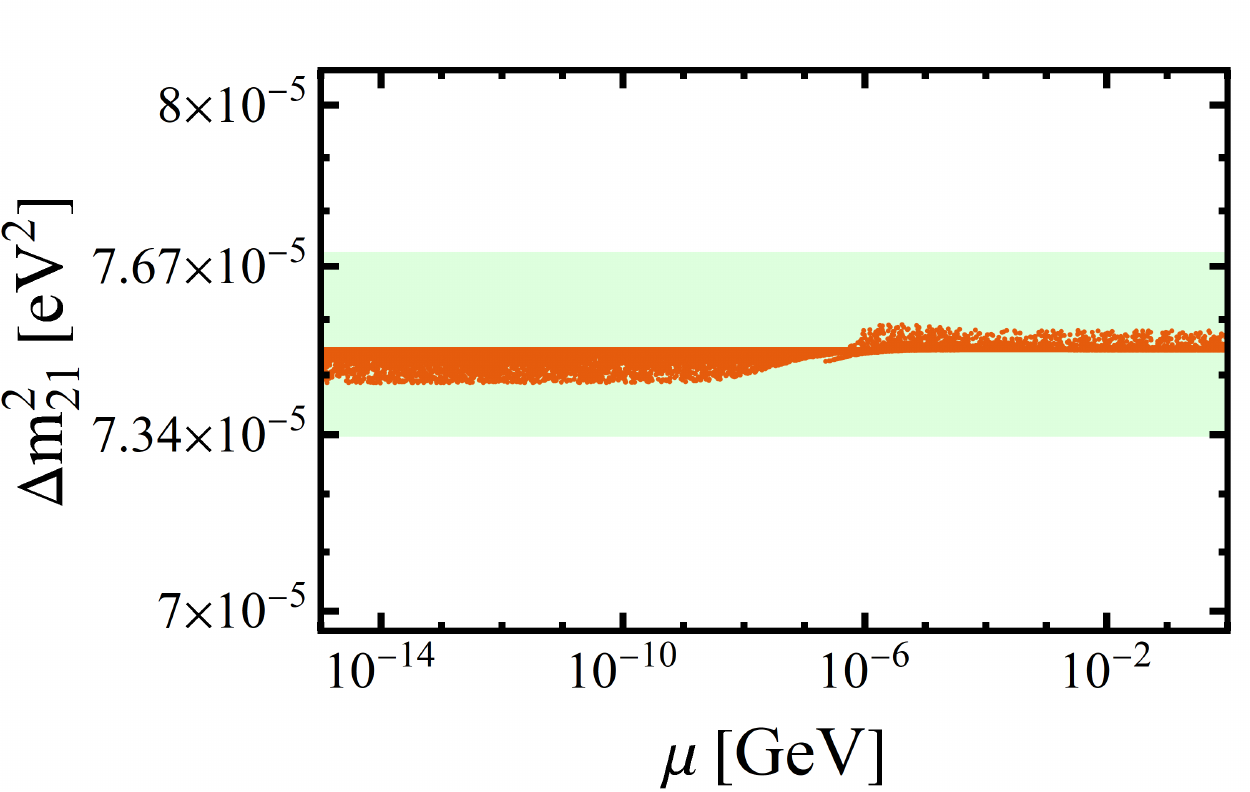}
}\hspace{0.5cm}
\subfloat	{
  \includegraphics[width=0.46\linewidth]{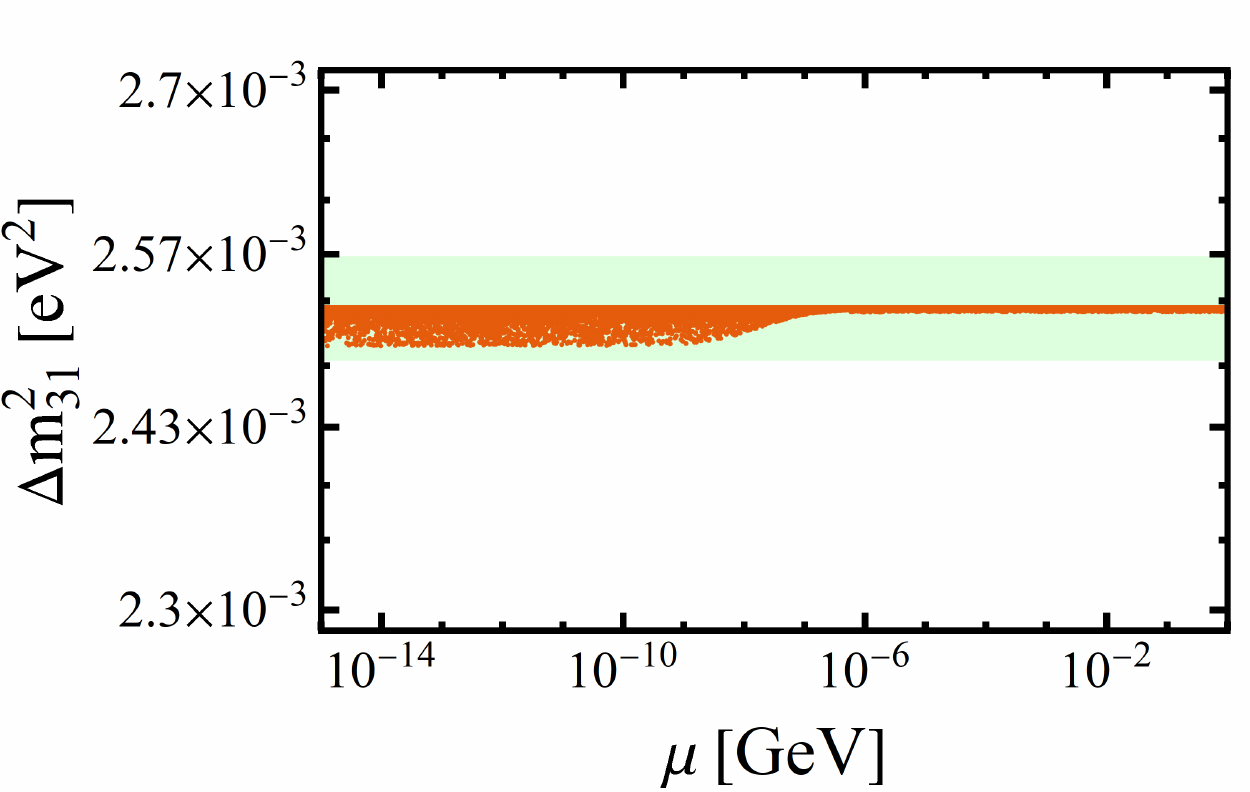}
}
\caption{{\bf Top panels:} Log-Log scale plots of the light neutrino mass splittings as a function of the linear parameter $m_L$ in the ISS+LSS, for normal ordering and parameter choices as described in the text. The coloured bands correspond to the current $1\sigma$ experimental limits on the allowed mass splittings. For regions $m_L \lesssim \, 10^{-9}$ GeV, the generated Dirac matrix $m_{D}$ grows too large and the block diagonalisation assumptions in \ref{inversemasslight} break down. 
{\bf Bottom panels:} Log-Log scale plots of the light neutrino mass splittings as a function of $\mu$ \textit{when} $m_L > 10^{-9} \rm{\,GeV}$. Note that much smaller values of $\mu$ are allowed if $m_L$ is turned on compared to the pure ISS scenario.}
\label{linearlimits}
\end{figure}

A \textit{minimal flavour violating hypothesis} first employed in Ref.~\cite{Forero:2011pc} is used where both the lepton number violating parameters,\footnote{The Majorana mass term $\mu$ can be taken to be diagonal without loss of generality, as can $\mu$ and $m_L$, but not $\mu,\, m_L,\, M_{R}$ simultaneously.} $m_L$ and $\mu$, are taken to be diagonal such that $m_D$ is the only source of flavour violation in the theory. These parameters are randomly scanned with the ranges,
\begin{eqnarray}
\,\,\,\,\left(\mu\right)_{ii} &\sim \mathcal{O}(10^{-8} - 1)\, \rm{GeV}
\end{eqnarray}
for the pure ISS and,
\begin{eqnarray}
\left(\mu\right)_{ii} &\sim & \mathcal{O}(10^{-17} - 1)\, \rm{GeV}\nonumber\\
\left(m_L\right)_{ii} &\sim & \mathcal{O}(10^{-9} - 1)\, \rm{GeV}
\end{eqnarray}
in the ISS+LSS scenario. The reasons for these choices are shown in Fig.~\ref{inverselimits} and Fig.~\ref{linearlimits}. Figure~\ref{inverselimits} shows the light neutrino mass splittings as a function of $\mu$ in the inverse seesaw model. Once $\mu < 10^{-8}$~GeV the Dirac mass matrix $m_D$ generated using the Casas-Ibarra method becomes too large and the block-diagonalisation assumptions used in  Eq.~\ref{inversemasslight} break down. Figure~\ref{linearlimits} shows the light neutrino mass splittings in the ISS+LSS scenario. The top panels demonstrate that $m_L$ should be greater than $10^{-9}$~GeV. The bottom panels show the neutrino mass splitting dependence on $\mu$, where we only select points that satisfy $m_L > 10^{-9}$~GeV. Clearly in this case we can achieve much lower values of $\mu$ than in the ISS model.

Similarly, the matrix $M_{R}$ which will fix the overall mass scale of the SNs is taken to be diagonal with values given by
\begin{eqnarray}
\label{mr}
\left(\frac{M_{R}^{\rm{deg}}}{1\, \rm{TeV}}\right) &= \begin{pmatrix} 
 1& 0 & 0\\
 0 & 1 & 0\\
0& 0 & 1
\end{pmatrix} \nonumber\\ 
\left(\frac{M_{R}^{\rm{hier}}}{1\, \rm{TeV}}\right) &= \begin{pmatrix} 
 1 & 0 & 0\\
 0 & 5 & 0\\
0& 0 & 5 
\end{pmatrix}
\end{eqnarray}
in the degenerate and hierarchical cases, respectively, as justified below.

As stated earlier, the Casas-Ibarra $R$ matrix is kept real in order to explore leptogenesis purely driven by the Dirac phase. It was varied according to
\begin{equation}
x,\, y,\, z \sim \left[ 0,2\pi\right]
\end{equation}
in the case of the pure ISS, where $x,y$ and $z$  are the elements of  the $R$ matrix as defined in Eq.~\eqref{inverseR}. In the case of the ISS+LSS, it was varied according to
\begin{equation}
r_{1,2,3} \sim \mathcal{O}(10^{-2} - 1)\
\end{equation}
with the $R$ matrix satisfying Eq.~\eqref{Rmatrix}.

\begin{figure}[t]
\captionsetup[subfigure]{justification=centering}
\centering
\subfloat{
  \includegraphics[width=0.47\textwidth]{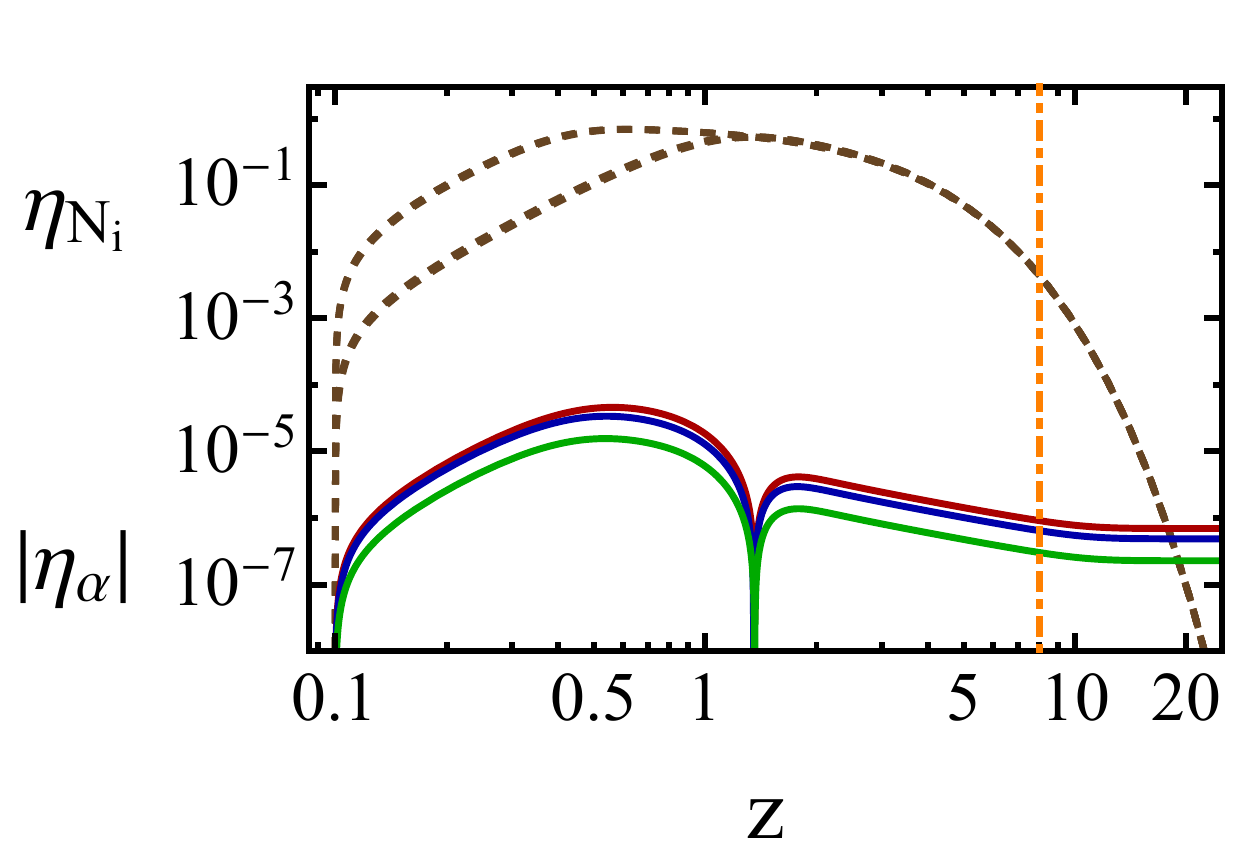}
  }
\subfloat{
  \includegraphics[width=0.47\textwidth]{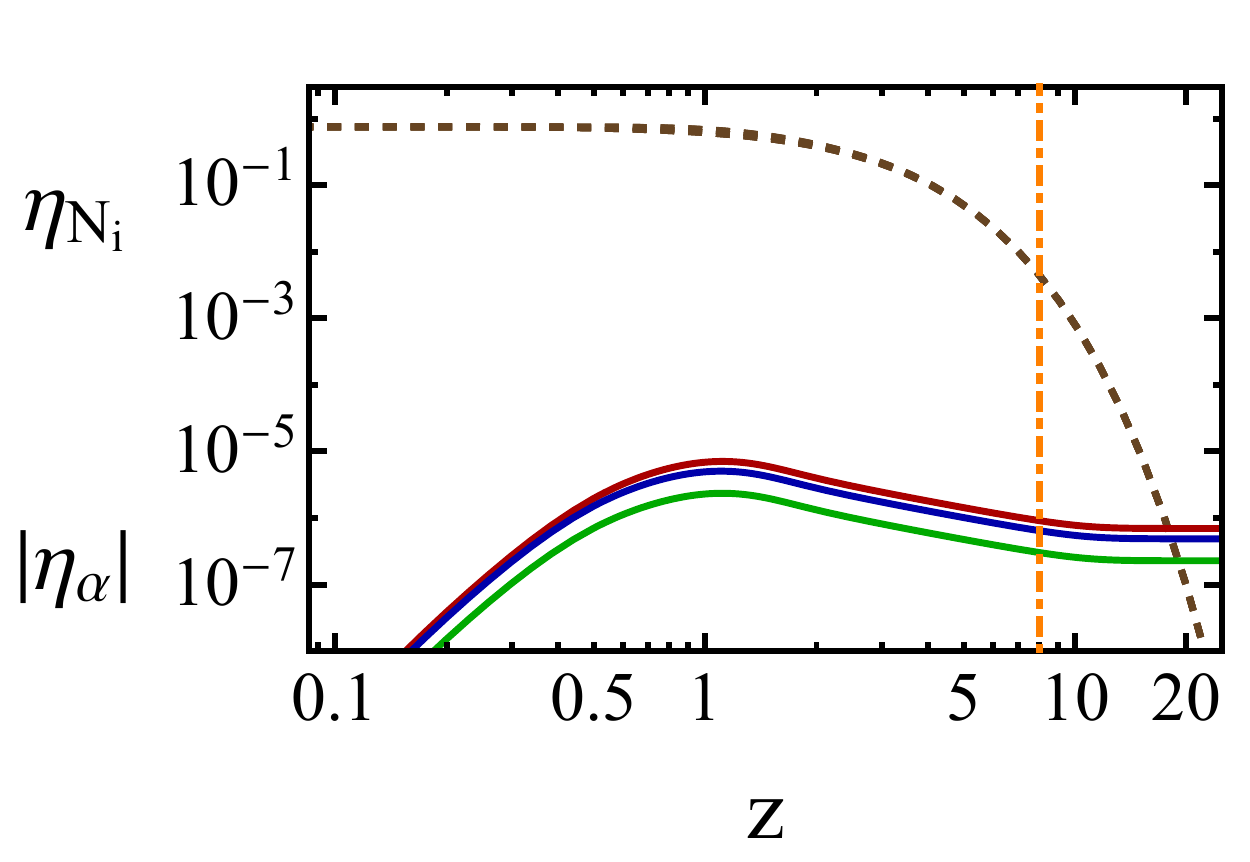}
  }
\caption{Solutions of the Boltzmann equations for the same set of parameters solved with $\eta_{N_i}(z_{in})=0$ (\textbf{Left Panel}) and $\eta_{N_i}(z_{in}) = \eta_{N_i}^{\rm{eq}}$ (\textbf{Right Panel}). The brown, dotted lines correspond to the population of the six SNs $\eta_{N_i}$ and the coloured, full lines to $\left|\eta_\alpha\right|$. The orange, dot-dashed line is the critical temperature, $z_c \simeq 8.02$ at which the lepton asymmetry is read off $\eta_\alpha (z_c)$. In both cases the same asymmetry is generated, $\eta_B = -\frac{28}{1377}\sum_\alpha \eta_\alpha = 7.229\times10^{-10}$ which is expected for strong washout thermal leptogenesis. Any asymmetry generated while thermally populating the sterile neutrinos is quickly washed out once a thermal population is reached. The parameters chosen in this example are: $m_L = 1.84\times 10^{-4}\,[\rm{GeV}],\,\mu=6.148\times10^{-8}\,[\rm{GeV}],\, r_1=3.869\times10^{-2},\,r_2=4.676\times 10^{-2},\,r_3 = 6.989\times 10^{-2}$.}
\label{initialconditions}
\end{figure}

In solving the Boltzmann equations we considered the initial conditions $\eta_{\alpha}(z_{in}) = 0$ for the lepton asymmetry and $\eta_{N_i}(z_{in}) = \eta_{N_i}^{\rm{eq}}$ for the SN population. Due to the strong washout nature of this low-scale theory there is an insensitivity to any primordial lepton asymmetry (which gets washed out immediately). Likewise the theory is insensitive to the initial abundance of SNs in the thermal bath. Due to the relatively low-scale masses of the SN, it should however be expected that a thermal abundance of them should be produced at temperatures well before an asymmetry is generated, thus motivating the initial conditions used. The insensitivity to the initial abundance and initial asymmetry is illustrated in Fig.~\ref{initialconditions}. These show the solutions of the Boltzmann equations for the same set of parameters as a function of $z$, using different initial conditions. In the left panel we take  $\eta_{N_i}(z_{in})=0$ and in the right we take  $\eta_{N_i}(z_{in}) = \eta_{N_i}^{\rm{eq}}$. The upper, dotted lines show the population abundances of the sterile neutrinos, and the lower, solid lines the active neutrino asymmetries. The vertical dot-dashed line shows the value, $z_C$ from Eq.~\ref{zc},  at which we read off the lepton asymmetry. The plots show that this is independent of the initial conditions used. The parameters used are detailed in the figure caption.
%\FloatBarrier

%%%%%%%%%%%%%%%%%%%%%%%%%
\subsection{Hierarchical Limit}
%%%%%%%%%%%%%%%%%%%%%%%

\begin{figure}[t]
\captionsetup[subfigure]{justification=centering,labelformat=empty}
\centering
\subfloat[1][\quad\quad\quad\quad$\mu \simeq 3.5 \times 10^{-5} \, \rm{(GeV)}$,\\\quad\quad\quad$x \simeq 1.54\pi$, $y \simeq 1.72\pi$, $z \simeq 0.21\pi$.]{
  \includegraphics[width=0.42\textwidth]{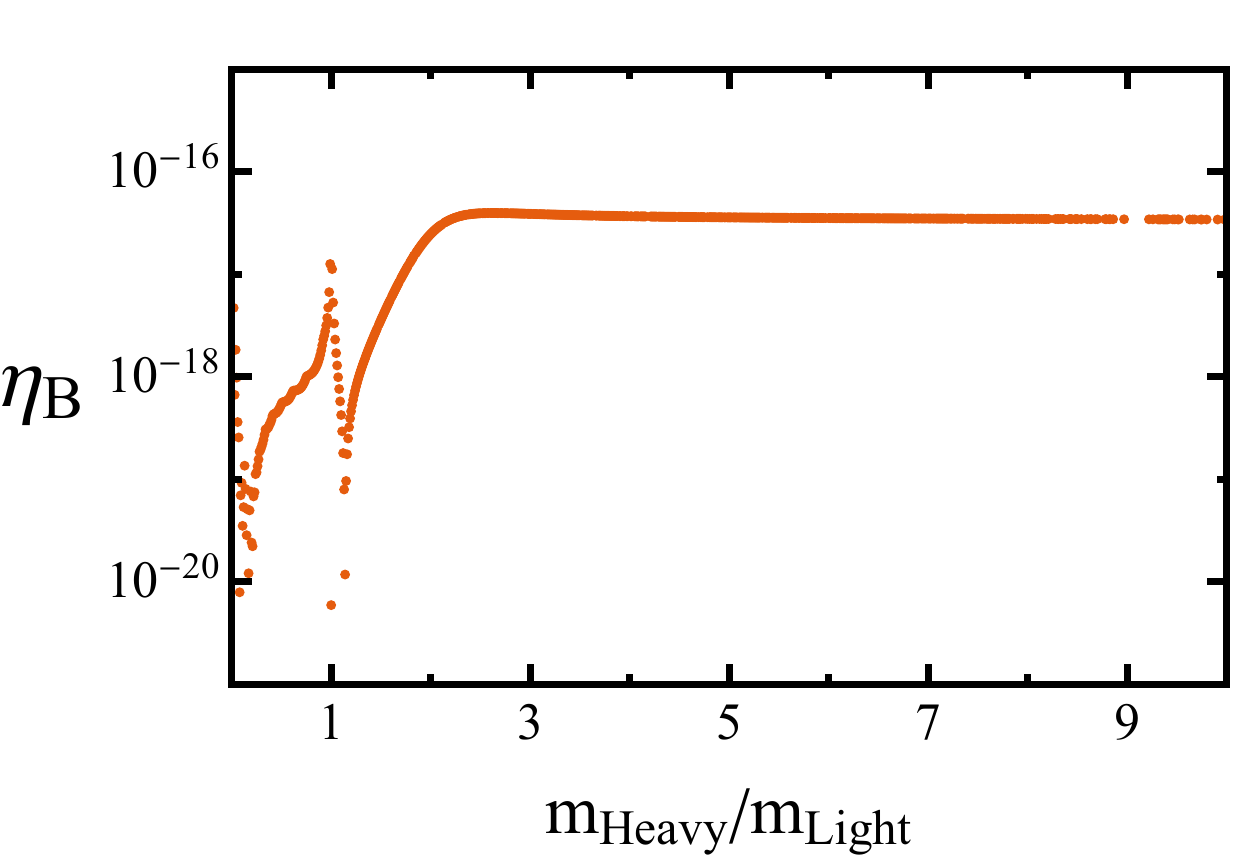}
  }
\subfloat[2][\quad\quad\quad\quad$\mu \simeq 1.7 \times 10^{-2} \, \rm{(GeV)}$,\\\quad\quad\quad$x \simeq 1.48\pi$, $y \simeq 1.37\pi$, $z \simeq 0.45\pi$.]{
  \includegraphics[width=0.42\textwidth]{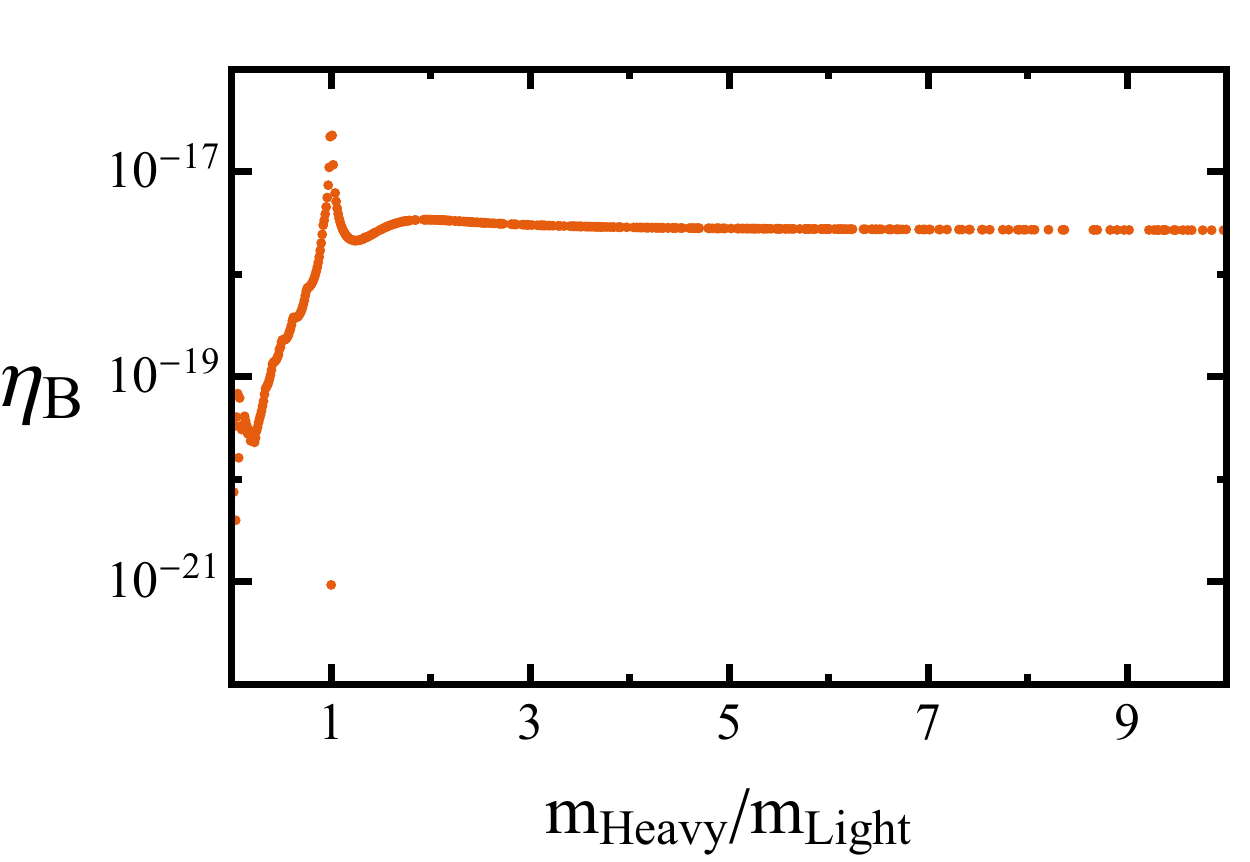}
  }
\newline
\centering
\subfloat[3][\quad\quad\quad\quad$\mu \simeq 6.9 \times 10^{-1} \, \rm{(GeV)}$,\\\quad\quad\quad$x \simeq 1.21\pi$, $y \simeq 0.25\pi$, $z \simeq 0.40\pi$.]{
 \includegraphics[width=0.42\textwidth]{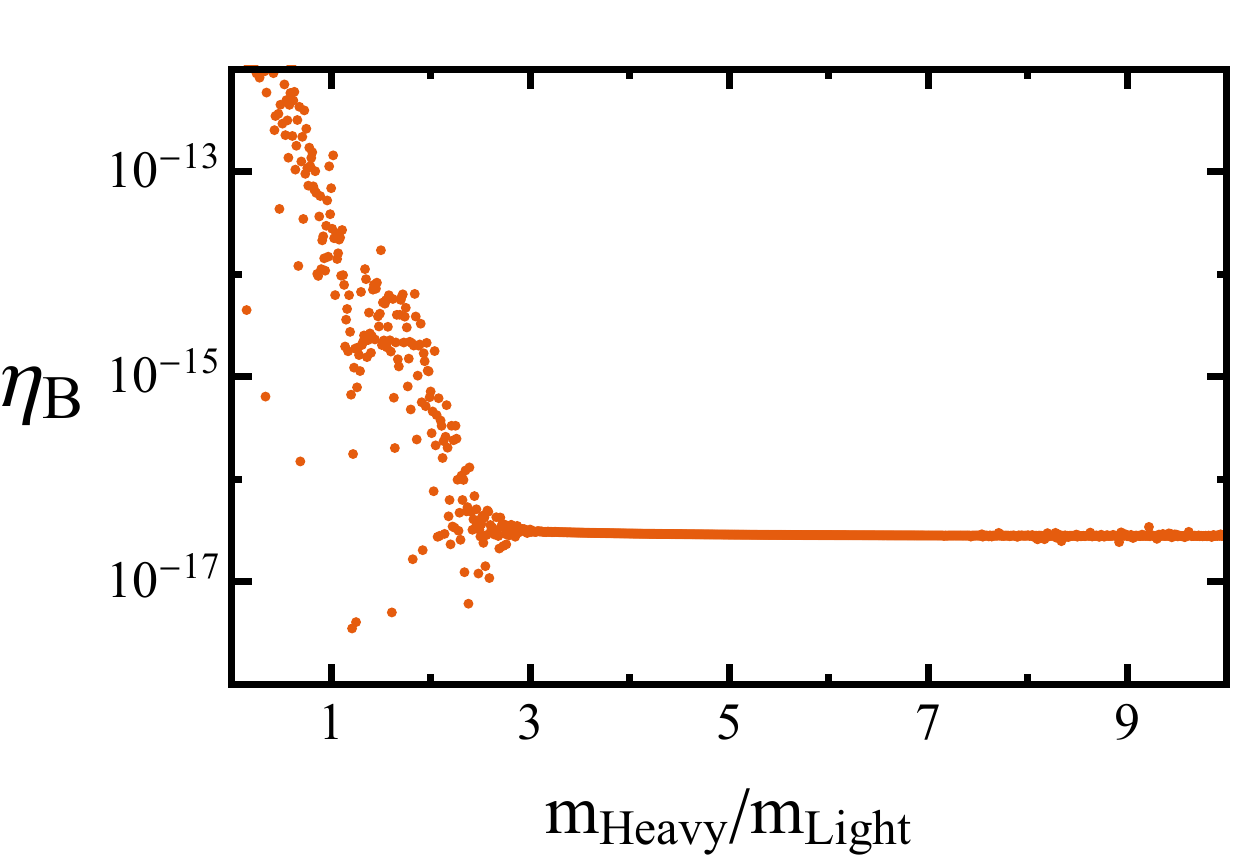}
}
\subfloat[4][\quad\quad\quad\quad$\mu \simeq 1.4 \times 10^{-3} \, \rm{(GeV)}$,\\\quad\quad\quad$x \simeq 0.93\pi$, $y \simeq 0.15\pi$, $z \simeq 0.09\pi$.]{
  \includegraphics[width=0.42\textwidth]{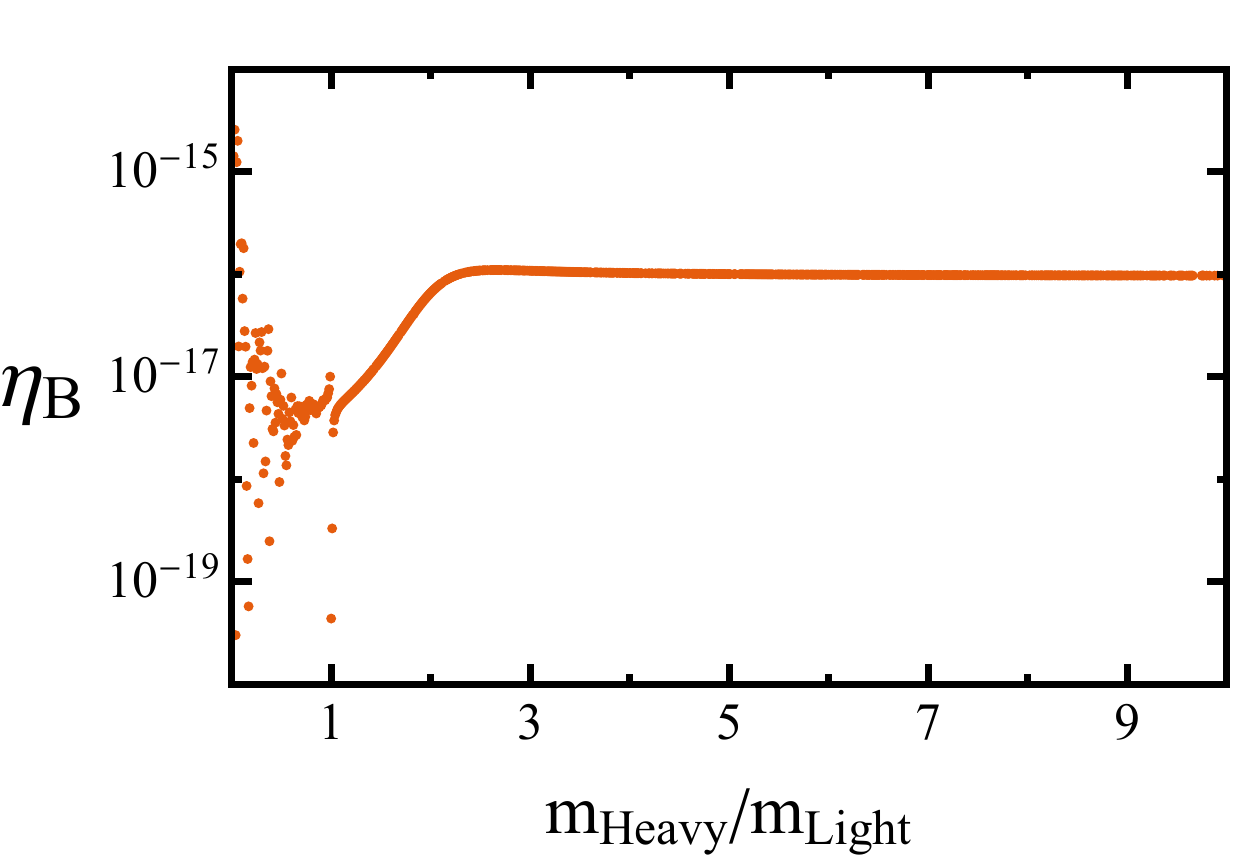}
}
\newline
\centering
\subfloat[5][\quad\quad\quad\quad$\mu \simeq 2.9 \times 10^{-4} \, \rm{(GeV)}$,\\\quad\quad\quad$x \simeq 0.74\pi$, $y \simeq 0.29\pi$, $z \simeq 0.91\pi$.]{
 \includegraphics[width=0.42\textwidth]{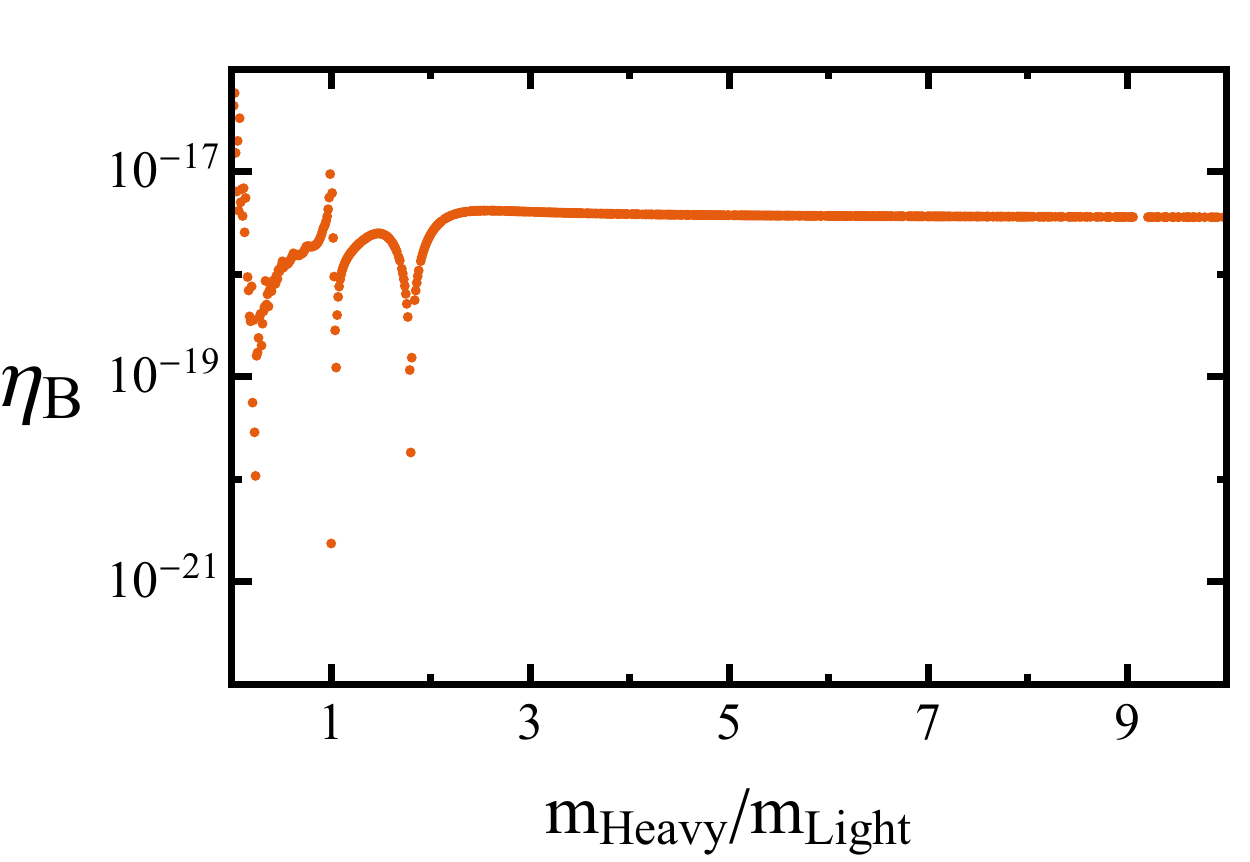}
}
\subfloat[6][\quad\quad\quad\quad$\mu \simeq 3.8 \times 10^{-3} \, \rm{(GeV)}$,\\\quad\quad\quad$x \simeq 1.52\pi$, $y \simeq 1.62\pi$, $z \simeq 0.68\pi$.]{
  \includegraphics[width=0.42\textwidth]{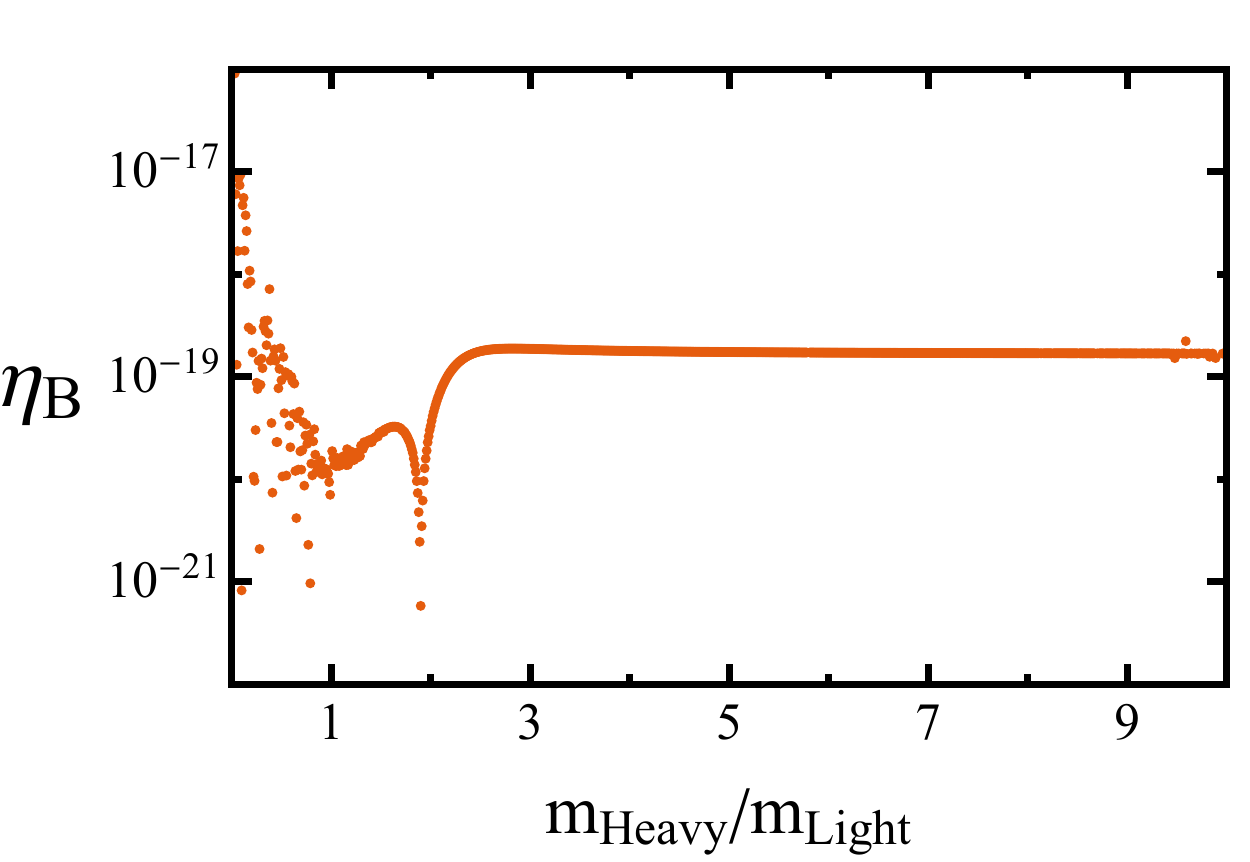}
}
\caption{Log-Log plots of final baryon asymmetry generated as a function of ratio of heavy SN masses to light SN masses (which remains fixed at 1 TeV) for some sample points. For computational convenience, only the ISS scenario is plotted and therefore none of the points generate the correct asymmetry (as discussed in the text). However, the plots serve to illustrate the freeze out of the asymmetry as some of the sterile neutrino masses get heavier. 
By $m_\text{Heavy}/m_\text{Light} \sim 5$, the asymmetry is almost completely insensitive to the heavy masses.}
\label{hierlimit}
\end{figure}

The hierarchical limit is taken to be the point at which the heavier sterile neutrinos' contribution to the final asymmetry is frozen out; the decays of these heavier SNs occur in a region in which the lighter SNs remain in thermal equilibrium and whose interactions wash out the generated asymmetry. In order to quantify the point at which the heavy SN contributions freeze out, a scan was performed over the masses of the heavier SNs whilst leaving all other parameters fixed.\footnote{The active neutrinos are still taken to have a quasi-degenerate mass spectrum since, due to the strong washout nature of the theory, the hierarchical limit as defined above is achieved before a very large mass hierarchy is formed amongst the heavy SNs.} The hierarchical limit is chosen by hand at a point in which the final baryon asymmetry is largely insensitive to the heavier SN masses. Based on the scan of a few hundred sample parameters it was found that this decoupling was generally comfortably achieved by $m_{\rm{heavy}}/m_{\rm{light}} \sim 5$. Sample plots are shown in Fig.~\ref{hierlimit} in order to illustrate this behaviour and justify the choice made in Eq.~\ref{mr}. These figures all show that the value of $\eta_B$ has stabilised once $m_{\mathrm{heavy}}/m_{\mathrm{light}}\sim 3$, so that $5$ is a conservative choice.

%\begin{figure}[!htbp]
%\FloatBarrier

%%%%%%%%%%%%%%%%%%%%%%%%%%%%
\subsection{Inverse Seesaw Case}
\label{inverse}
%%%%%%%%%%%%%%%%%%%%%%%%%%%%%%%%%

Before considering the full $3+3$ generation case, firstly we estimate the washout associated within the pure ISS scenario by considering the one generation $1+1$ limit.\footnote{As mentioned previously, we need at least two $\nu_R$ states in order to satisfy neutrino oscillation data, but it is convenient to examine this unrealistic but simple case in order to illustrate how $\mu$ influences the washout.}
In this case the associated mass matrix becomes,
\begin{equation}
\label{onegeninv}
M_{\text{inv},1g} = \begin{pmatrix}
0 & m_D & 0 \\
m_D & 0 & M_R \\
0 & M_R & \mu \end{pmatrix}.
\end{equation}
Employing Eq.~\ref{blockrotation} to rotate into the mass basis for the heavy sterile neutrinos (to first order)
\begin{equation}
\label{onegeninvrot}
M'_{\text{inv},1g} \simeq \begin{pmatrix}
0 & i m_{D} \left(1+\frac{\mu}{4 M_R} \right) & m_D \left( 1 - \frac{\mu}{4 M_R} \right) \\
i m_{D} \left(1+\frac{\mu}{4 M_R} \right) & M_R-\frac{\mu}{2} & 0 \\
m_D \left( 1 - \frac{\mu}{4 M_R} \right) & 0 & M_R + \frac{\mu}{2} \end{pmatrix} 
\end{equation}
and substituting the Dirac mass term, which in the one generation case is exactly
\begin{equation}
m_{D} = \frac{m_{n}^{1/2} M_R}{ \mu} \, ,
\end{equation}
leads to an estimate of the washout parameter as
\begin{equation}
\label{onegeninvwash1}
K^{1gen}_{N_i} \simeq \frac{\Gamma_{N_i}}{H(T=M_{N_i})}\simeq\frac{\sqrt{45} M_p  M_R  m_n}{16 \pi^{5/2} v^2 g_{*}^{1/2} \mu} \left| 1 \pm \frac{\mu}{4 M_R} \right|^2.
\end{equation}

\begin{figure}[t]
\captionsetup[subfigure]{justification=centering}
\centering
{
  \includegraphics[width=0.45\linewidth]{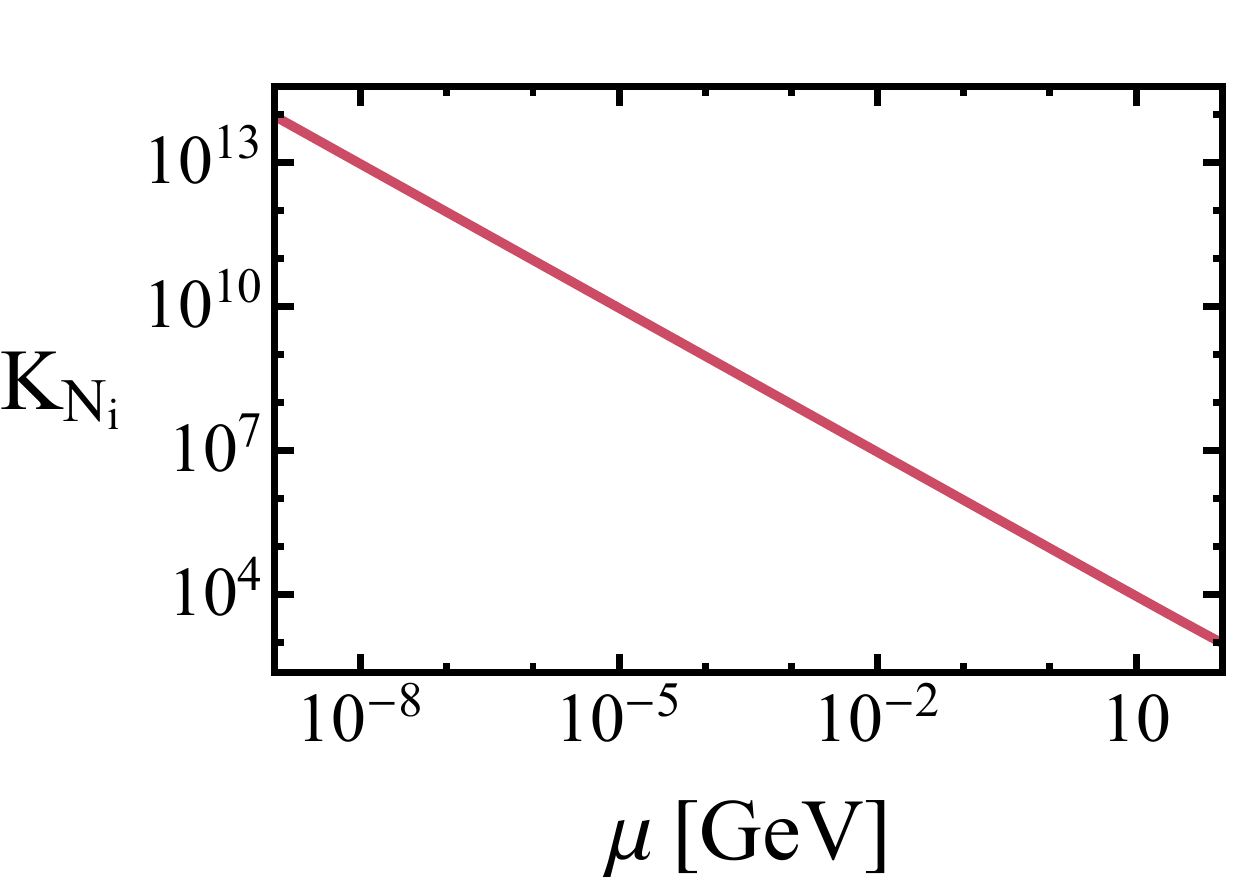}
  }
{
  \includegraphics[width=0.45\linewidth]{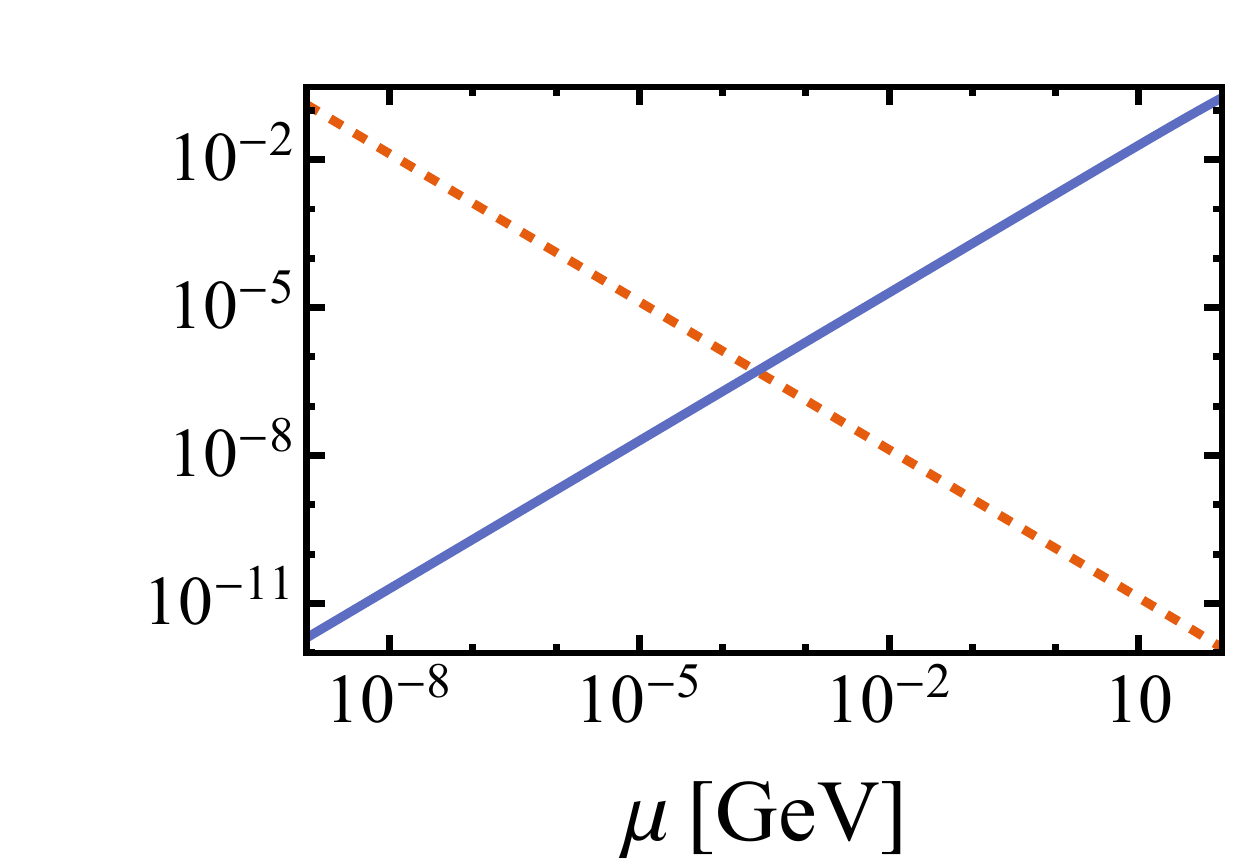}
  }
\caption{{\bf Left panel:}\  Log-Log plot of the washout parameter as a function of $\mu$. As $\mu$ is decreased, the Yukawa coupling must increase in order to account for the correct active neutrino masses thereby increasing the total washout.
{\bf Right panel:}\  Log-Log plot of the mass splitting, $(1-x_{ij})$ in blue (solid) against $\frac{1}{8\pi}\left(h_{\nu}^{\dag}h_{\nu}\right)_{jj}$ in orange (dashed) as a function of $\mu$. For the one generation case, $\mu \sim 10^{-3.5}\ \text{GeV}$ satisfies the resonance condition.
}
\vspace{0.5cm}
\label{invonegen}
\end{figure}

In the left panel of Fig.~\ref{invonegen}, the washout parameter is plotted against $\mu$, with $M_R$ fixed to 1~TeV and $m_n = 0.1\ \text{eV}$. For the region applicable to the inverse seesaw, $\mu < M_{R}$, where the SNs are pseudo-Dirac, the theory is squarely in the strong-washout regime. Unsurprisingly, the washout parameter increases as the lepton-number violating parameter $\mu$ decreases, since larger Yukawa couplings are required  to account for the fixed active neutrino masses. The right panel of Fig.~\ref{invonegen} shows the mass splitting between the two sterile neutrinos 
\begin{equation}
\label{res2}
1 - x_{ij}= 1 - \left(\frac{m_{N_j}}{m_{N_i}}\right)^{2} = 1 - \frac{\left(M_R \pm \frac{1}{2} \mu\right)^{2}}{\left(M_R \mp \frac{1}{2} \mu \right)^{2}}
\end{equation}
as a function of $\frac{1}{8\pi}\left(h_{\nu}^{\dag} h_{\nu} \right)_{jj}$, the two parameters relevant for the resonance condition of Eq.~\ref{resonantcondition}. Resonance can be achieved in the theory, allowing for the potential generation of a net asymmetry, even in this regime of strong washout. It is important to note that in the pure ISS, the parameter which determines the strength of the washout also determines the mass splitting between the heavy sterile neutrinos; in other words, it also affects the potential resonance contributions.
 Figure~\ref{invonegen} shows the washout parameter $K_{N_i}$ (left panel) and the mass splitting $(1-x_{ij})$ and $\frac{1}{8\pi}(h^\dagger_\nu h_\nu)_{jj}$ which contribute to the resonance condition.
The resonance condition is satisfied where the two lines in the right-hand panel cross, at $\mu\sim 10^{-3.5}$~GeV. We observe that the region of resonance corresponds to an extremely high value of the washout parameter: $K \sim 10^7$, shown in the left-hand panel as a function of $\mu$.

For TeV scale sterile neutrinos, Eq.~\ref{onegeninvwash1} can be used to estimate the required Majorana mass $\mu$ in order to move to the weak washout regime, $K_{N_i} < 1$:
\begin{equation}
\mu \sim 4\ \rm{TeV},
\end{equation} 
which is completely outside the region of applicability of the ISS.\footnote{We note that Ref.~\cite{Abada:2015rta} proposed a weak-washout scenario by setting the SN mass scale to $1$ GeV and relying on zero initial abundance of neutrinos close to the sphaleron decoupling point in order to successfully satisfy the out-of-equilibrium condition for asymmetry generation.}

\begin{figure}[t]
\centering
\includegraphics[width=0.7\linewidth]{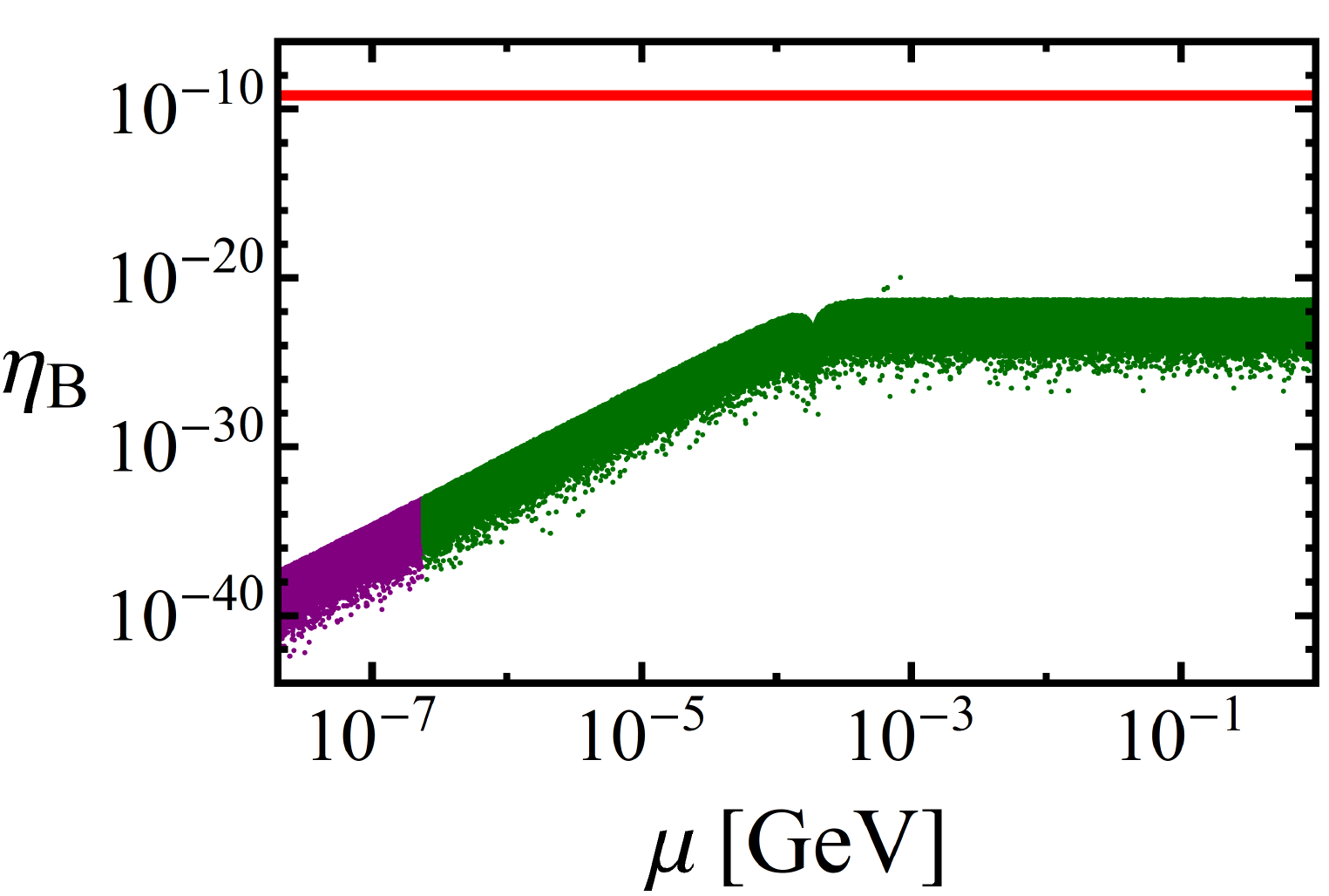}
\caption{The generated baryon asymmetry as a function of the lepton number violating parameter $\mu$ for the degenerate case. The points in purple, which correspond to larger values of $m_D$, are excluded by unitarity violation limits for the PMNS matrix. In red is the required BAU, $\eta_B \sim 6.21 \times 10^{-10}$. For the pure ISS scenario, the washout is extremely strong, even in regions of resonance and therefore a sizeable asymmetry cannot be generated in the Dirac phase case. The thickness of the green band is due to a scan over the Casas-Ibarra matrix $R$.}
\label{3geninvasym}
\end{figure}

Moving on to the full $3+3$ scenario, Fig.~\ref{3geninvasym} plots the generated baryon asymmetry from numerically solving the Boltzmann equations as a function of $\mu/\rm{GeV}$ for the degenerate case. All points generate the correct active neutrino mass differences; points in purple at low $\mu$ are excluded by unitarity violation. As we are specifically considering the scenario of Dirac-phase leptogenesis, all points generate a baryon asymmetry of the correct sign for $\delta_{CP} \sim 3\pi/2$, though obviously the magnitude is too small by many orders of magnitude. The thickness of the green band in Fig.~\ref{3geninvasym} and for most figures that follow is due to a scan over the Casas-Ibarra matrix $R$.

\begin{figure}[t]
\centering
%\subfloat[]
{
  \includegraphics[width=0.45\linewidth]{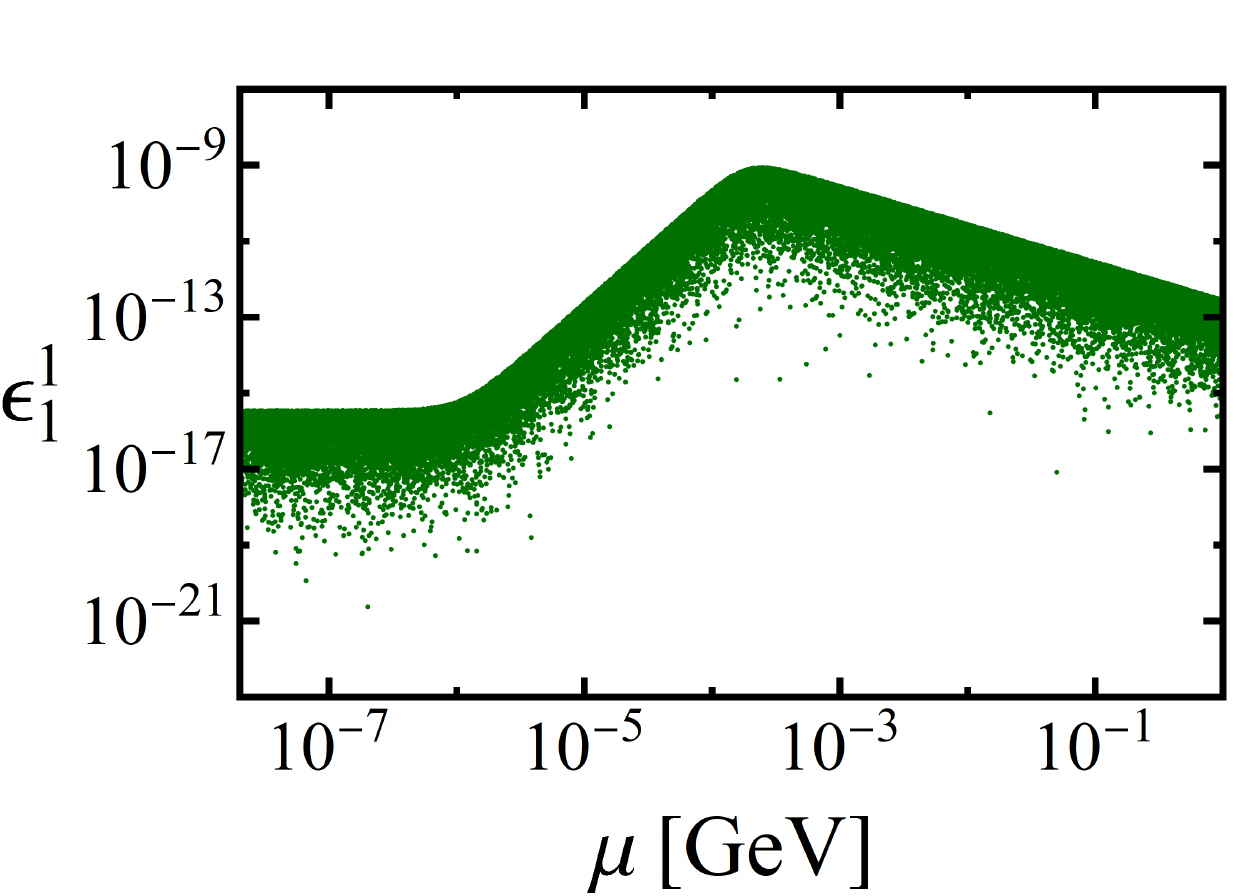}
}
%\subfloat[]
{
  \includegraphics[width=0.45\linewidth]{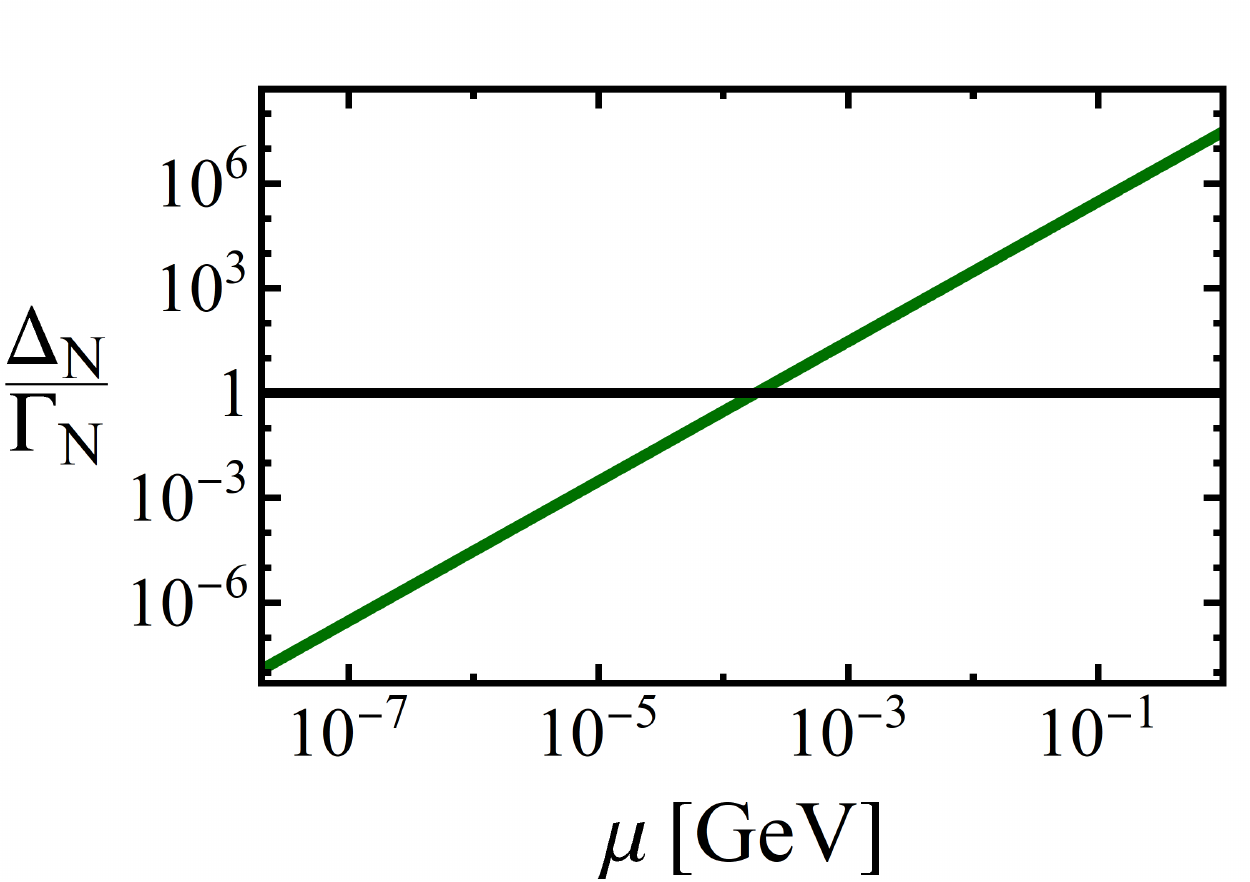}
}
\caption{Estimating the resonance contribution in 3+3 low-scale ISS models. {\bf Left panel:}\ Plot of the CP-asymmetry parameter $\epsilon^{\alpha}_{i}$ (in this case $\alpha=1$ and $i=1$) as a function of $\mu$ in GeV. A clear resonance peak appears in the regime: $\mu \sim 10^{-3}-10^{-4}$. The thickness of the green band is due to a scan over the Casas-Ibarra matrix $R$. {\bf Right panel:}\ Plot of $\Delta_{N_i}/\Gamma_{N_i}$ against the same parameter where $\Delta_{N}/\Gamma_{N} = (1-x_{ij})/\frac{1}{8\pi}\left(h_{\nu}^{\dag}h_{\nu}\right)_{jj}$ where $1-x_{ij}$ is defined in Eq.~\eqref{res2}. The horizontal lines corresponds to where the resonance condition is satisfied. Unlike in traditional resonant models, the mass splitting between the heavy sterile states cannot be adjusted independently of the required Yukawa couplings, because both depend on $\mu$. Nevertheless, a region satisfying the resonance condition of Eq.~\ref{resonantcondition} is possible.}
\label{resinv3}
\end{figure}

The reason is that, in scenarios of TeV-scale ISS, the generated asymmetry is heavily washed out. Figure~\ref{resinv3} demonstrates that a resonant enhancement occurs in the theory,\footnote{As the asymmetry is purely generated by the Dirac phase $\delta_{CP}$, for this scenario $\sum_{\alpha} \epsilon^{\alpha}_{i} = 0$ and as such only one of the 18 CP-asymmetry parameters is plotted, in this case $\epsilon^{1}_{1}$. Each CP-asymmetry parameter $\epsilon^{\alpha}_{i}$ has similar behaviour due to the degenerate limit considered.} in line with expectations from the one generation scenario. The resonant enhancement can clearly be seen in the left-hand plot at $\mu\sim 10^{-3}-10^{-4}$~GeV. The right-hand plot demonstrates where the resonance condition is satisfied where $\Delta_{N}/\Gamma_{N} = (1-x_{ij})/\frac{1}{8\pi}\left(h_{\nu}^{\dag}h_{\nu}\right)_{jj}$ and $1-x_{ij}$ is defined in Eq.~\eqref{res2}.

However, while the net CP asymmetry generated per sterile neutrino decay may increase, this is accompanied by an increased rate of washout, ultimately preventing the generation of a sufficiently large final baryon asymmetry. This behaviour is obviously unaffected by the hierarchy among the SN masses and therefore will be the same for the hierarchical case. The strength of the washout could potentially be lowered in some pure ISS scenarios by resorting to special textures of $m_D,\,\mu$ or $M_R$. As an example if a mass splitting is introduced by hand in $M_R$ (in the degenerate scenario) an additional resonant contribution will exist independent of $\mu$ without generating an increased washout along with it. Such textures would require additional symmetries in the theory such as a discrete flavour symmetry in order to justify their existence.

\begin{figure}
\centering
\includegraphics[width=0.7\linewidth]{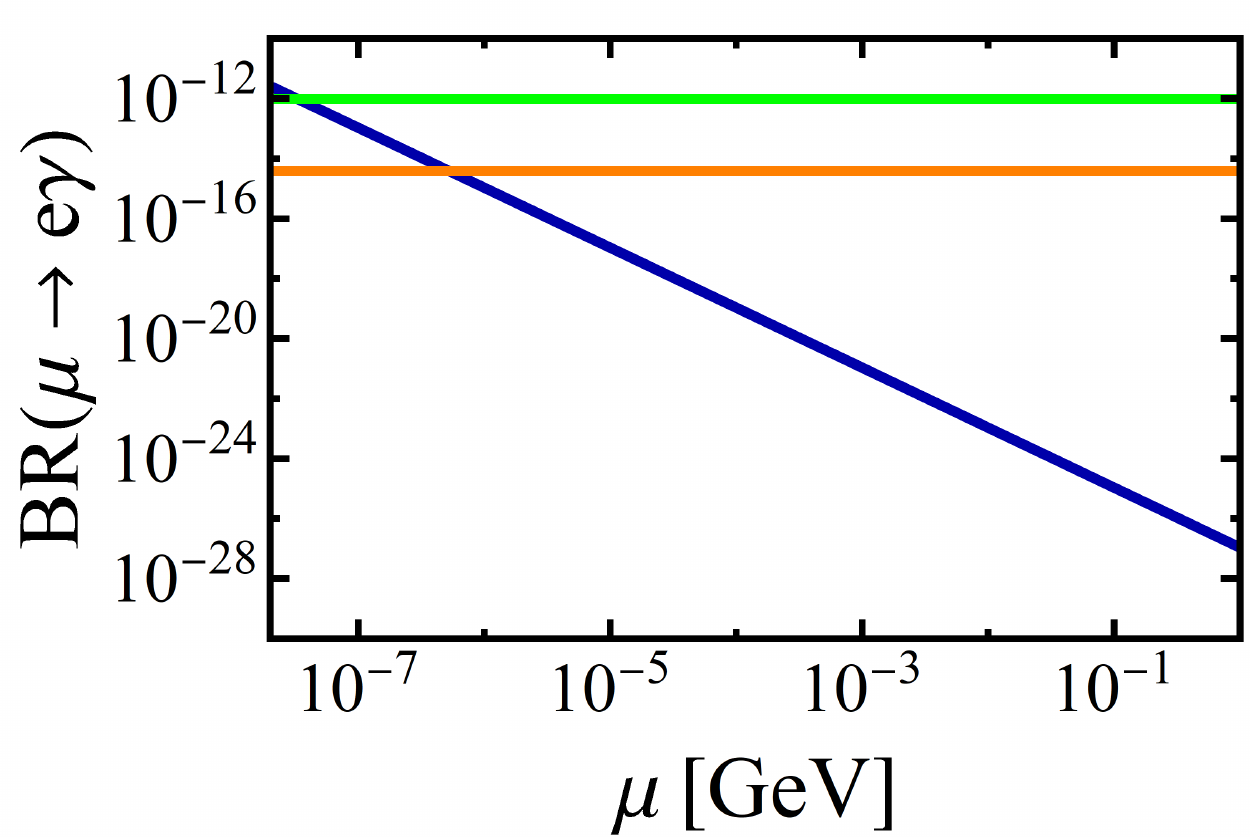}
\caption{A plot of the branching ratio of $\mu \rightarrow e \gamma$ in the low-scale ISS model as a function of $\mu$ (blue line). The green line is the current upper limit, while the orange line is the combined future limit expected from MEGII and the $\mu \to e$ conversion experiments, in particular COMET. While currently unconstrained, future cLFV experiments will begin to probe the parameter space of these low-scale models. Due to the minimal flavour-violating assumption taken, the size of the LFV is tied to the size of $m_D$.}
\label{muegaminv}
\vspace{0.5cm}
\end{figure}

Finally, though unable to account for baryogenesis, the discovery potential of the ISS through cLFV is estimated in Fig.~\ref{muegaminv}, which plots $\rm{BR}(\mu \rightarrow e \gamma)$.\footnote{As discussed above, in regimes where the photon diagram dominates, $\mu \rightarrow e \gamma$ and $\mu-e$ conversion are complementary.} The current limits placed on cLFV processes involving electrons does not constrain the ISS model. The parameter space which MEGII, COMET and Mu2E is predicted to probe is already constrained from unitarity violation experiments. Due to the minimal flavour-violating scenario considered, the cLFV effects are generated through the Casas-Ibarra parametrisation of $m_D$ (rather than imposing large cLFV effects through non-diagonal entries in $M_R$ or $\mu$) which will therefore have the largest contributions in regions of small $\mu$ for fixed $M_R$.

%\FloatBarrier

%%%%%%%%%%%%%%%%%%%%%%%%%%%%%%%%%%
\subsection{Inverse + Linear Seesaw Case}
%%%%%%%%%%%%%%%%%%%%%%%%%%%%%%%%%%%%

As in  Sec.~\ref{inverse}, we first consider a one-generation example to estimate how inclusion of the LSS term influences the strength of the washout relevant for thermal leptogenesis. Due to the linear nature of the LSS term of Eq.~\ref{linearmasslight} compared to the ISS term of Eq.~\ref{inversemasslight}, its contribution dominates in the light neutrino mass generation through $m_D$. The washout efficiency of the ISS+LSS scenario can therefore be estimated by considering the pure $1+1$ LSS case,
\begin{equation}
\label{onegenlin}
M_{\rm{lin,1g}} = \begin{pmatrix}
0 & m_D & m_L \\
m_D & 0 & M_R \\
m_L & M_R & 0 \end{pmatrix}.
\end{equation}
In this limiting case, the (non-unique) matrix rotation into the mass basis of the heavy sterile neutrinos above the critical temperature can be performed exactly and takes the form
\begin{equation}
\label{onegenlinrot}
M_\text{lin} = U_{\rm{lin}}^T M_{\rm{lin}}  U_{\rm{lin}} =  \begin{pmatrix}
0 & \left(\frac{1}{2}-\frac{i}{2}\right)\left(m_D+i m_L\right) & \left(\frac{1}{2}+\frac{i}{2}\right)\left(m_D-i m_L\right) \\
\left(\frac{1}{2}-\frac{i}{2}\right)\left(m_D+i m_L\right) & M_R & 0 \\
\left(\frac{1}{2}+\frac{i}{2}\right)\left(m_D-i m_L\right) & 0 & M_R \end{pmatrix},
\end{equation}
with
\begin{equation}
\label{Ulinrot}
U_{\rm{lin}} = \begin{pmatrix}
1 & 0 & 0 \\
0 & \left( \frac{1}{2}-\frac{i}{2}\right) & \left( \frac{1}{2}+\frac{i}{2}\right) \\
0 & \left( \frac{1}{2}+\frac{i}{2}\right) & \left( \frac{1}{2}-\frac{i}{2}\right) \end{pmatrix}.
\end{equation}
The neutrino oscillation data is recovered with a Dirac mass matrix of the form
\begin{equation}
m_D = -\frac{m_n M_R}{2 m_L} 
\end{equation}
which can be used to estimate the strength of the washout through the parameter,
\begin{equation}
\label{onegenlinwash}
K^{1gen}_{N_i} = \frac{\Gamma_{N_1}}{H(T=M_{N_i})} = \frac{\sqrt{45} M_p}{16\pi^{5/2} v^2 g_{*}^{1/2} M_R}\left| \left(\frac{1}{2}\mp \frac{i}{2}\right)\left(\frac{2 m_n M_R}{m_L} \pm i m_L\right) \right|^2.
\end{equation}

\begin{figure}[b]
\centering
\includegraphics[width=0.7\linewidth]{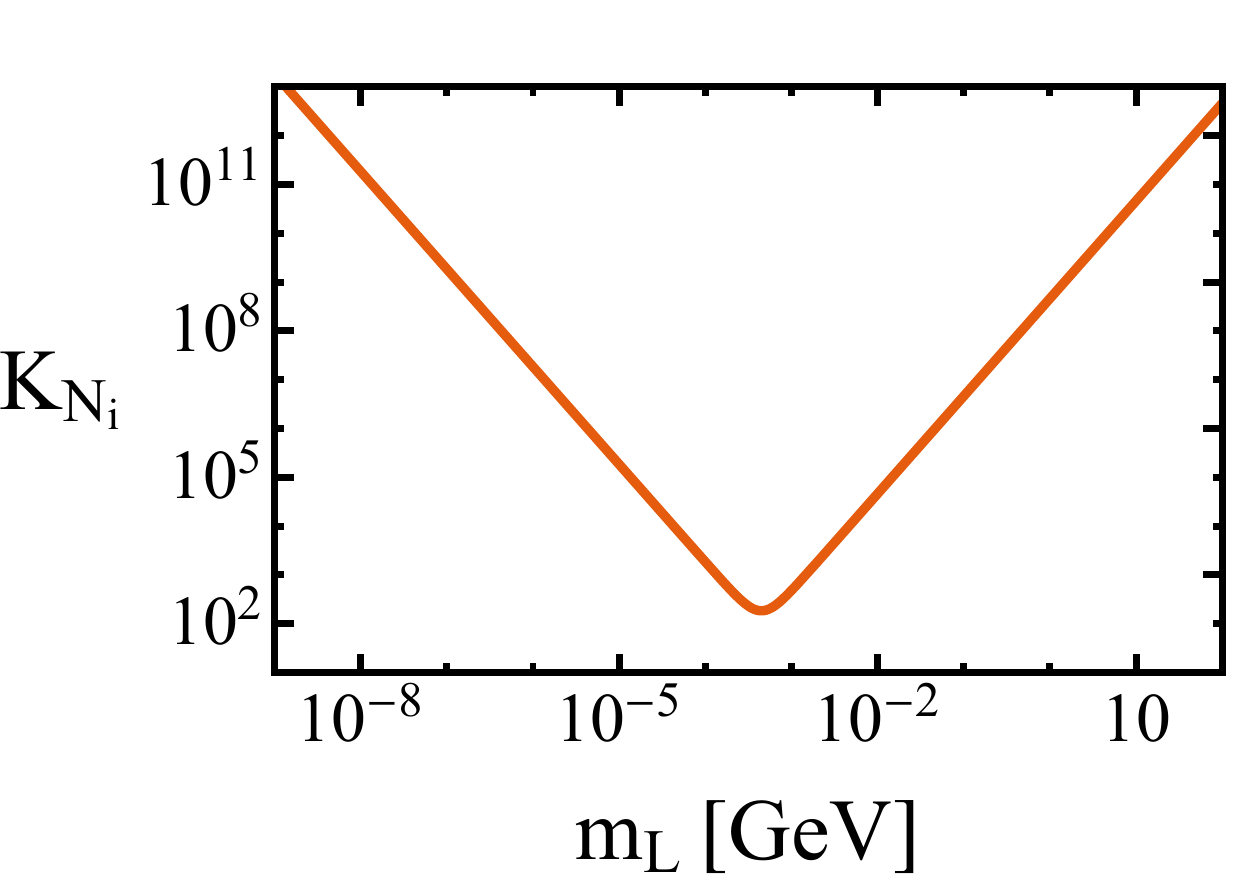}
\caption{Log-Log plot of the washout parameter as a function of  $m_L$ in the one generation LSS. The washout is minimised in this case for $m_L \sim 4.5\times 10^{-4}\,\rm{[GeV]}$ with $K_{N_i} \sim 100$, which is much lower than in the pure ISS case.}
\label{onegenwash-lin}
\end{figure}

Figure~\ref{onegenwash-lin} plots the washout parameter in the pure one-generation LSS as a function of $m_L$. Similarly to the case of the pure ISS, for $\mathcal{O}(\rm{TeV})$ scale sterile neutrinos, the washout remains strong. However, as a result of the mixing between $\nu_R$ and $S_L$, lower values of the washout exist. More importantly, the parameter determining the strength of the washout, $m_L$, does not impact the mass of the heavy sterile neutrinos (above the critical temperature), by contrast with the case of the ISS. Therefore when both $\mu$ and $m_L$ are turned on, $m_L$ will largely be responsible in determining the neutrino masses and couplings (and therefore largely responsible in determining the washout where $\mu$ will be sub-dominant), whereas $\mu$ will introduce a mass splitting between the heavy sterile neutrinos. Therefore when both parameters are switched on, resonant Dirac-phase leptogenesis is possible at relatively low values of strong washout.
%\FloatBarrier

%%%%%%%%%%%%%%%%%%%%%%%%%%%%%%%
\subsubsection{Degenerate Limit}
%%%%%%%%%%%%%%%%%%%%%%%%%%%%%%

\begin{figure}[t]
\centering
\includegraphics[width=0.45\linewidth]{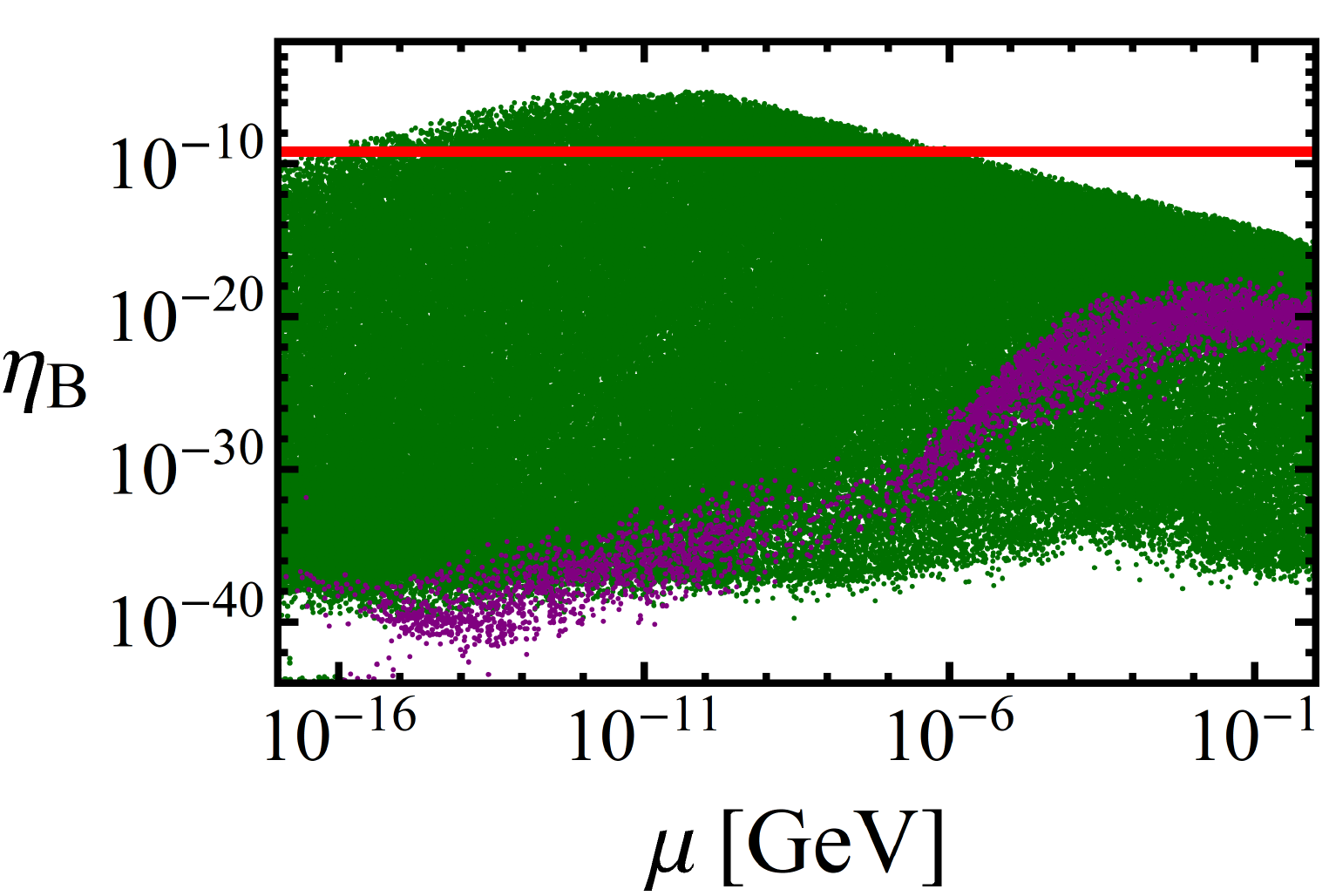}\hspace{0.5cm}
\includegraphics[width=0.45\linewidth]{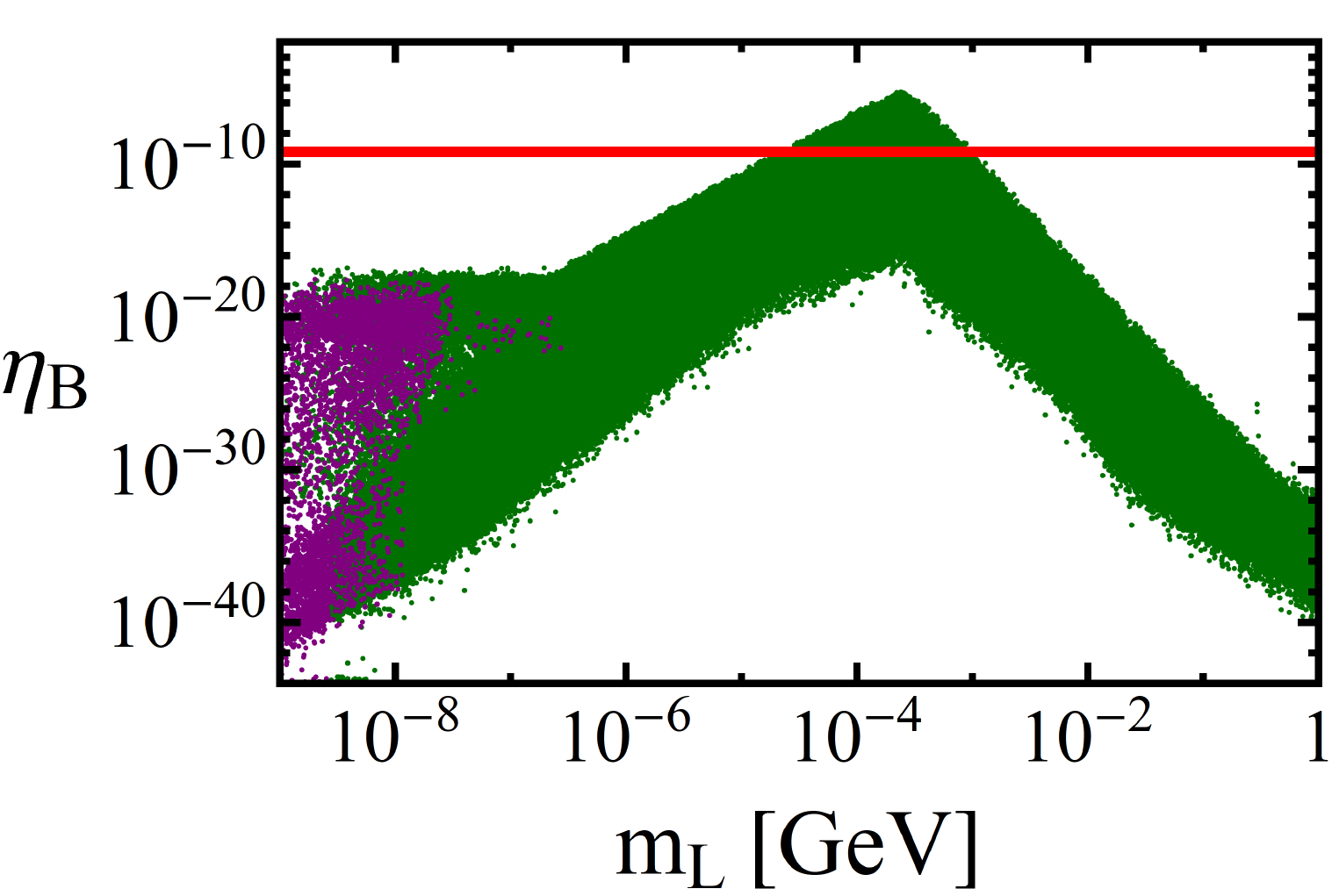}
\caption{The generated baryon asymmetry for the degenerate case as a function of the Majorana mass $\mu$ (\textbf{Left Panel}) and the lepton number violating parameter $m_L$ (\textbf{Right Panel}). The points in purple, which correspond to large values of $m_D$, are excluded from unitarity violation limits on the PMNS matrix. In red is the required BAU, $\eta_B \sim 6.21 \times 10^{-10}$. Generation of the correct asymmetry is possible in regions close to where the washout is minimised (but still in the strong regime) and requires $10^{-5} \lesssim m_{L}/\rm{GeV} \lesssim 10^{-3}$ along with $10^{-16} \lesssim \mu/\rm{GeV} \lesssim 10^{-6}$ where resonance occurs due to the small mass splitting.}
\label{3gendeginvasym}
%\vspace{0.5cm}
\end{figure}

Figure~\ref{3gendeginvasym} plots the baryon asymmetry generated as functions of $\mu$ and $m_L$ for the degenerate case where all $M_R$ masses equal $1\ \text{TeV}$, as per Eq.~\ref{mr}. Similarly to the pure ISS case, all points generate an asymmetry of the correct sign for $\delta_{CP} = 3\pi /2$. The points in purple are ruled out by non-unitarity of the PMNS matrix and are characterised by low values of $m_L$, as can be seen in the right-hand plot. In order to generate the correct asymmetry for these TeV scale sterile neutrinos via the Dirac phase, two ingredients are necessary: the LSS term, $m_L$, needs to be close to the region in which the washout is minimised, $ 10^{-5} \lesssim m_L/\text{GeV} \lesssim 10^{-3}$, and the ISS term, $\mu$, is required to be in the broad range $10^{-16} \lesssim \mu/\rm{GeV} \lesssim 10^{-6}$ in order for the mass splittings amongst the heavy sterile neutrinos to provide a resonant contribution to the asymmetry generation.

\begin{figure}[t]
\captionsetup[subfigure]{justification=centering,labelformat=empty}
\centering
\subfloat[1][\quad\quad\quad$m_L<10^{-8}$]{
  \includegraphics[width=0.32\textwidth]{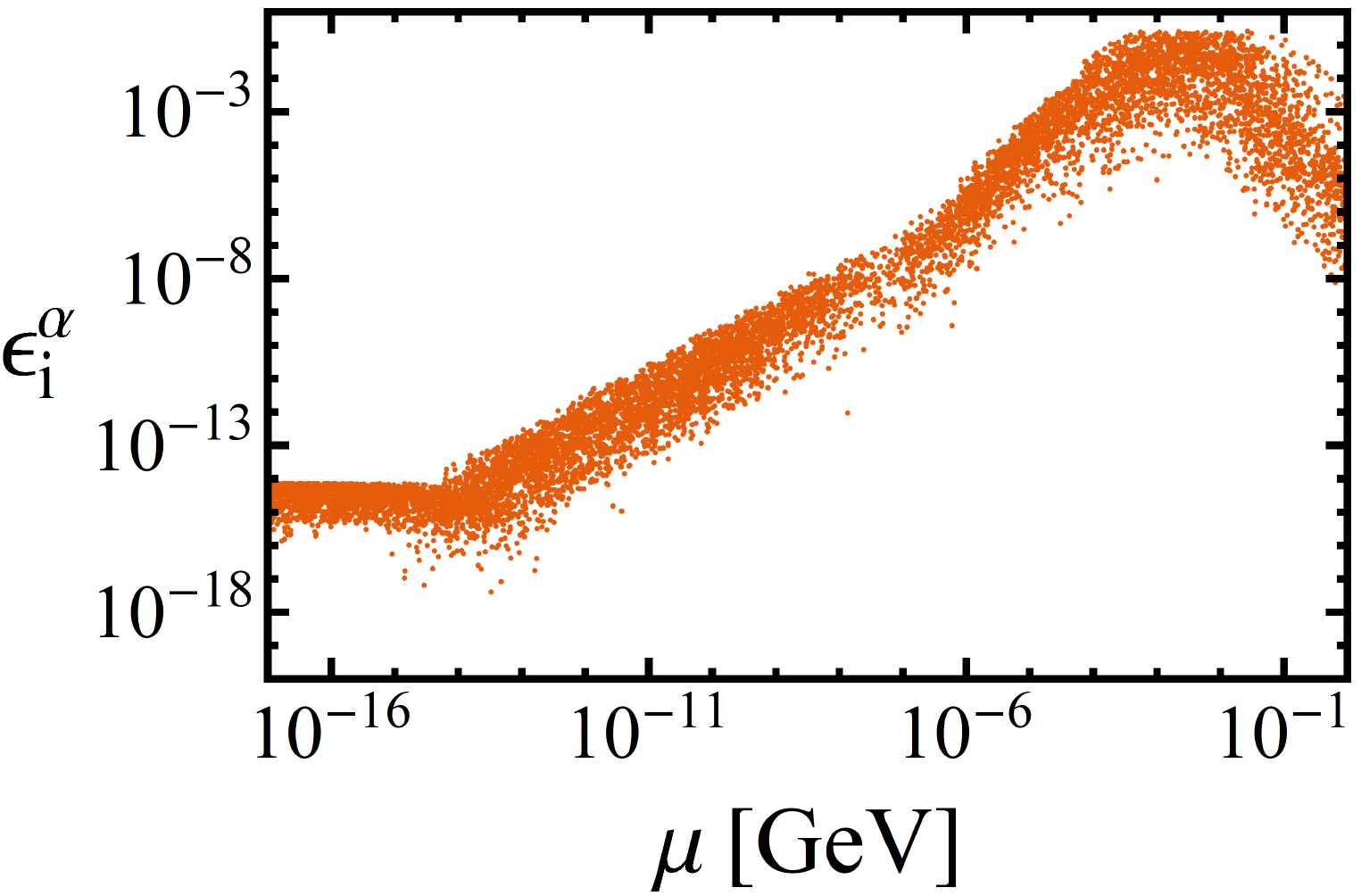}
  }
\subfloat[2][\quad\quad\quad$10^{-8}<m_L<10^{-7}$]{
  \includegraphics[width=0.32\textwidth]{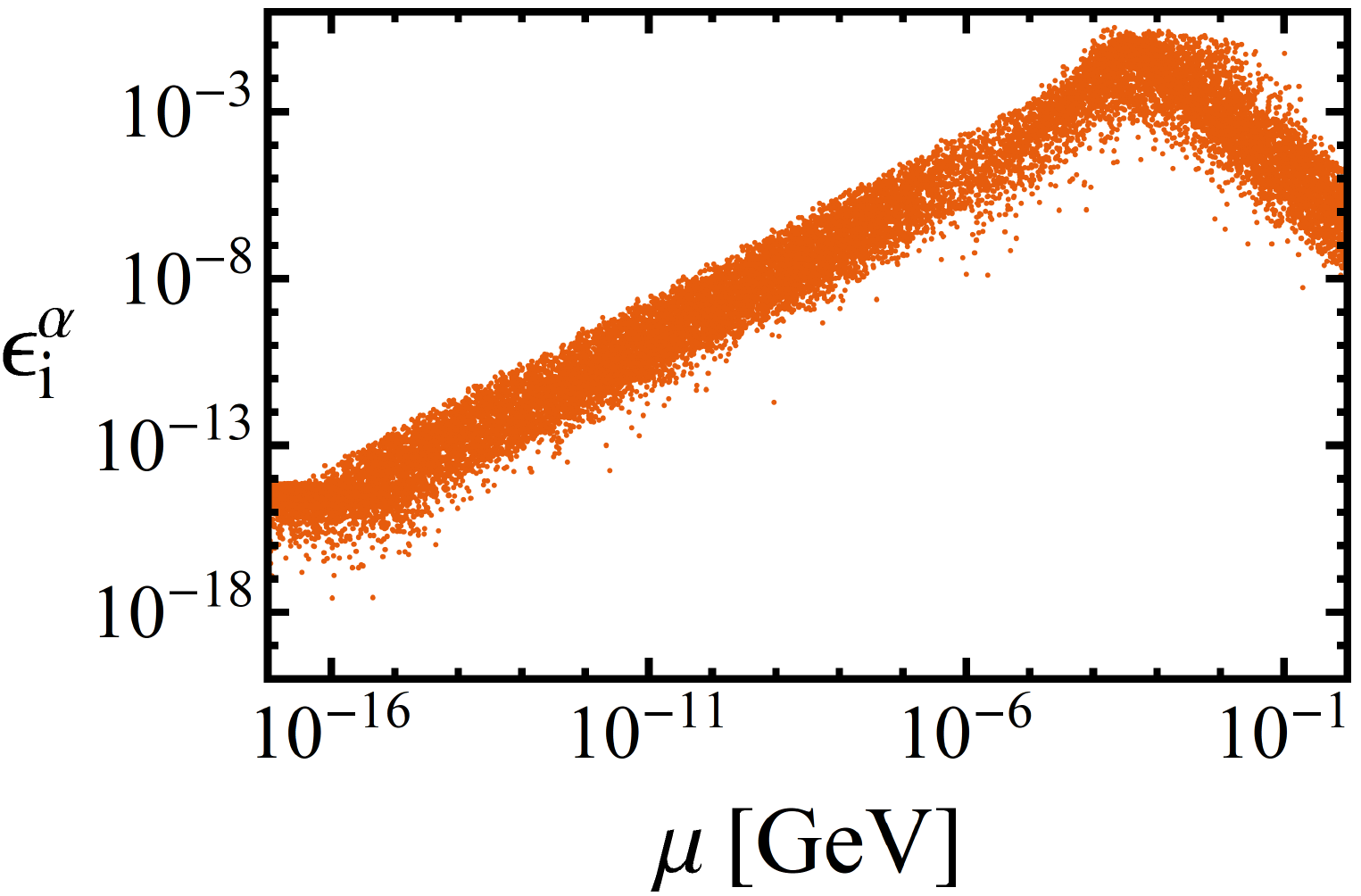}
  }
\subfloat[3][\quad\quad\quad$10^{-7}<m_L<10^{-6}$]{
  \includegraphics[width=0.32\textwidth]{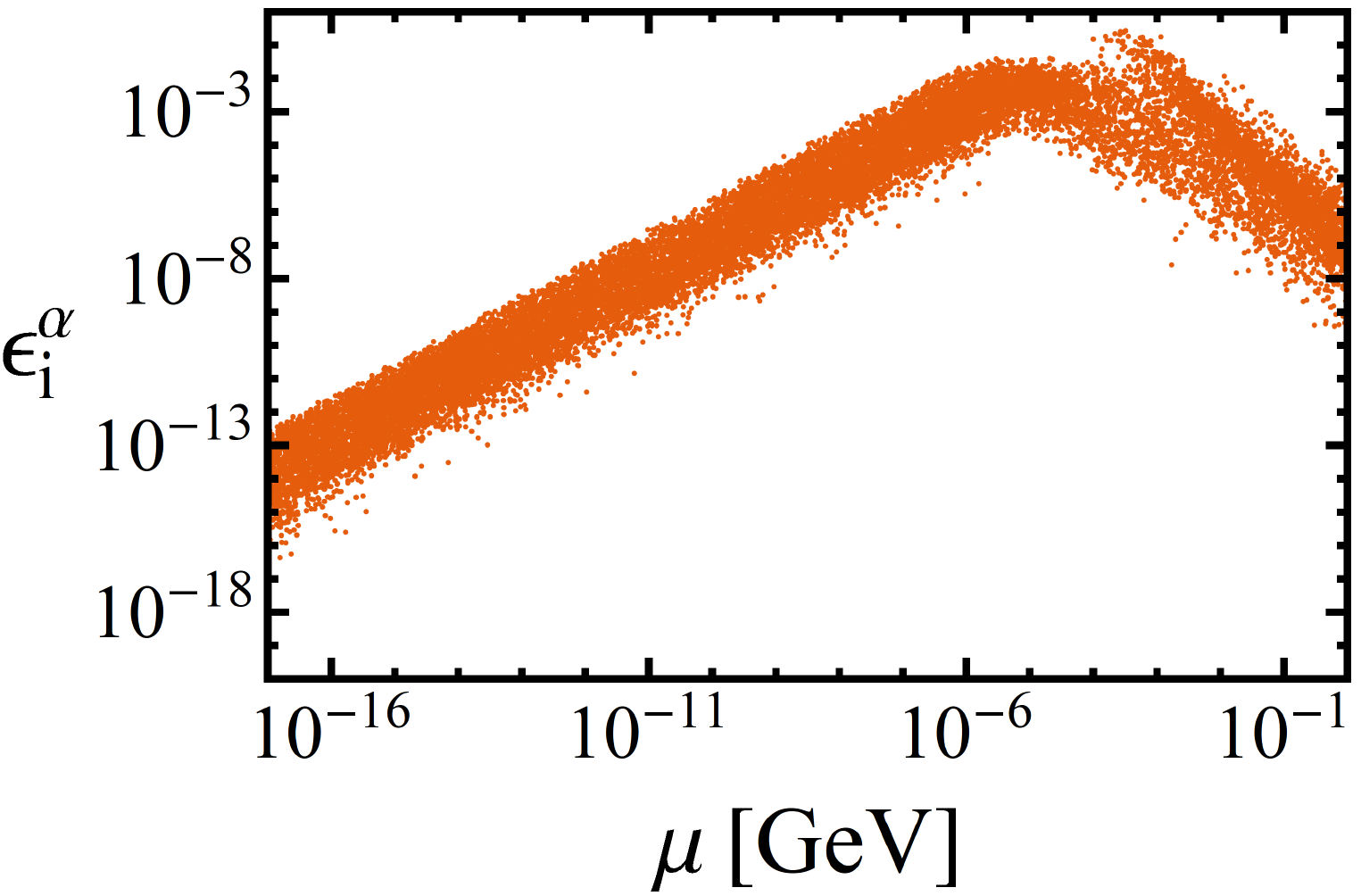}
  }  
\newline
\centering
\subfloat[4][\quad\quad\quad$10^{-6}<m_L<10^{-5}$]{
  \includegraphics[width=0.32\textwidth]{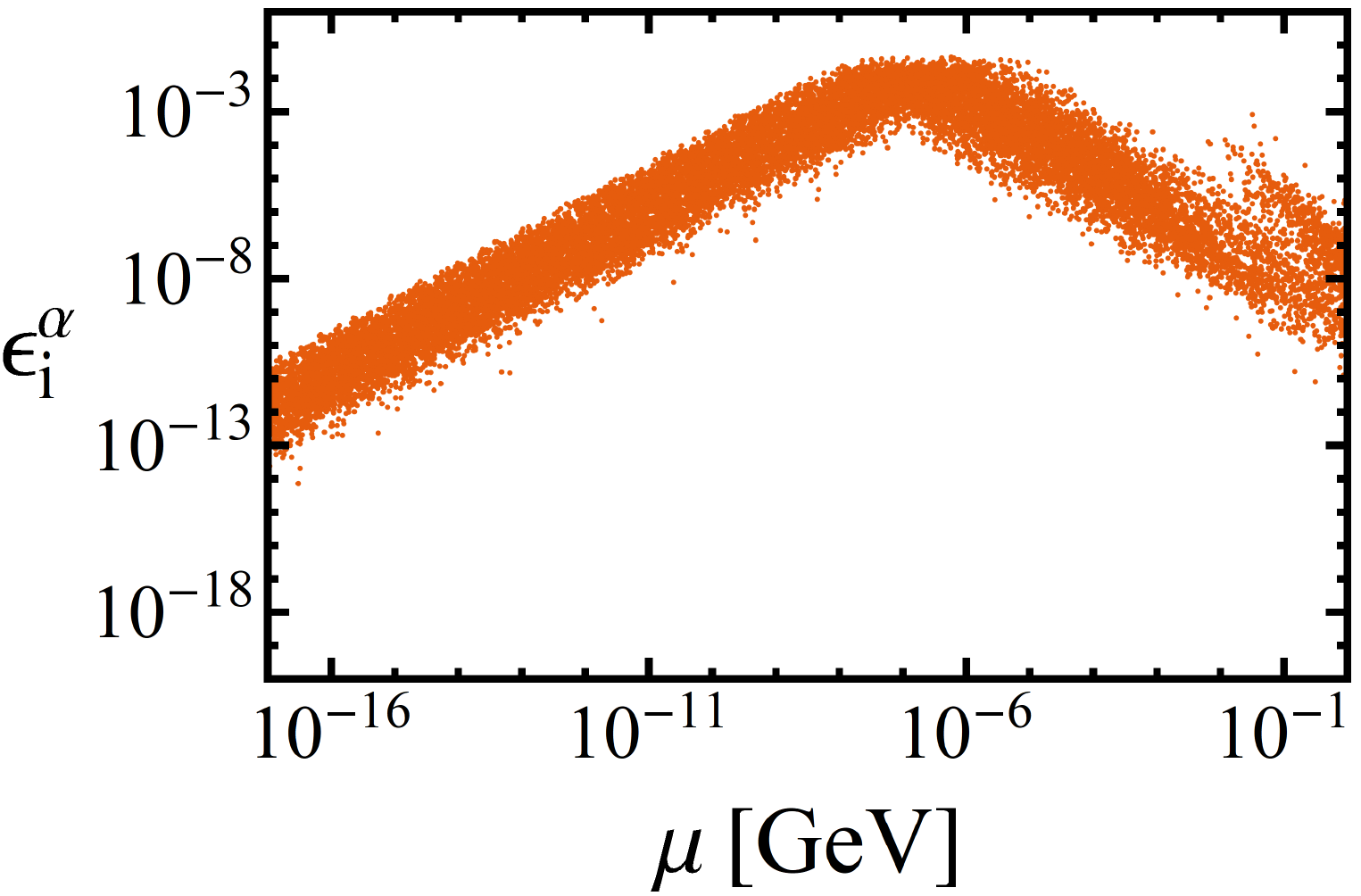}
  }
\subfloat[5][\quad\quad\quad$10^{-5}<m_L<10^{-4}$]{
  \includegraphics[width=0.32\textwidth]{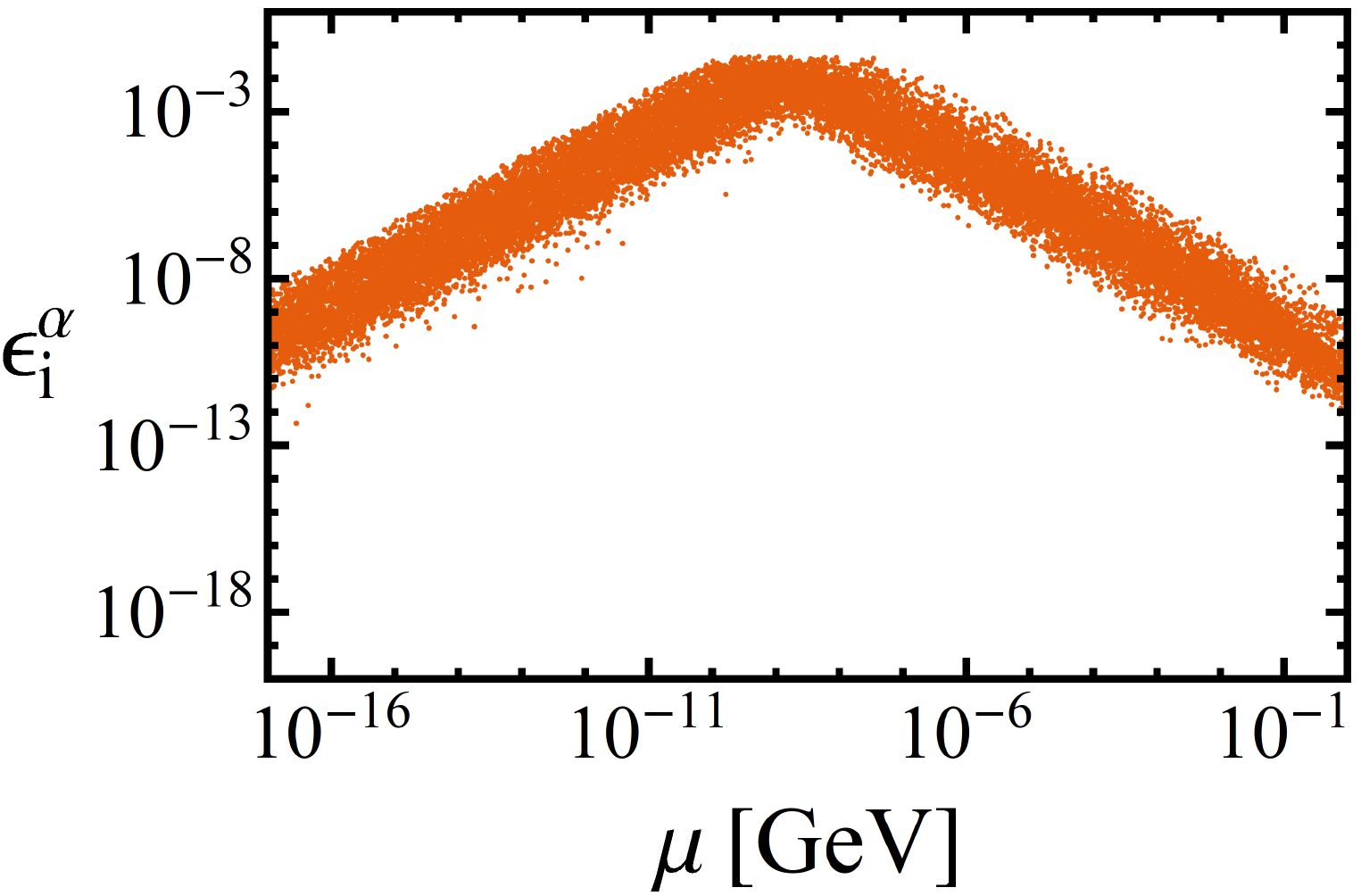}
  }
\subfloat[6][\quad\quad\quad$10^{-4}<m_L<10^{-3}$]{
  \includegraphics[width=0.32\textwidth]{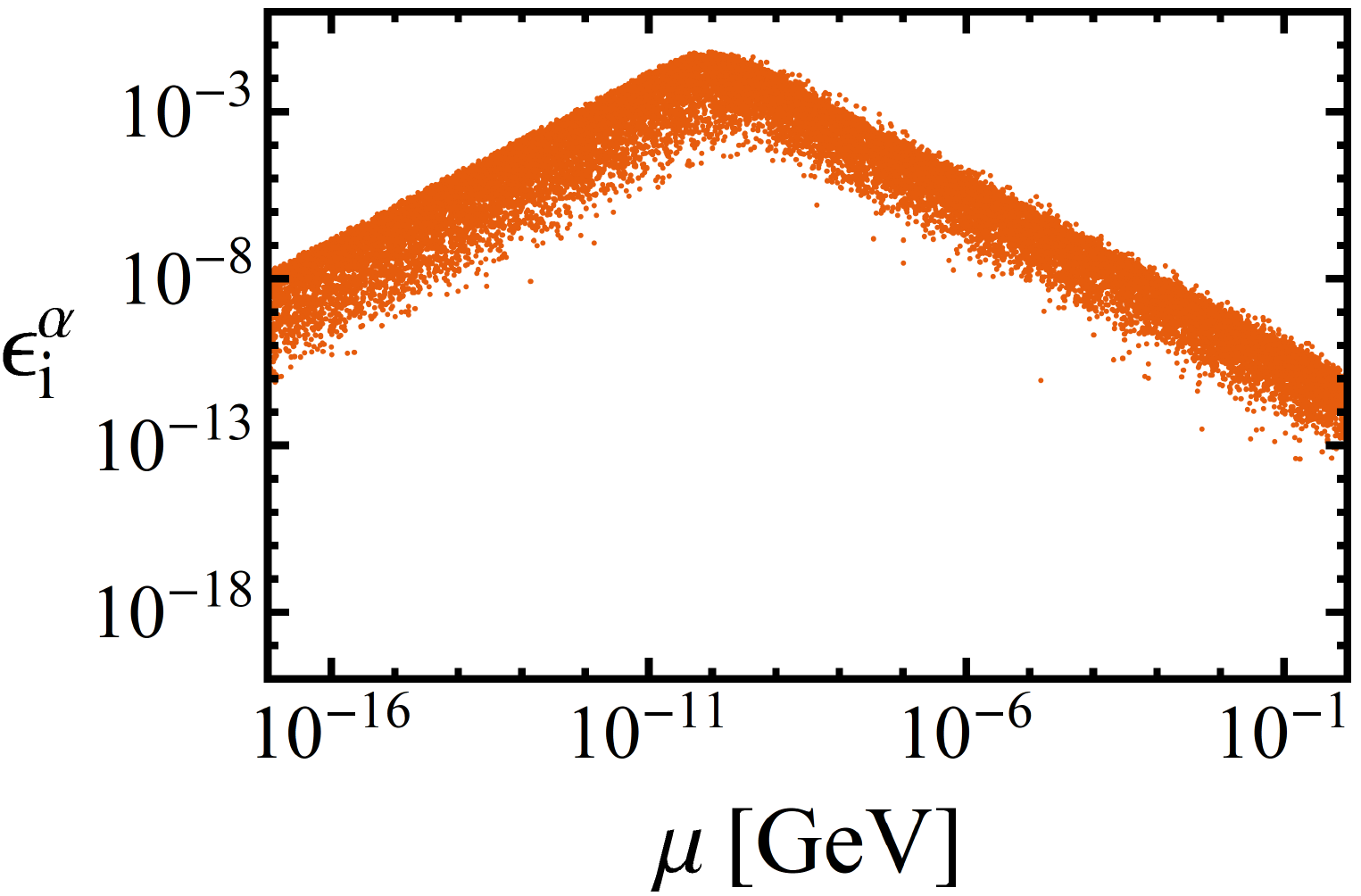}
  }  
\newline
\centering
\subfloat[7][\quad\quad\quad$10^{-3}<m_L<10^{-2}$]{
  \includegraphics[width=0.32\textwidth]{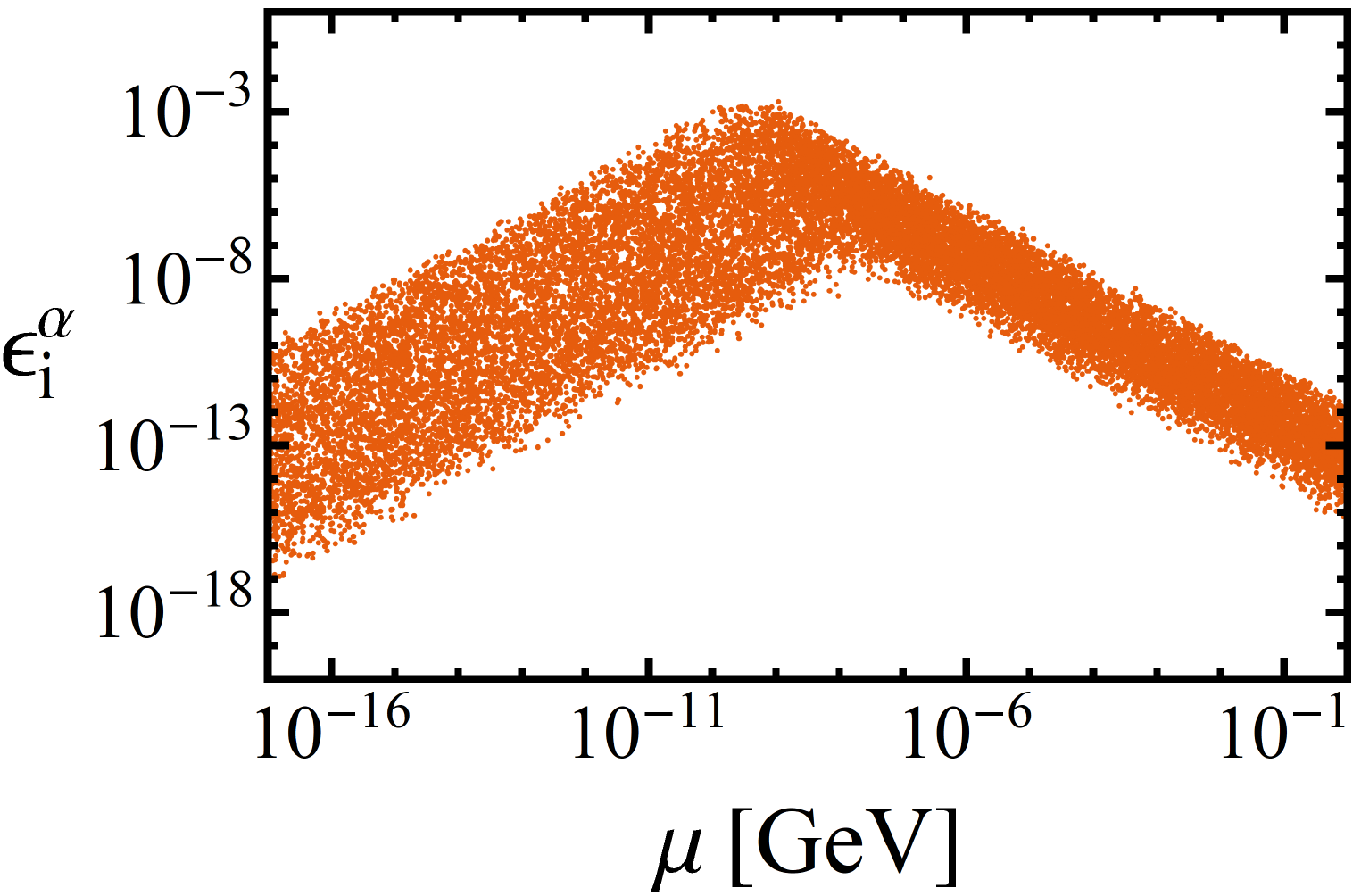}
  }
\subfloat[8][\quad\quad\quad$10^{-2}<m_L<10^{-1}$]{
  \includegraphics[width=0.32\textwidth]{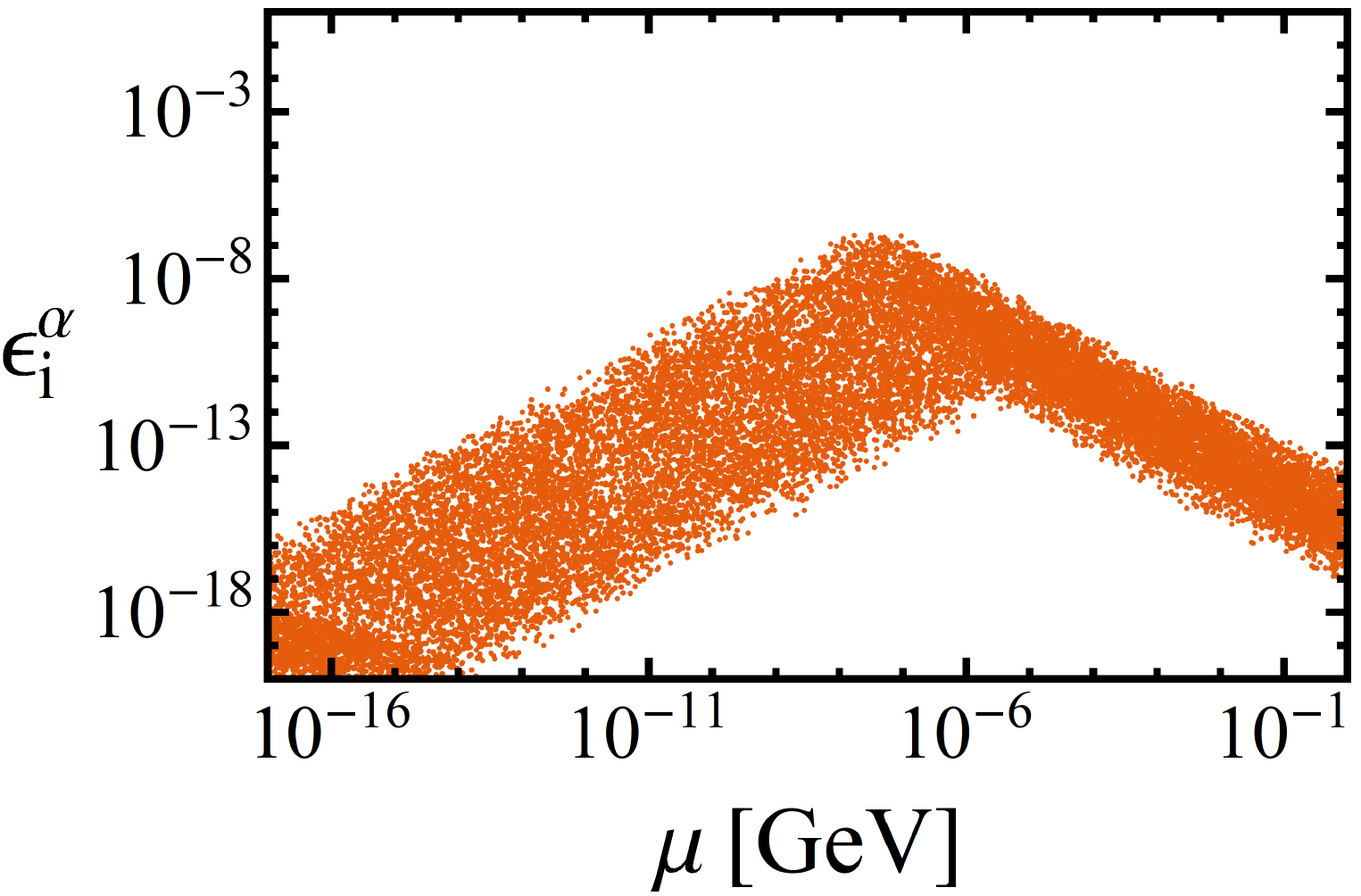}
  }
\subfloat[9][\quad\quad\quad$10^{-1}<m_L$]{
  \includegraphics[width=0.32\textwidth]{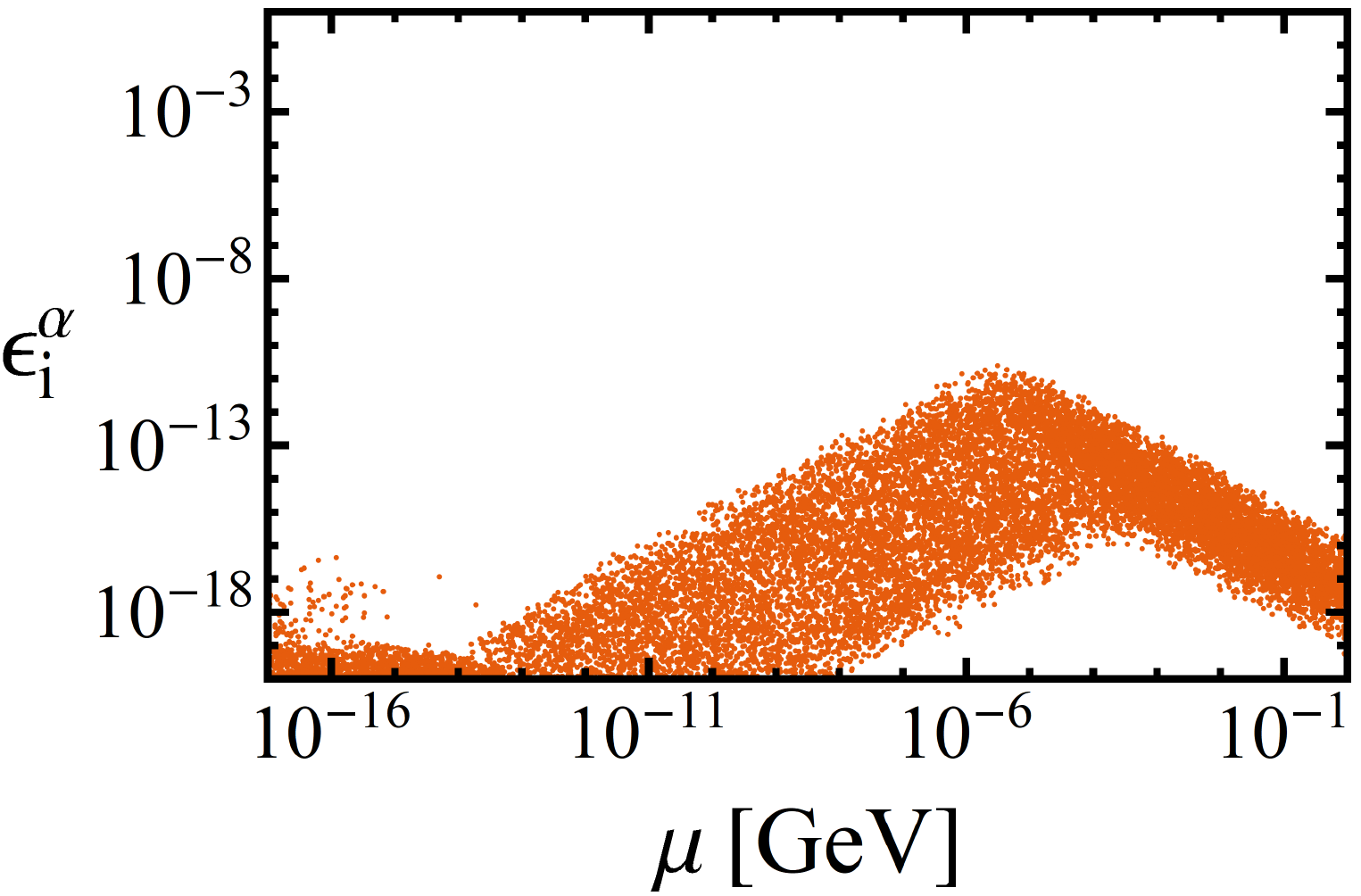}
  }  
\caption{Log-Log plots of CP asymmetry parameter $\epsilon_{i}^{\alpha}$ as a function of $\mu$ for fixed values of $m_L$ described above. In each case a resonant enhancement is noticeable as $\mu$ for a particular mass splitting amongst the SNs.}
\label{epsdeg}
\end{figure}

Figure~\ref{epsdeg} plots snapshots as a function of $m_L$ of the CP-asymmetry parameter $\epsilon_{i}^{\alpha}$ as a function of $\mu$ for fixed bands of $m_L$. While a clear resonance exists as $\mu$ is lowered for all snapshots, it is only for regions close to where the washout is minimised that the resonance effect is strong enough to generate a sufficient asymmetry. This serves to illustrate that for $\mathcal{O}$(TeV) scale extended type-I seesaw models, both the LSS and ISS contributions are required in order for thermal leptogenesis to be possible, potentially providing a unique experimental signature of pseudo-Dirac sterile fermions with leptonic decay channels.

%\begin{figure}[!htbp]

\begin{figure}[t]
\centering
\includegraphics[width=0.7\linewidth]{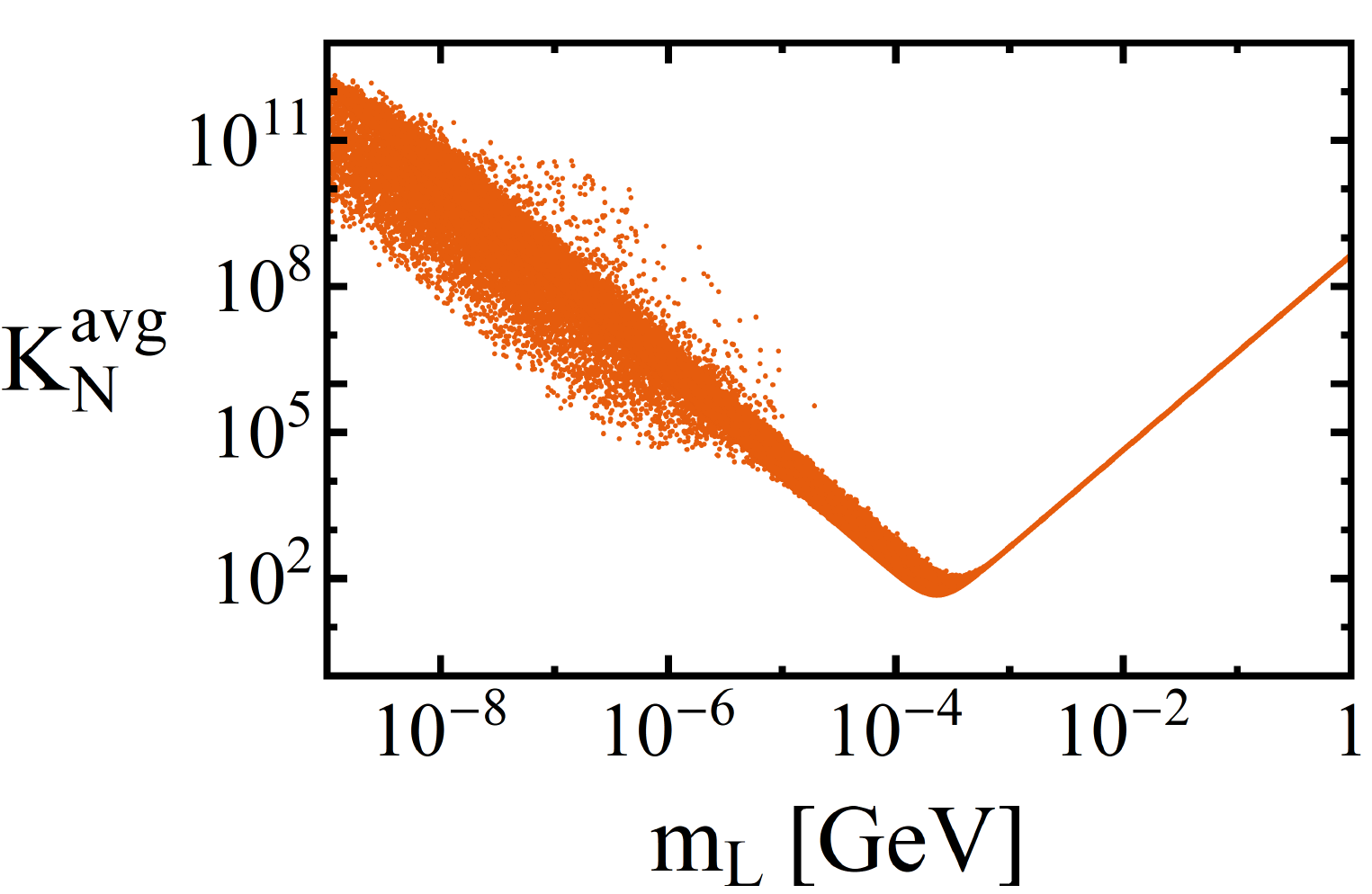}
\caption{Log-log plot of the average washout parameter $K_{N}^{\rm{avg}} = \frac{1}{6} \sum_{i} K_{N_i}$ in the three-generation case as a function of $m_L/\rm{GeV}$. 
Comparison of this with the one-generation case of the pure LSS as in Fig.~\ref{onegenwash-lin} justifies the claim that the ISS term is sub-dominant to the LSS in the generation of $m_D$.}
\label{washdublin3}
\end{figure}

Figure~\ref{washdublin3} plots the average washout parameter in the three generation case, which closely resembles the pure LSS one-generation plot of Fig.~\ref{onegenwash-lin}. From Fig.~\ref{washdublin3}, it is evident that contributions to the washout (and therefore the couplings $y_{\nu}$) from $\mu$ are sub-dominant compared to $m_L$, justifying the assumption that $m_L$ is almost solely responsible for determining the strength of the washout, even at low values.

\begin{figure}[b]
\centering
\includegraphics[width=0.7\linewidth]{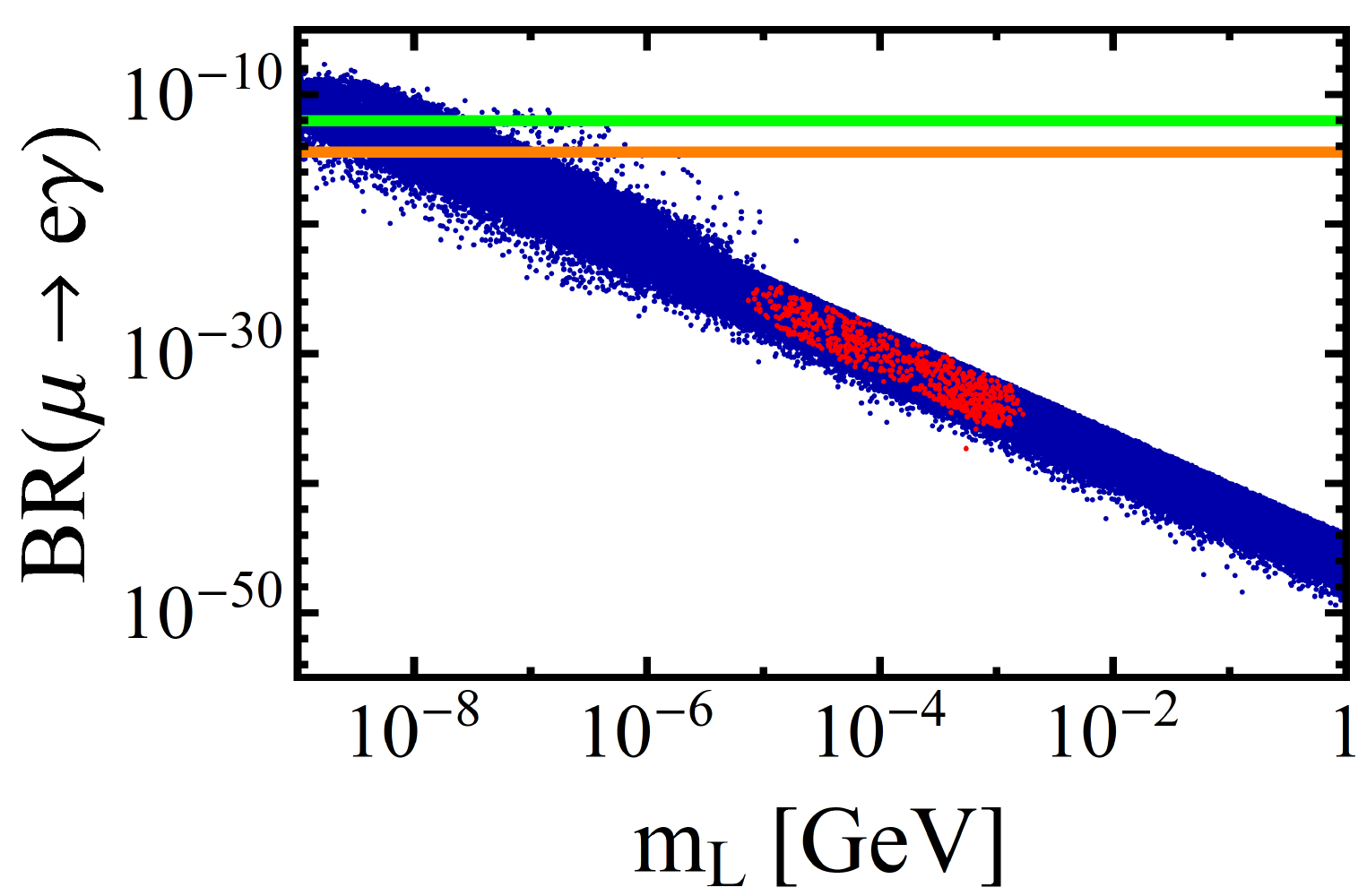}
\caption{A plot of contribution to $\mu \rightarrow e \gamma$ in the low-scale ISS+LSS model for the degenerate case. The horizontal green line is the current experimental upper limit and the orange line is the 
combined future limit expected from MEGII and the $\mu \to e$ conversion experiments, in particular COMET, assuming the penguin diagram contribution dominates. 
Regions of small $m_L$ in the ISS+LSS model are currently constrained. The points in red generate a baryon asymmetry in the 
required region, $4\times 10^{-10} < \eta_{B} < 8 \times 10^{-10}$. This region is currently unconstrained and will remain so in the future. The points in blue give rise to asymmetry values that are too small.}
\label{muegamdubdeglin}
\end{figure}

In the case of Dirac-phase leptogenesis, a region of parameter space exists for both $m_L$ and $\mu$ for the well-motivated choice $\delta_{CP} = 3\pi/2$. Figure~\ref{muegamdubdeglin} plots $\rm{BR}(\mu \rightarrow e \gamma)$ as a function of $m_L$. Similarly to the pure ISS case, current and near-future cLFV experiments will not probe regions where Dirac-phase leptogenesis occurs.

%\FloatBarrier

%%%%%%%%%%%%%%%%%%%%%%%%%%%%%%%%
\subsubsection{Hierarchical Limit}
%%%%%%%%%%%%%%%%%%%%%%%%%%%%%%%%%

\begin{figure}[t]
\centering
\includegraphics[width=0.45\linewidth]{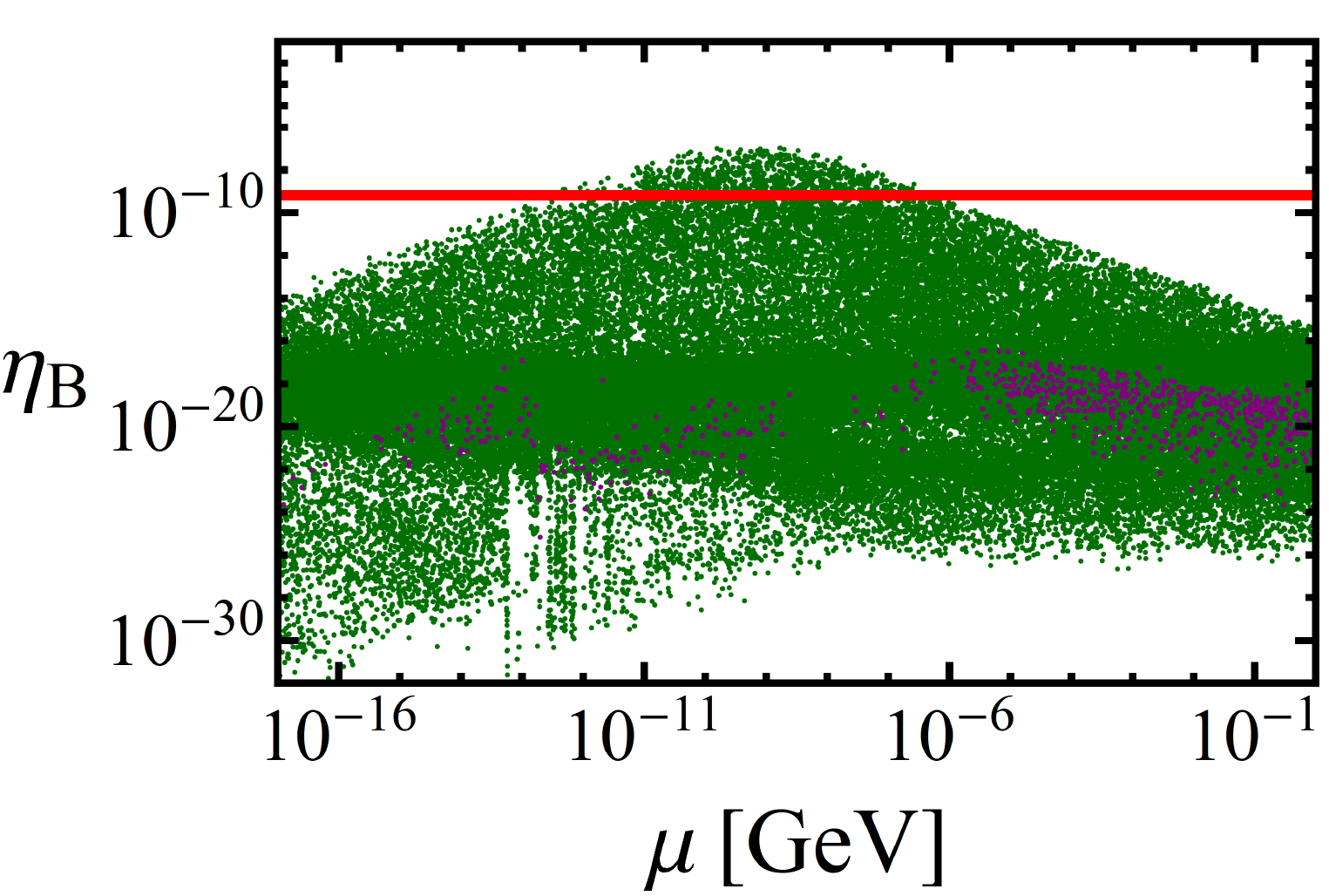}
\includegraphics[width=0.45\linewidth]{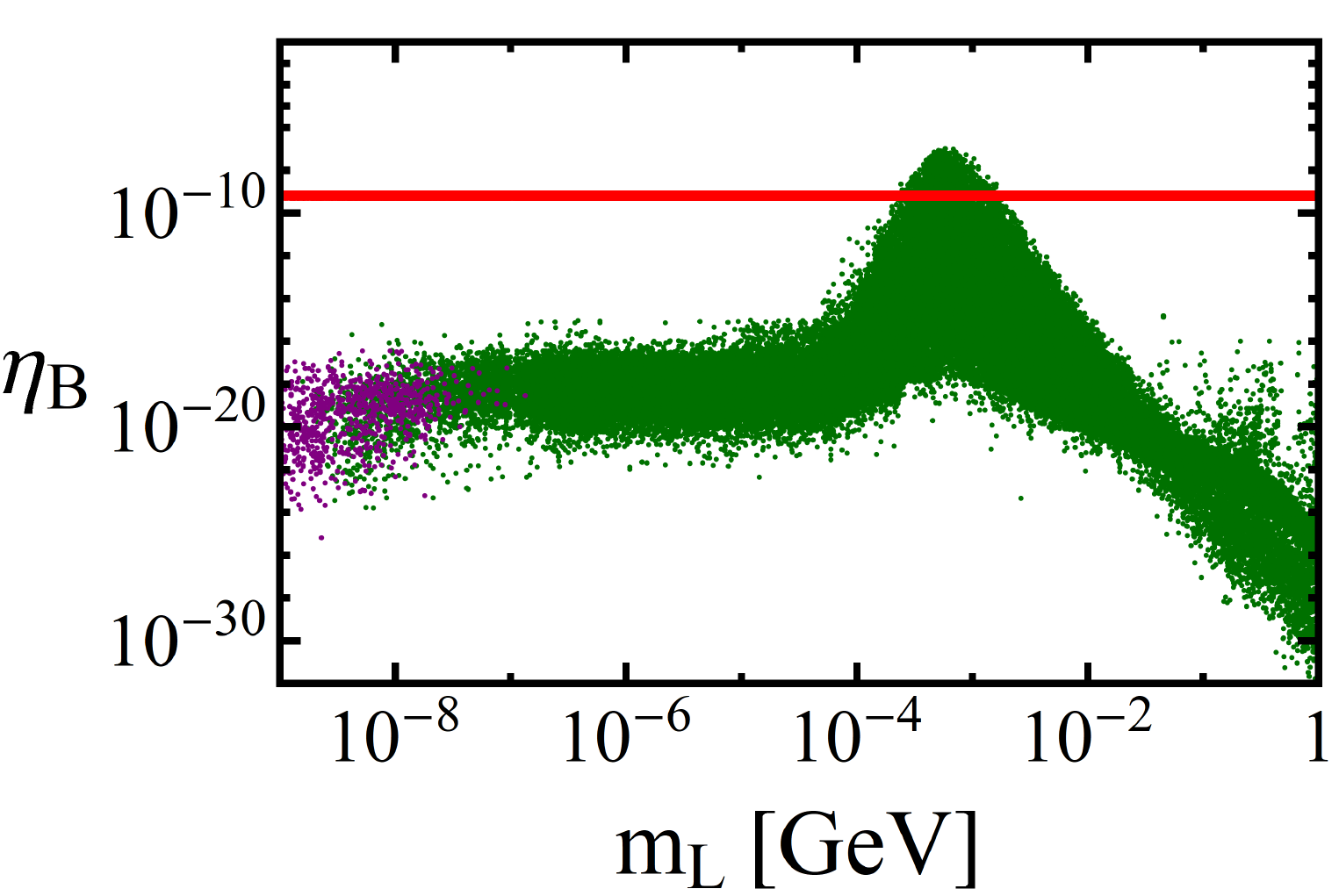}
\caption{The generated baryon asymmetry as a function of the Majorana mass $\mu$ (\textbf{Left Panel}) and the lepton number violating parameter $m_L$ (\textbf{Right Panel}) in the hierarchical case. The hierarchical case 
behaves similarly to the degenerate case with slightly more constrained allowed couplings, $5 \times 10^{-3} \lesssim m_{L}/\rm{GeV} \lesssim 5 \times 10^{-2}$ along with $10^{-12} \lesssim \mu/\rm{GeV} \lesssim 10^{-6}$ compared to the degenerate case. The points in purple are excluded by unitarity violation limits on the PMNS matrix.}
\label{3genhierlinasym}
\end{figure}

In the hierarchical case, in which the heavier SNs are set to 5 TeV as justified in Sec.~\ref{hierlimit}, Dirac-phase leptogenesis is also possible, as illustrated in Fig.~\ref{3genhierlinasym}. The asymmetry is produced via the same mechanism as in the degenerate case: the relevant area of parameter space is where $m_L$ minimises the strength of the washout, and the resonant enhancement is due to mass splittings driven by a small $\mu$. Unsurprisingly, this occurs in a smaller region of $\mu$ and $m_L$ than before, $5 \times 10^{-3} \lesssim m_L/\rm{GeV} \lesssim 5 \times 10^{-2}$ and $10^{-12} \lesssim \mu/\rm{GeV} \lesssim 10^{-6}$, compared to the degenerate case, making the hierarchical case more constrained. This result is similar to that found for the type-I Dirac-phase case in Ref.~\cite{Anisimov:2007mw}.

\begin{figure}[t]
\centering
\includegraphics[width=0.6\linewidth]{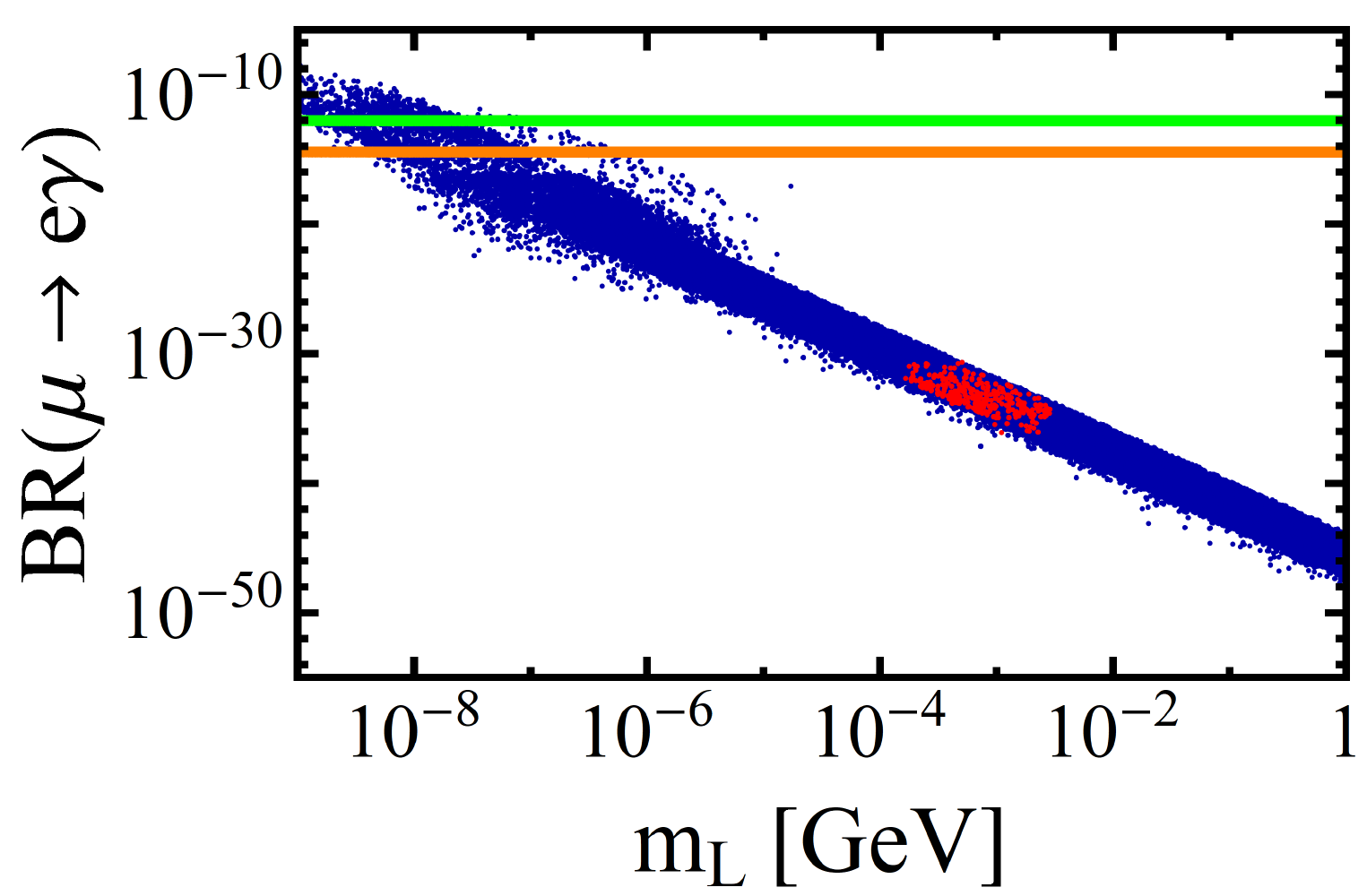}
\caption{A plot of contribution to $\mu \rightarrow e \gamma$ in the low-scale hierarchical ISS+LSS model. In green is the current limit from MEG and in orange is the combined future limit from MEGII and the $\mu \to e$ conversion experiments, in particular COMET, assuming the penguin diagram contribution dominates. Regions of small $m_L$ in the ISS+LSS model are currently probed and constrained by MEG and future experiments will serve to slightly improve the sensitivity on $m_L$. The points in red generate a baryon asymmetry in the region $4\times 10^{-10} < \eta_{B} < 8 \times 10^{-10}$. This parameter region is unconstrained by current experiments and beyond the reach of future cLFV experiments. The points in blue give rise to asymmetry values that are too small.}
\label{muegamdubhierlin}
\end{figure}

The constraints from current and future cLFV are of similar strength as in the degenerate situation: low scale, viable Dirac-phase leptogenesis cannot be constrained by unitarity violation or cLFV processes as demonstrated in Fig.~\ref{muegamdubhierlin}. Singlet neutrinos may be produced at particle colliders however the classic signal of same-sign dileptons~\cite{Antusch:2016ejd} is suppressed due to the pseudo-Dirac nature of the SNs. Trilepton production has been studied in~\cite{Das:2014jxa} where constraints have been derived on the mixing between the active and sterile states $\left| B_{l N} \right|$. However, in our model the region of successful Dirac-phase leptogenesis has  $\left| B_{l N} \right| \sim 10^{-12}$, far too small to be constrained at the LHC.

%%%%%%%%%%%%%%%%%%%%%%%%%%%
\section{Conclusion}
%%%%%%%%%%%%%%%%%%%%%%%%%%%%%%

We have studied the thermal leptogenesis implications of the low-scale $3+3$ inverse-seesaw (ISS) and extended inverse-seesaw (ISS+LSS) models. Both scenarios provide a framework for TeV scale sterile neutrinos, with much larger Yukawa couplings required in order to account for current active neutrino oscillation data, in contrast to the vanilla type-I seesaw mechanism. The implications of purely Dirac-phase leptogenesis were considered, as values of $\delta_{CP} \neq 0,\pi$ will necessarily contribute to the asymmetry generation in any low-scale seesaw model, and $\delta_{CP}$ remains the only phase that can be measured by neutrino oscillation experiments. Currently there is a suggestive hint that it may be maximally violating, which might suggest that it has some role to play in generating the matter--anti-matter asymmetry. 

In the pure ISS scenario, the washout is too strong for asymmetry generation even when incorporating resonance effects where no special textures are assumed in the theory beyond diagonal couplings. In the ISS+LSS scenario, successful leptogenesis can be achieved due to the mass splitting between the sterile neutrinos providing a resonant enhancement of the asymmetry, a regime which is naturally motivated in the model if the LNV parameters are taken to be small from technical-naturalness arguments. Degenerate and hierarchical cases were considered, and in both situations predictions on the parameters $m_L$ and $\mu$ are made for successful Dirac-phase leptogenesis. Unsurprisingly, the hierarchical case appears slightly more constrained than the degenerate case.

The discovery potential of these models from precision tests was considered, particularly through probes of unitarity violation in the PMNS matrix and charged-lepton flavour-violating processes such as $\rm{BR}(\mu \rightarrow e \gamma)$ and $\mu \rightarrow e$ conversion. It was found that regions of successful Dirac-phase leptogenesis do not coincide with the current or future reach of such cLFV experiments or unitarity tests. Collider tests were also briefly considered, but the active-sterile mixing in regions of successful Dirac-phase leptogenesis is too low for near-future discovery. It is worth noting, however, that an extension to a left-right symmetric model, well-motivated by the neutrino mass basis choice we have employed, would increase the experimental signatures of such a scenario.

\acknowledgments
This work was supported in part by the Australian Research Council. TPD thanks Stephen J. Lonsdale and Thomas Hugle for helpful discussion along with Apostolos Pilaftsis and Avelino Vicente for valuable email correspondence.
\newpage
\appendix

\section{Decay and Scattering Rates}
\label{sec:Appendix A}

Following the literature, the following rescaled variables are used:
\begin{equation}\
\label{rescaled}
z = \frac{m_{N_1}}{T}\,, \;\;x = \frac{s}{m_{N_1}^2}\,, \;\; 
a_i = \left(\frac{m_{N_i}}{m_{N_1}}\right)^2\,, \;\; 
a_r = \left(\frac{m_{\rm IR}}{m_{N_1}}\right)^2 \simeq 10^{-5}\,, \;\;
c_i\ =\ \left(\,\frac{\Gamma_{N_i}}{m_{N_1}}\,\right)^2\; .
\end{equation}
where  $s$ is the Mandelstam variable, and $m_{\rm{IR}} = m_{\Phi} / M_{N_1}$ is an
infrared mass regulator for the t-channel whose value is set to $10^{-5}$ as discussed in Refs.~\cite{HahnWoernle:2009qn,Luty:1992un}.

The total decay width $\Gamma_{N_i}$ of the SNs is given by
\begin{equation}
\Gamma_{N_i}\ =\ \sum_{l=1}^{3} \Gamma^{\alpha}_{N_i} =\ \frac{m_{N_i}}{8\pi}\
\sum_{l=1}^{3} h_{li}^{*} h_{li} .
\end{equation}

The collision terms for $1\to 2$ and $2\to 2$ processes which appear in Eqs.~\ref{boltzn} and \ref{boltzl} are calculated in Ref.~\cite{Pilaftsis:2003gt} for the unflavoured case 
and additional flavour-specific processes are calculated in Ref.~\cite{Pilaftsis:2005rv}. These scattering rates will be summarised below without suppressed flavour indices.
The rate for a generic process $X\to   Y$ and its conjugate counterpart $\overline{X}\to \overline{Y}$ is defined as $\gamma^X_Y$.
For   the  $1\to   2$   process,\   $N_i\to   L\Phi$  or   $N_i\to
L^C\Phi^\dagger$, 
$\gamma^{X}_{Y}$ is given by
\begin{eqnarray}
\gamma^{N_i}_{L_\alpha \Phi}\ = \frac{m_{N_1} m_{N_i}^{2} \Gamma_{N_{i}}^{\alpha} }{\pi^2\, z}\ K_1(z \sqrt{a_i})\,,
\end{eqnarray}
in terms of the rescaled variables of Eq.~(\ref{rescaled}) where	 $K_n(z)$ is an
$n$th-order modified Bessel function.
The $2 \to 2$ processes can be divided into $\Delta L = 0,1,2$ cases, each contributing to the washout of lepton flavour at different rates.
For a generic $2 \to 2$ process the collision term is calculated through
\begin{equation}
  \label{2t2g}
\gamma^{X}_{Y} = \frac{m^4_{N_1}}{64\,\pi^4 z} \int\limits_{x_{\rm thr} }^\infty\! dx \sqrt{x} \; K_1(z\sqrt{x}) \; \sigma^{X}_{Y}(x)\ ,
\end{equation}
where $x_{\rm thr}$ corresponds to $\rm{Min}(m(X),m(Y))$ such that the process is kinematically allowed. Each process is therefore simply calculated by substituting each reduced cross section $\sigma^{X}_{Y}$, numerically interpolating \ref{2t2g} over a range of z values and including in the numerical Boltzmann Equations.

The relevant $\Delta L = 1$ processes and cross-sections are,
\begin{eqnarray}
  \label{del1-1}
\sigma^{N_i L_\alpha}_{Qu^C} & = & \frac{3 y_{t}^{2}}{4\pi}   \left( h_{\alpha i}^{*} h_{\alpha i} \right)\left(\frac{x-a_i}{x}\right)^2\;,\nonumber\\
  \label{del1-2}
\sigma^{N_iu^C}_{L_\alpha Q^C} & = & \sigma^{N_iQ}_{L_\alpha u}\nonumber\\ 
&=& \frac{3 y_{t}^{2}}{4\pi}  \left( h_{\alpha i}^{*} h_{\alpha i}  \right)\left(1-\frac{a_i}{x}+\frac{a_i}{x}
\ln\left(\frac{x-a_i+a_r}{a_r}\right)\right),\nonumber\\
  \label{del1-3}
\sigma^{N_iV_\mu}_{L_\alpha \Phi} \!&=&\! 
\frac{3 g^2}{8\pi\,x} \,
\left(h_{\alpha i}^{*} h_{\alpha i}  \right)
\left(\frac{(x+a_i)^2}{x-a_i + 2 a_r}\,
\ln\left(\frac{x-a_i + a_r}{a_r}\right)\right),\nonumber\\
\label{del1-4}
\sigma^{N_i L_\alpha}_{\Phi^\dagger V_\mu} \!&=&\!
\frac{3 g^2}{16\pi\, x^2}\left(h_{\alpha i}^{*} h_{\alpha i}  \right)\left(5x-a_i)\,(a_i -x) +
2(x^2+xa_i - a^2_i)\,
\ln\left(\frac{x-a_i + a_r}{a_r}\right)\right),\nonumber\\
\label{del1-5}
\sigma^{N_i \Phi^\dagger}_{L_\alpha V_\mu} \!&=&\!
\frac{3 g^2}{16\pi\, x^2}\left(h_{\alpha i}^{*} h_{\alpha i}  \right) (x-a_i) \left( x-3a_i + 4a_i\,
\ln\left(\frac{x-a_i+a_r}{a_r}\right)\right)\,.
\end{eqnarray}
For the scatterings involving gauge bosons, only the $SU(2)_{L}$ processes were considered as the $U(1)_{Y}$ processes are expected to be subdominant.

The relevant $\Delta L = 2$ processes and cross-sections are,
\begin{eqnarray}
\label{del2-1}
\sigma^{ L_\alpha \Phi}_{L_\beta^C\Phi^\dagger} = 2 \sum_{i,j=1}^{6} \textrm{Re}\Big\lbrack\left(h_{\alpha j} h_{\beta j} h_{\alpha i}^{*} h_{\beta i}^{*}\right) \mathcal{A}^{ss}_{ij} & + &  \left(h_{\alpha j} h_{\beta j} h_{\alpha i}^{*} h_{\beta i}^{*}\right) \mathcal{A}^{tt}_{ij}\nonumber\\
& + & \left(h_{\alpha i}^{*} h_{\beta i} h_{\alpha j}^{*} h_{\beta j} + h_{\alpha i} h_{\beta i}^{*} h_{\alpha j} h_{\beta j}^{*}\right)\mathcal{A}^{(st)*}_{ij}\Big\rbrack\nonumber\\
\label{del2-2}
\sigma^{L_{\alpha} L_{\beta} }_{\Phi^\dagger\Phi^\dagger} \ = \
\sum_{i,j=1}^{6} {\rm Re} \left(h_{\alpha j} h_{\beta j} h_{\alpha i}^{*} h_{\beta i}^{*} \right) \mathcal{B}_{ij}\;,
\end{eqnarray}
where
\begin{eqnarray}
\label{aij}
\mathcal{A}^{(ss)}_{ij} \!\!&=&\!\! \left\{
\begin{array}{cc}
 0 &\quad (i=j)\,,\\
\frac{\displaystyle x\sqrt{a_i\,a_j}}{\displaystyle 
4\pi P^*_i P_j}\ &\quad (i\neq j)\,,
\end{array} \right. \nonumber\\
\mbox{}\nonumber\\
\mathcal{A}^{(tt)}_{ij} \!\!&=&\!\! \left\{
\begin{array}{cc}
\frac{a_i}{2\pi x}\Bigg[\, \frac{x}{a_i}\ -\
\ln\bigg(\frac{x+a_i}{a_i}\bigg)\, \Bigg], &\quad (i=j)\,,\nonumber\\
\frac{\sqrt{a_i\,a_j}}{2\pi x\:(a_i-a_j)}\left[\,
(x+a_j)\,\ln\bigg(\frac{x+a_j}{a_j}\bigg)\ -\
(x+a_i)\,\ln\bigg(\frac{x+a_i}{a_i}\bigg)\, \right], &\quad (i\neq j)\,,
\end{array} \right. \nonumber\\
\mbox{}\nonumber\\
\mathcal{A}^{(st)}_{ij} \!\!&=&\!\! 
\frac{\sqrt{a_i\,a_j}}{2\pi P_i}\left[\, 1\ -\ \frac{x+a_j}{x}\,
\ln\bigg(\frac{x+a_j}{a_j}\,\bigg)\right]\,,\nonumber\\
\mbox{}\nonumber\\
\mathcal{B}_{ij} \!\!&=&\!\! \left\{
\begin{array}{cc}
 \frac{1}{2\pi}\left[\,\frac{x}{x+a_i}\: +\: \frac{2\,a_i}{x+2a_i}
\ln\left(\frac{x+a_i}{a_i}\right)\,\right], &\quad (i=j)\,,\nonumber\\
\frac{\sqrt{a_i\,a_j}}{2\pi}
\left[\, \frac{1}{a_i-a_j}
\ln\left(\frac{a_i(x+a_j)}{a_j(x+a_i)}\right)\: +\:
\frac{1}{x+a_i+a_j}
\ln\left(\frac{(x+a_i)(x+a_j)}{a_i\,a_j}\right)\,\right]. &\quad (i\neq j)\,,
\end{array} \right.  \nonumber\\
\end{eqnarray}
and
\begin{equation}
  \label{Pi}
P^{-1}_i (x) \ =\ \frac{1}{x-a_i+i\sqrt{a_i c_i}}\ .
\end{equation}

Finally as flavoured leptogenesis is being considered there are also processes which do not violate lepton number but do violate lepton flavour \cite{Pilaftsis:2005rv}.
The relevant $\Delta L = 0$ processes are,
\begin{eqnarray}
\sigma^{L_\alpha \Phi}_{L_\beta \Phi} \!& = &\! 
\sum_{i,j=1}^6 \left(h_{\beta i}^* h_{\alpha i} h_{\beta j} h_{\alpha j}^* + 
h_{\beta i} h_{\alpha i}^* h_{\beta j}^* h_{\alpha j}\right)\mathcal{C}_{i j}\nonumber\\
\sigma^{L_\alpha \Phi^\dagger}_{L_\beta \Phi^\dagger} \!& = &\!
\sum_{i,j=1}^6 \mathrm{Re} \Big( h^{\nu
*}_{l\alpha}\,h^{\nu}_{k\alpha}\, 
h^{\nu}_{l\beta}\,h^{\nu *}_{k\beta} \Big)\,\mathcal{D}_{i j}\,,\nonumber\\
\sigma^{L_\alpha L^C_\beta}_{\Phi^\dagger \Phi} \!& = &\!
\sum_{i,j=1}^6 \mathrm{Re} 
\Big( h^{\nu *}_{l\alpha}\,h^{\nu}_{k\alpha}\, 
h^{\nu}_{l\beta}\,h^{\nu *}_{k\beta} \Big)\,\mathcal{E}_{i j}\,,
\end{eqnarray}
with
\begin{eqnarray}
\label{Cab}
\mathcal{C}_{i j} \!\!&=&\!\! \left\{
\begin{array}{lc}
0         &\quad (i = j)\\
\frac{\displaystyle x\sqrt{a_i a_j}}{\displaystyle 
4\pi P^*_i P_j}\ &\quad (i\neq j)
\end{array} \right.\nonumber\\
\mathcal{D}_{i j} \!\!&=&\!\! \left\{
\begin{array}{lc}
\frac{a_i}{\pi x}\:
\Bigg[\,\frac{x}{a_i} - \ln\Bigg(\frac{x+a_i}{a_i}\Bigg)         &\quad (i = j)\\
\frac{\sqrt{a_i a_j}}{\pi
x (a_i - a_j)}\: 
\Bigg[\, (x+a_j)\ln\Bigg(\frac{x+a_j}{a_j}\Bigg)-(x+a_i)
\ln\Bigg(\frac{x+a_i}{a_i}\Bigg)\, \Bigg] &\quad (i\neq j)
\end{array} \right.\nonumber\\
\mathcal{E}_{i j} \!\!&=&\!\! \left\{
\begin{array}{lc}
\frac{x}{\pi (x+a_i)}         &\quad (i = j)\\
\frac{\sqrt{a_i a_j}}{\pi
(a_i - a_j)}\:
\ln\Bigg(\frac{a_i (x+a_j)}{a_j (x+a_i)}\Bigg) &\quad (i \neq j)
\end{array} \right.
\end{eqnarray}

The scattering rates used in Eqs.~\ref{boltzn} and \ref{boltzl} are then calculated as functions of the collision terms above,
\begin{eqnarray}
\Gamma^{D(i \alpha)} & = & \frac{1}{n_\gamma}\
\gamma^{N_{i}}_{L_\alpha\Phi}\nonumber\\
\widehat{\Gamma}^{D(i \alpha)} & = &
\frac{1}{n_\gamma}\left(1+\frac{4}{21}\,\frac{\eta_{\Delta L}}{\eta_{\Delta L_\alpha}} \right)\,
\gamma^{N_{i}}_{L_\alpha\Phi}\nonumber\\
\widetilde{\Gamma}^{D(i \alpha)} & = &
\frac{1}{n_\gamma}\left(1+\frac{4}{21}\,\frac{\eta_{\Delta L}}{\eta_{\Delta L_\alpha}} \right)\,
\gamma^{N_{i}}_{L_\alpha\Phi}\nonumber\\
\Gamma^{S (i \alpha)}_{\rm Y} & = & \frac{1}{n_\gamma}\
\left(\, \gamma^{N_{i} L_\alpha}_{Q u^C} +  \gamma^{N_{i}
u^C}_{L_\alpha Q^C} + \gamma^{N_{i} Q}_{L_\alpha u}\, \right),\nonumber\\
\widehat{\Gamma}^{S(i \alpha)}_{\rm Y} &=& \frac{1}{n_\gamma}\
\biggl[\left(-\frac{\eta_{N_{i}}}{\eta^{\rm eq}_{N_{i}}}
+\frac{4}{21}\,\frac{\eta_{\Delta L}}{\eta_{\Delta L_\alpha}}\right)
\gamma^{N_{i} L}_{Q u^C} +
\left(1+\frac{1}{9}\,\frac{\eta_{\Delta L}}{\eta_{\Delta L_\alpha}}
-\frac{5}{63}\,\frac{\eta_{N_{i}}}{\eta^{\rm eq}_{N_{i}}}
\frac{\eta_{\Delta L}}{\eta_{\Delta L_\alpha}}\right)
\gamma^{N_{\alpha} u^C}_{L Q^C}\nonumber\\
& & +\left(1+\frac{5}{63}\,\frac{\eta_{\Delta L}}{\eta_{\Delta L_\alpha}}
-\frac{1}{9}\,\frac{\eta_{N_{i}}}{\eta^{\rm eq}_{N_{i}}}
\frac{\eta_{\Delta L}}{\eta_{\Delta L_\alpha}}\right)
\gamma^{N_{i} Q}_{L u}\biggr]\nonumber\\
\widetilde{\Gamma}^{S(i \alpha)}_{\rm Y} &=& \frac{1}{n_\gamma}\
\biggl[\left(\frac{\eta_{N_{i}}}{\eta^{\rm eq}_{N_{i}}}
+\frac{4}{21}\,\frac{\eta_{\Delta L}}{\eta_{\Delta L_\alpha}}\right)
\gamma^{N_{i} L}_{Q u^C} +
\left(1+\frac{1}{9}\,\frac{\eta_{\Delta L}}{\eta_{\Delta L_\alpha}}
+\frac{5}{63}\,\frac{\eta_{N_{i}}}{\eta^{\rm eq}_{N_{i}}}
\frac{\eta_{\Delta L}}{\eta_{\Delta L_\alpha}}\right)
\gamma^{N_{i} u^C}_{L Q^C}\nonumber\\
& & + \left(1+\frac{5}{63}\,\frac{\eta_{\Delta L}}{\eta_{\Delta L_\alpha}}
+\frac{1}{9}\,\frac{\eta_{N_{i}}}{\eta^{\rm eq}_{N_{i}}}
\frac{\eta_{\Delta L}}{\eta_{\Delta L_\alpha}}\right)
\gamma^{N_{i} Q}_{L u}\biggr]\nonumber\\
\Gamma^{S(i \alpha)}_{\rm G} & = & \frac{1}{n_\gamma}
\biggl(\gamma^{N_{i} L_\alpha}_{\Phi^\dagger V_\mu} +
\gamma^{N_{i} V_\mu}_{L_\alpha\Phi} +
\gamma^{N_{i}\Phi^\dagger }_{L_\alpha V_\mu}\biggr)\nonumber\\
\widehat{\Gamma}^{S(i \alpha)}_{\rm G} &=& \frac{1}{n_\gamma}\ 
\biggl[\left(-\frac{\eta_{N_{i}}}{\eta^{\rm eq}_{N_{i}}}
+\frac{4}{21}\frac{\eta_{\Delta L}}{\eta_{\Delta L_\alpha}}\right)
\gamma^{N_{i} L}_{\Phi^\dagger V_\mu}
+\left(1+\frac{4}{21}\,\frac{\eta_{\Delta L}}{\eta_{\Delta L_\alpha}}\right)
\gamma^{N_{i} V_\mu}_{L \Phi}\nonumber\\
& & + \left(1-\frac{4}{21}\,\frac{\eta_{N_{i}}}{\eta^{\rm
eq}_{N_{i}}}
\frac{\eta_{\Delta L}}{\eta_{\Delta L_\alpha}}\right)
\gamma^{N_{i}\Phi^\dagger }_{L V_\mu}\biggr]\nonumber\\
\widetilde{\Gamma}^{S(i \alpha)}_{\rm G} &=& \frac{1}{n_\gamma}
\biggl[ \left( \frac{\eta_{N_{i}}}{\eta^{\rm eq}_{N_{i}}}+\frac{4}{21}\,\frac{\eta_{\Delta L}}{\eta_{\Delta L_\alpha}} \right) 
\gamma^{N_{i} L}_{\Phi^\dagger V_\mu}
 +\left(1+\frac{4}{21}\,\frac{\eta_{\Delta L}}{\eta_{\Delta L_\alpha}}\right)
\gamma^{N_{i} V_\mu}_{L \Phi}\nonumber\\
& & +\left(1+\frac{4}{21}\frac{\eta_{N_{i}}}{\eta^{\rm eq}_{N_{i}}}
\frac{\eta_{\Delta L}}{\eta_{\Delta L_\alpha}}\right)
\gamma^{N_{i}\Phi^\dagger }_{L V_\mu}\Biggr]\nonumber\\
\Gamma^{W(i \alpha)}_{\rm Y} & = & \frac{1}{n_\gamma}
\biggl[\left(2+\frac{4}{21}\frac{\eta_{\Delta L}}{\eta_{\Delta L_\alpha}}\right)
\gamma^{N_{i} L}_{Q u^C} +
\left(1+\frac{17}{63}\frac{\eta_{\Delta L}}{\eta_{\Delta L_\alpha}}\right)
\gamma^{N_{i} u^C}_{L Q^C}\nonumber\\
& & +\left(1+\frac{19}{63}\frac{\eta_{\Delta L}}{\eta_{\Delta L_\alpha}}\right)
\gamma^{N_{\alpha} Q}_{L u}\biggr]\nonumber\\
\Gamma^{W(i \alpha)}_{\rm G} & = & \frac{1}{n_\gamma}\ 
\biggl[\left(2+\frac{4}{21}\,\frac{\eta_{\Delta L}}{\eta_{\Delta L_\alpha}}\right)
\gamma^{N_{i} L}_{\Phi^\dagger V_\mu} +
\left(1+\frac{4}{21}\,\frac{\eta_{\Delta L}}{\eta_{\Delta L_\alpha}}\right)
\gamma^{N_{i} V_\mu}_{L \Phi}\nonumber\\
& & + \left(1+\frac{8}{21}\frac{\eta_{\Delta L}}{\eta_{\Delta L_\alpha}}\right)
\gamma^{N_{i} \Phi^\dagger }_{L V_\mu}\biggr]\nonumber\\
\Gamma^{\Delta L =2(\alpha \beta)}_{\rm Y} &=& \frac{1}{n_\gamma}\ 
\biggl[\left(1+\frac{4}{21}\frac{\eta_{\Delta L}}{\eta_{\Delta L_\alpha}}\right)
\left(\gamma^{L_\alpha\Phi}_{\,L_\beta^C\Phi^\dagger} + 
\gamma^{L_\alpha L_\beta}_{\Phi^\dagger\Phi^\dagger}\right)\biggr]\nonumber\\
\label{scatterings}
\Gamma^{\Delta L =0(\alpha \beta)}_{\rm Y} &=& \frac{1}{n_\gamma}\ 
\biggl[\left(1+\frac{4}{21}\,\frac{\eta_{\Delta L}}{\eta_{\Delta L_\alpha}}\right)
\gamma^{L_\alpha\Phi}_{\,L_\beta\Phi}  + 
\gamma^{L_\alpha\Phi^\dagger}_{L_\beta\Phi^\dagger} + 
\gamma^{L_\alpha L^C_\beta}_{\Phi \Phi^\dagger}\biggr].
\end{eqnarray}

\clearpage

\section{$\delta_{CP}$ dependence}

\begin{figure}[t]
\captionsetup[subfigure]{justification=centering,labelformat=empty}
\centering
\includegraphics[scale=0.6]{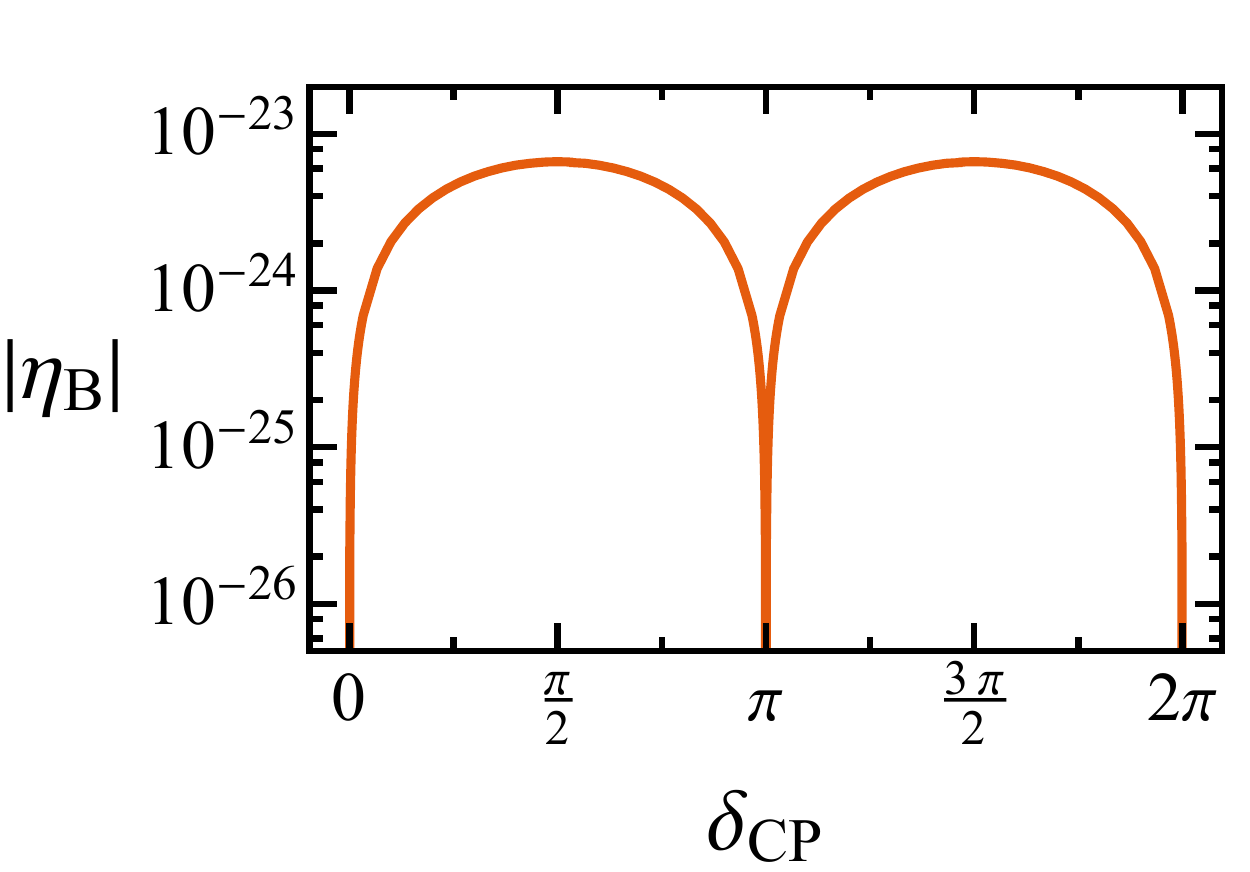}
\subfloat[1][\!\!\!\!\!\!\huge$\mathbf{\eta_e}$]{
  \includegraphics[width=0.3\textwidth]{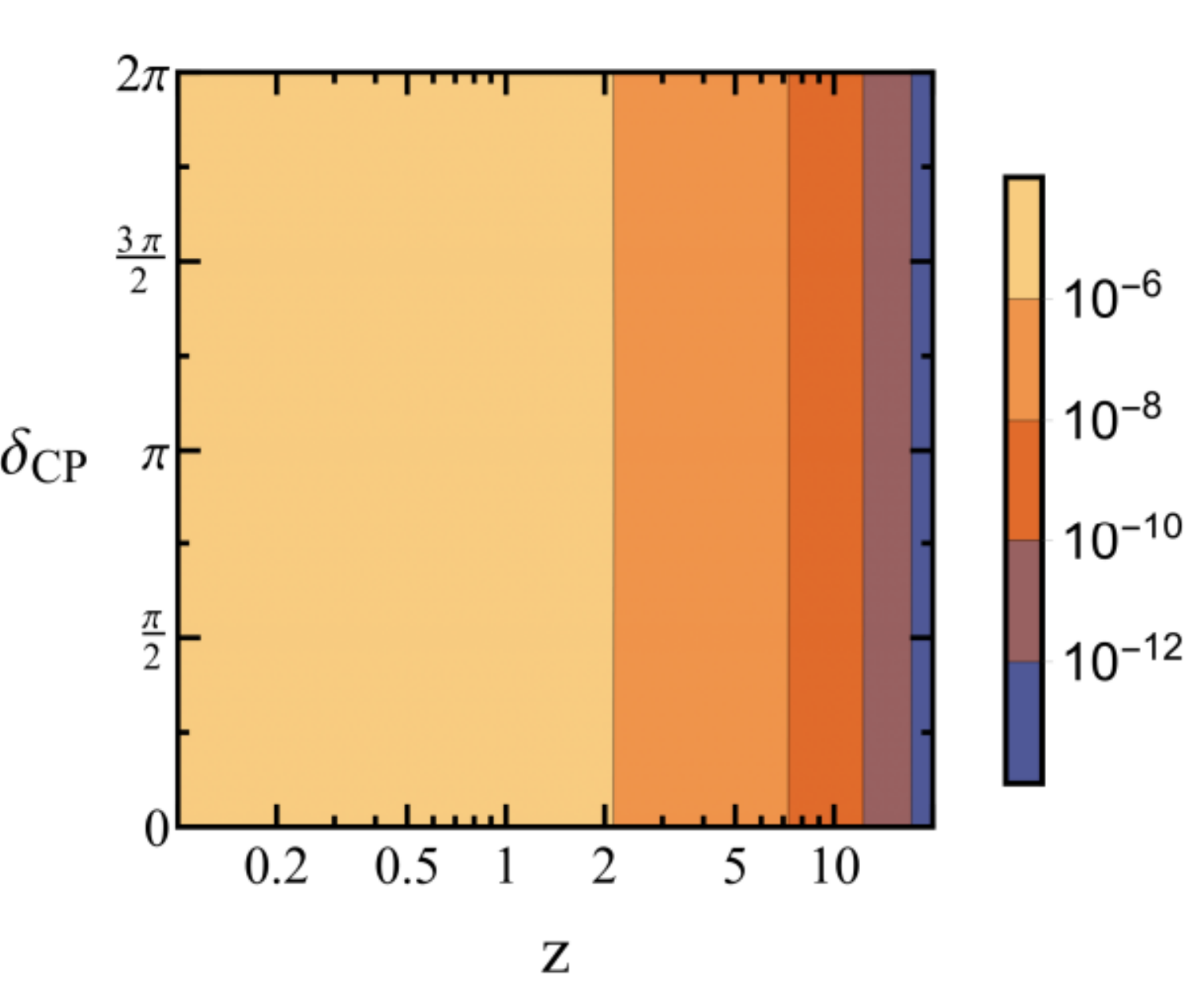}
  }
\subfloat[2][\!\!\!\!\!\!\huge$\mathbf{\eta_\mu}$]{
  \includegraphics[width=0.3\textwidth]{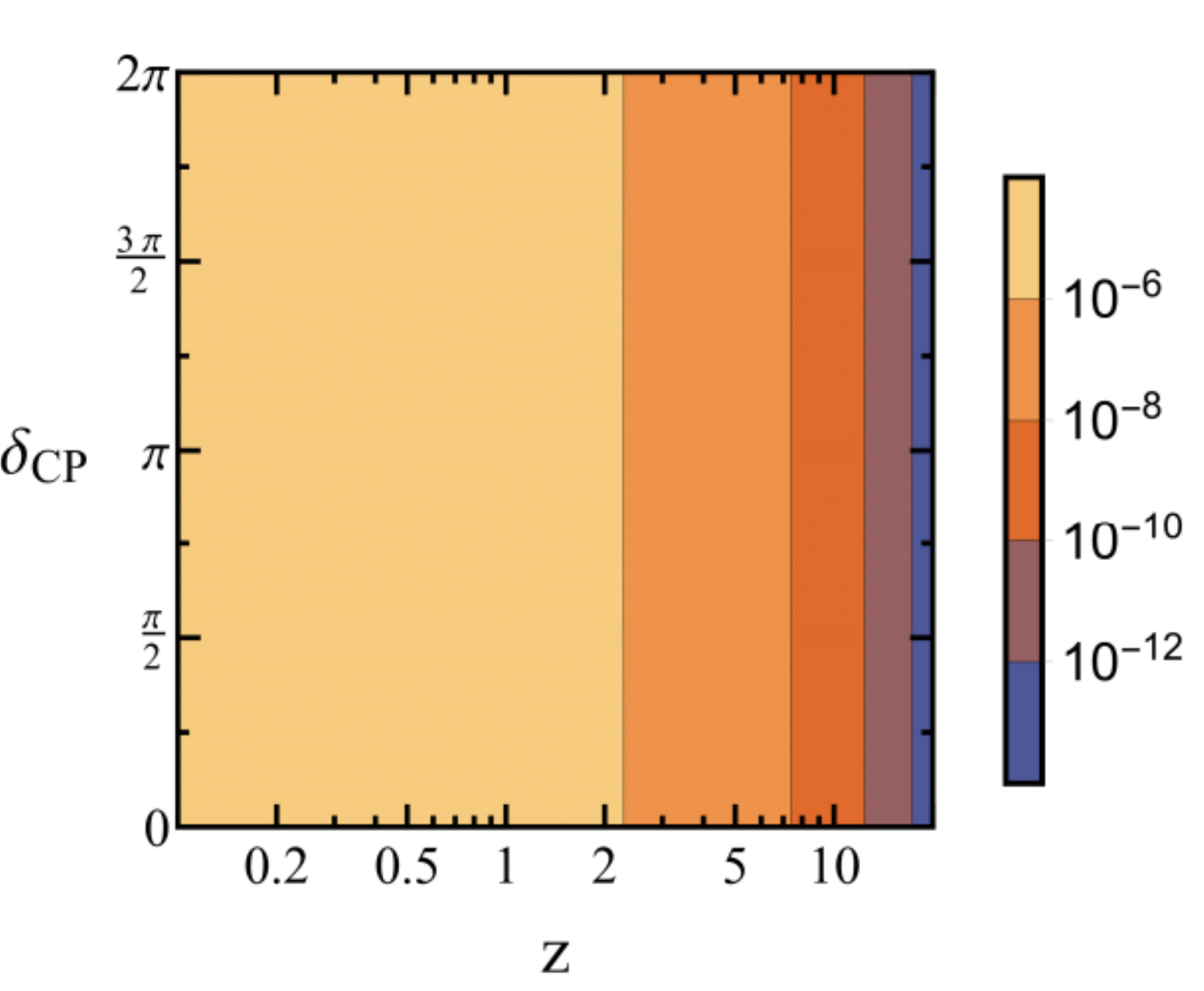}
  } 
\subfloat[7][\!\!\!\!\!\!\huge$\mathbf{\eta_\tau}$]{
  \includegraphics[width=0.3\textwidth]{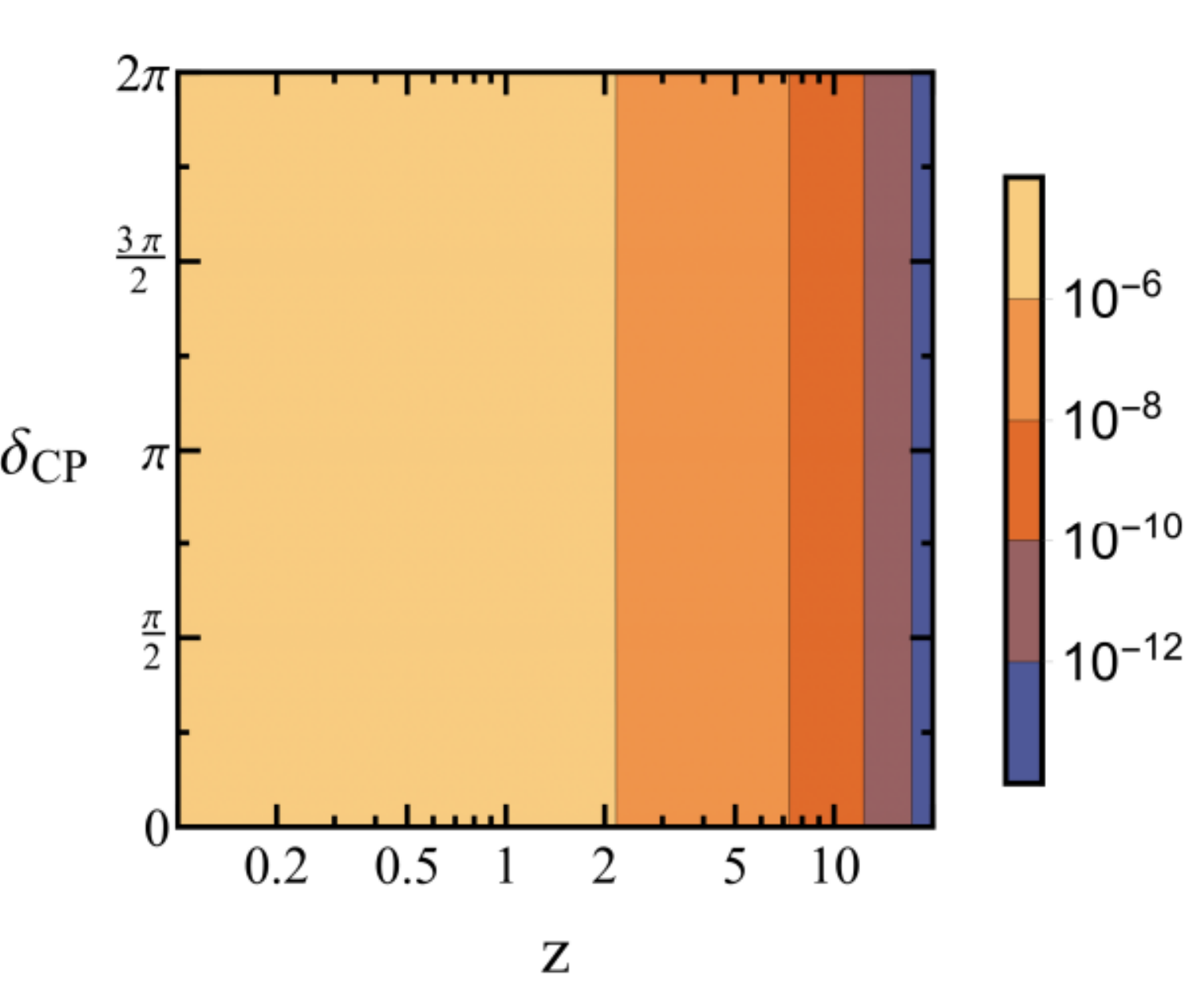}
  }
\caption{In scenarios where the asymmetry is dominated by terms in the scatterings which do not have a net dependence on $\delta_{CP}$ the net asymmetry is maximised when the CP violation is maximised.}
\label{deltaCP1}
%\end{figure}
%\begin{figure}[!htbp]
\captionsetup[subfigure]{justification=centering,labelformat=empty}
\centering
\includegraphics[scale=0.6]{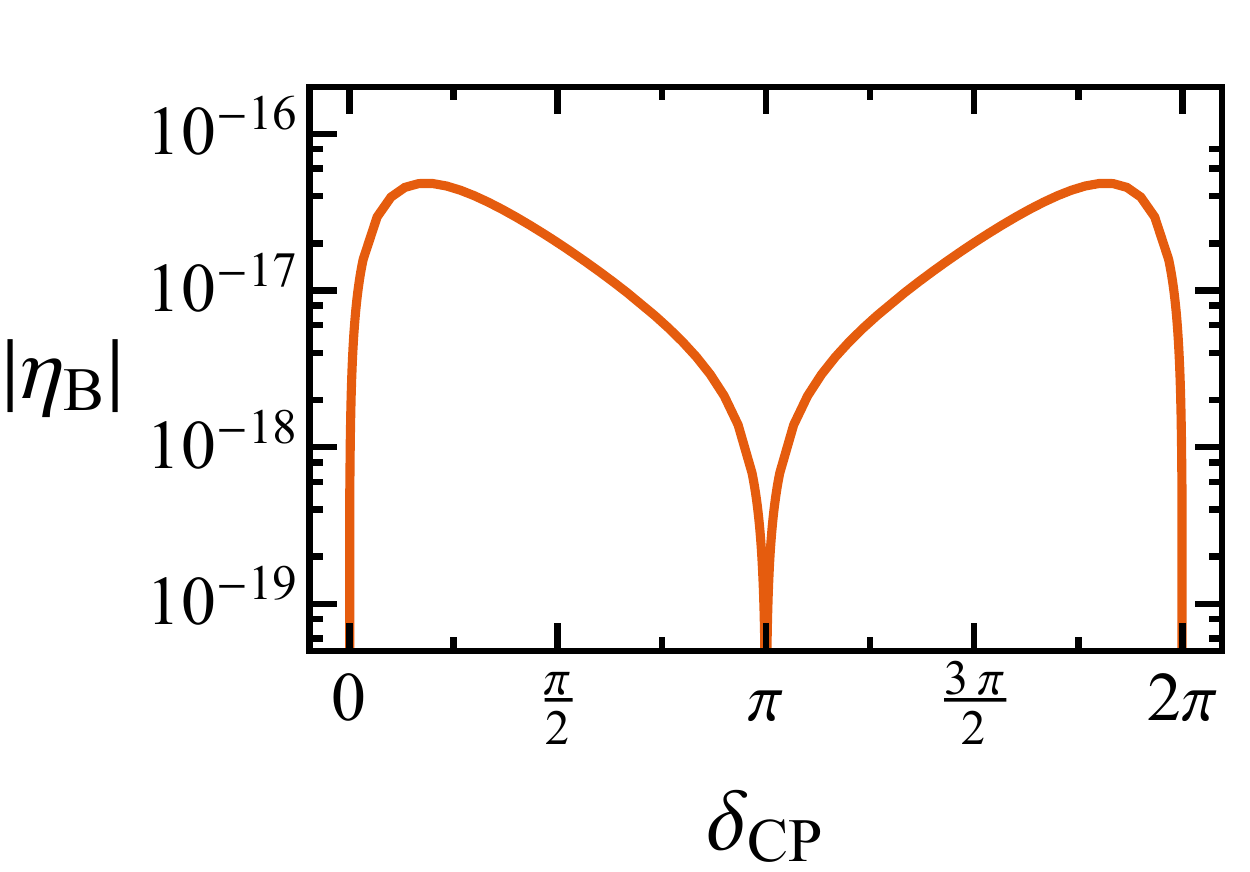}
\subfloat[1][\!\!\!\!\!\!\huge$\mathbf{\eta_e}$]{
  \includegraphics[width=0.3\textwidth]{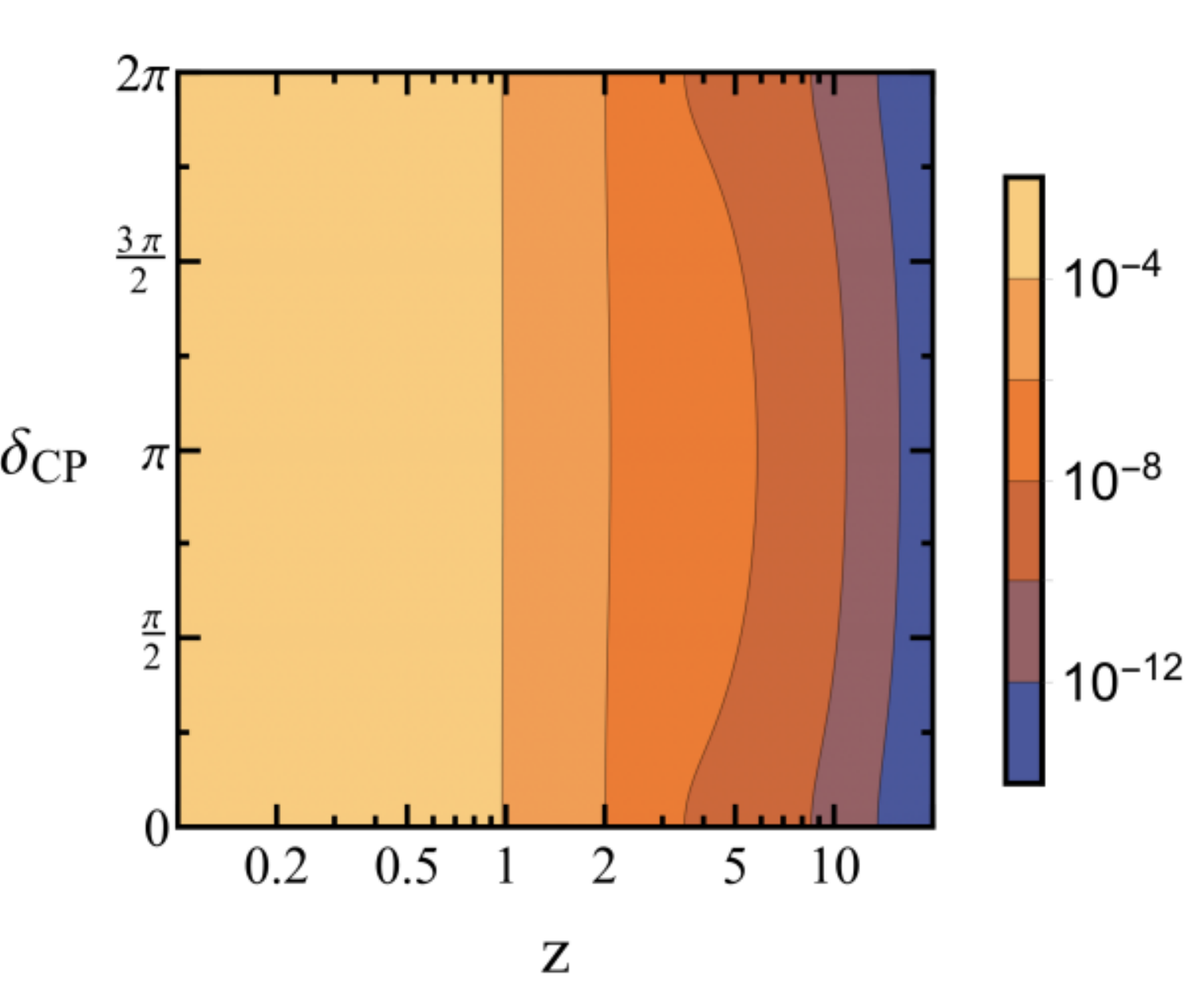}
  }
\subfloat[2][\!\!\!\!\!\!\huge$\mathbf{\eta_\mu}$]{
  \includegraphics[width=0.3\textwidth]{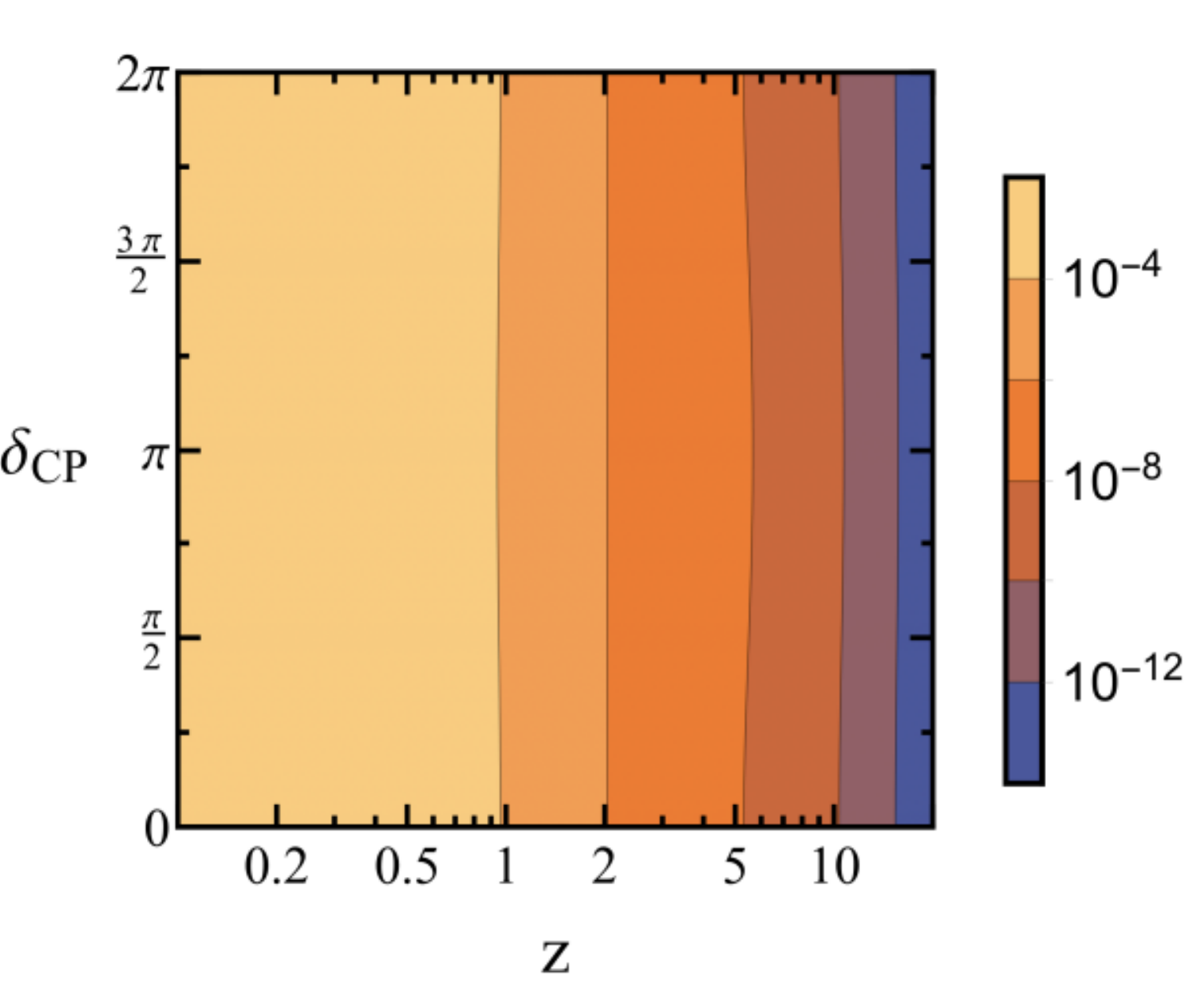}
  } 
\subfloat[7][\!\!\!\!\!\!\huge$\mathbf{\eta_\tau}$]{
  \includegraphics[width=0.3\textwidth]{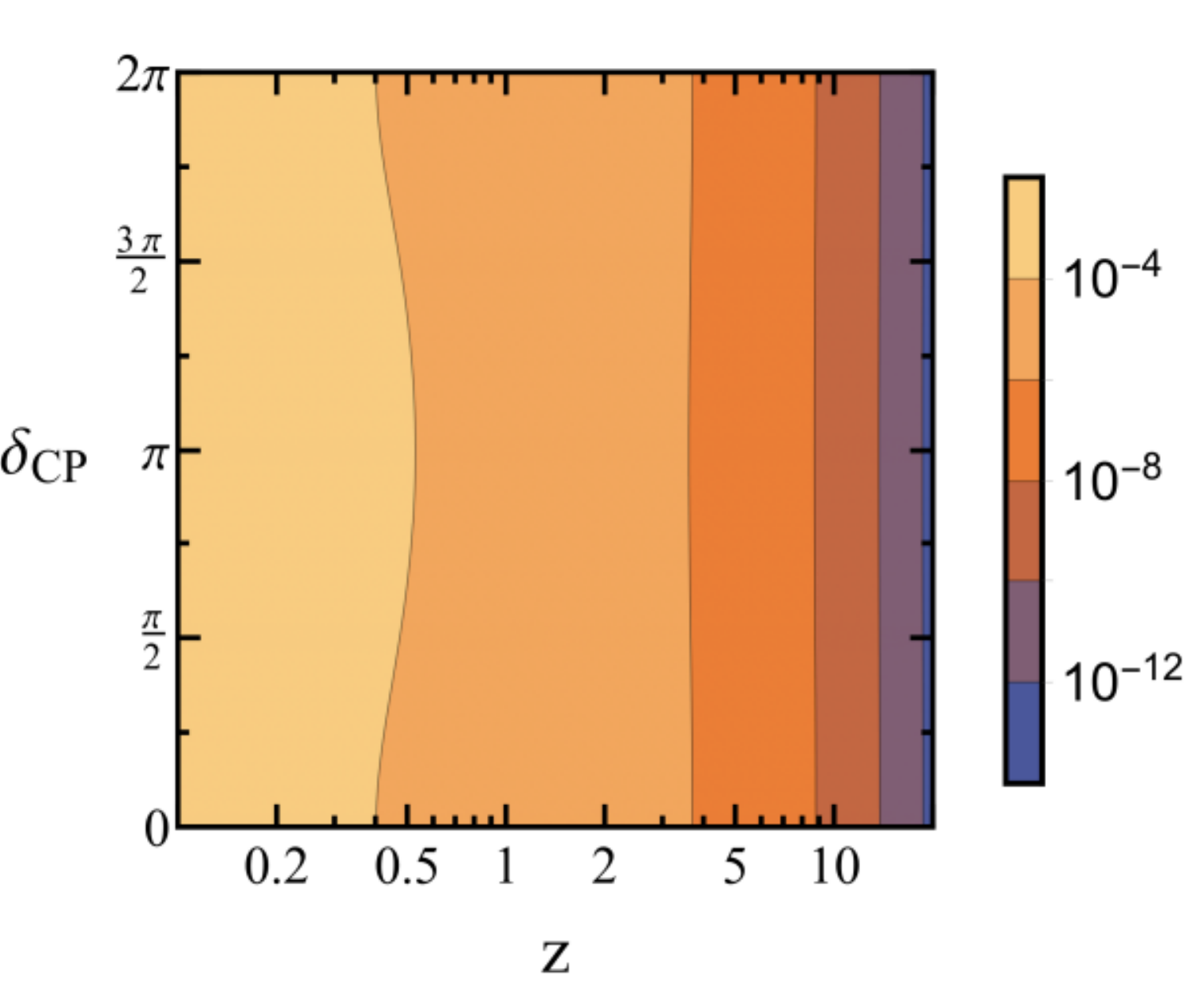}
  }
\caption{If the total scattering is maximised around $\delta_{CP} = \pi$, more asymmetry is washed out in this region and the baryon asymmetry will be maximised for values of $\delta_{CP}$ closer to 0 and $2\pi$. }
\label{deltaCP2}
\end{figure}

\begin{figure}[t]
\captionsetup[subfigure]{justification=centering,labelformat=empty}
\centering
\includegraphics[scale=0.6]{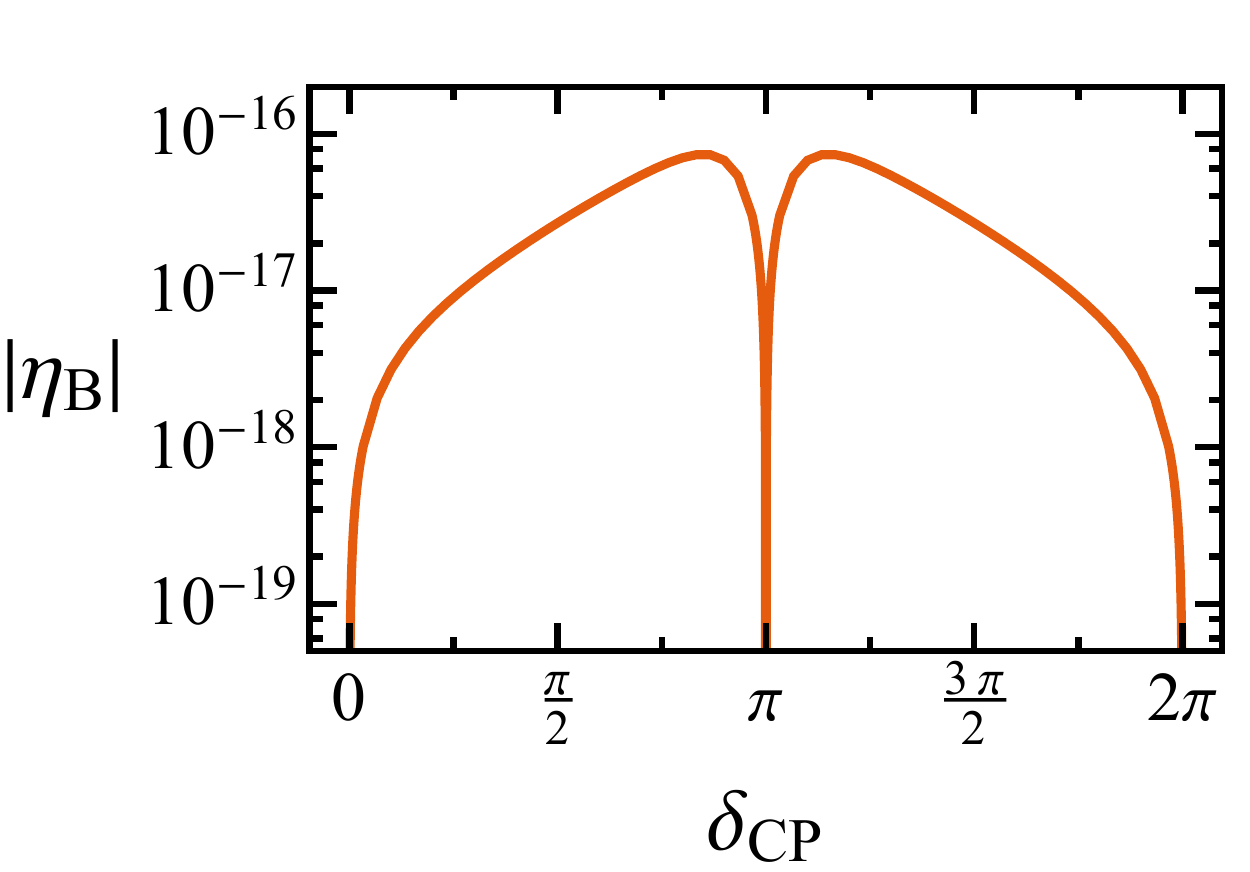}
\subfloat[1][\!\!\!\!\!\!\huge$\mathbf{\eta_e}$]{
  \includegraphics[width=0.3\textwidth]{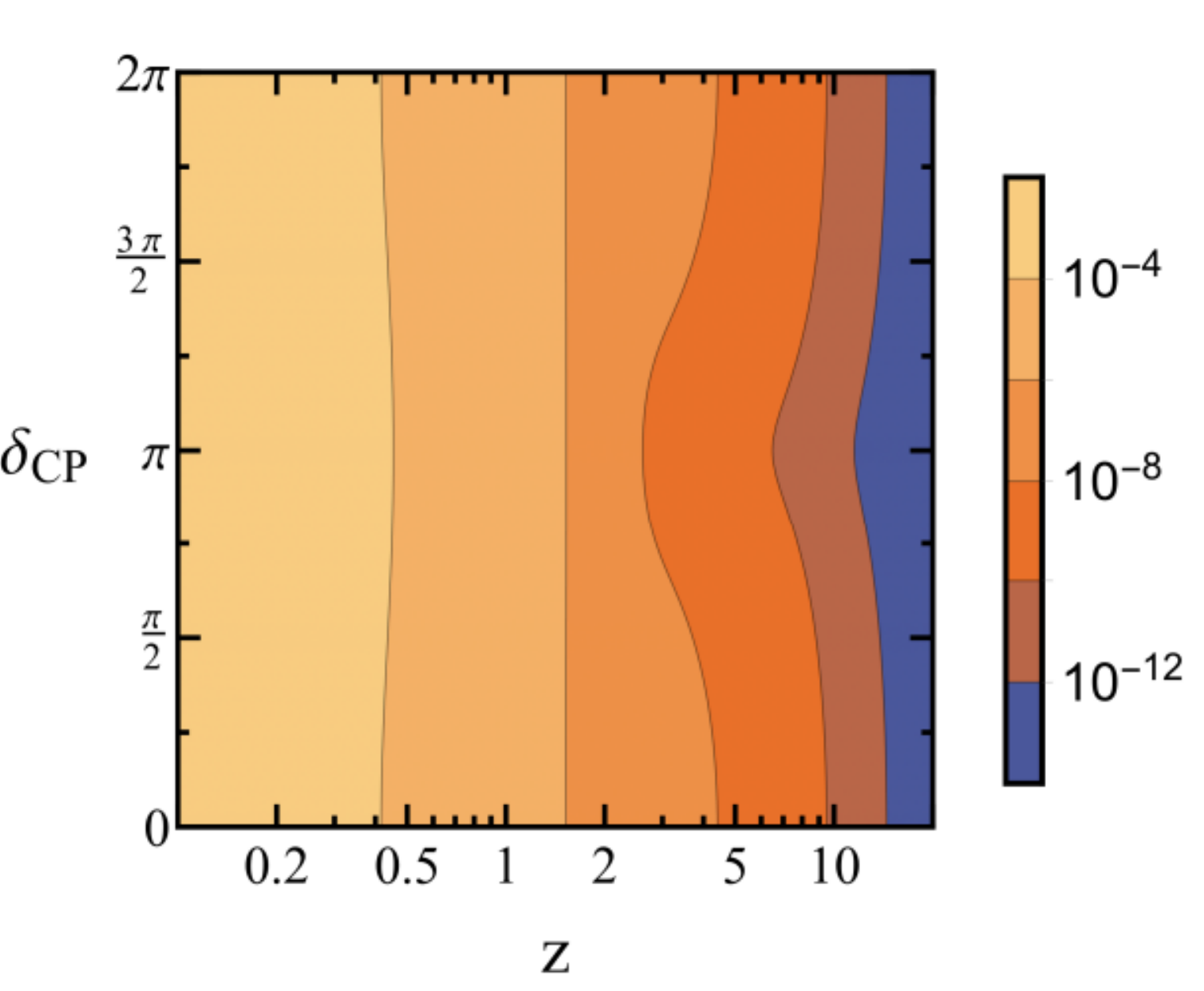}
  }
\subfloat[2][\!\!\!\!\!\!\huge$\mathbf{\eta_\mu}$]{
  \includegraphics[width=0.3\textwidth]{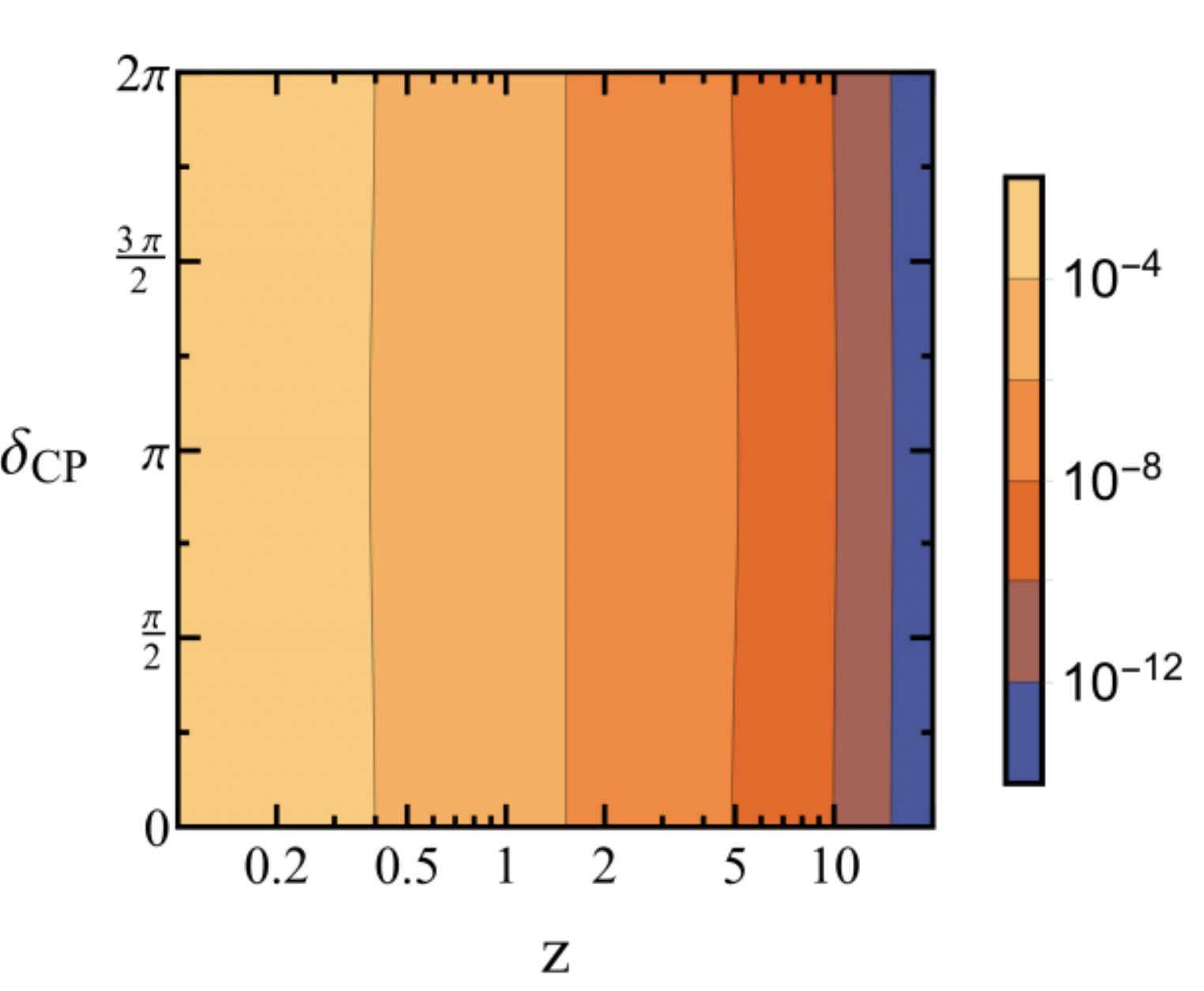}
  } 
\subfloat[7][\!\!\!\!\!\!\huge$\mathbf{\eta_\tau}$]{
  \includegraphics[width=0.3\textwidth]{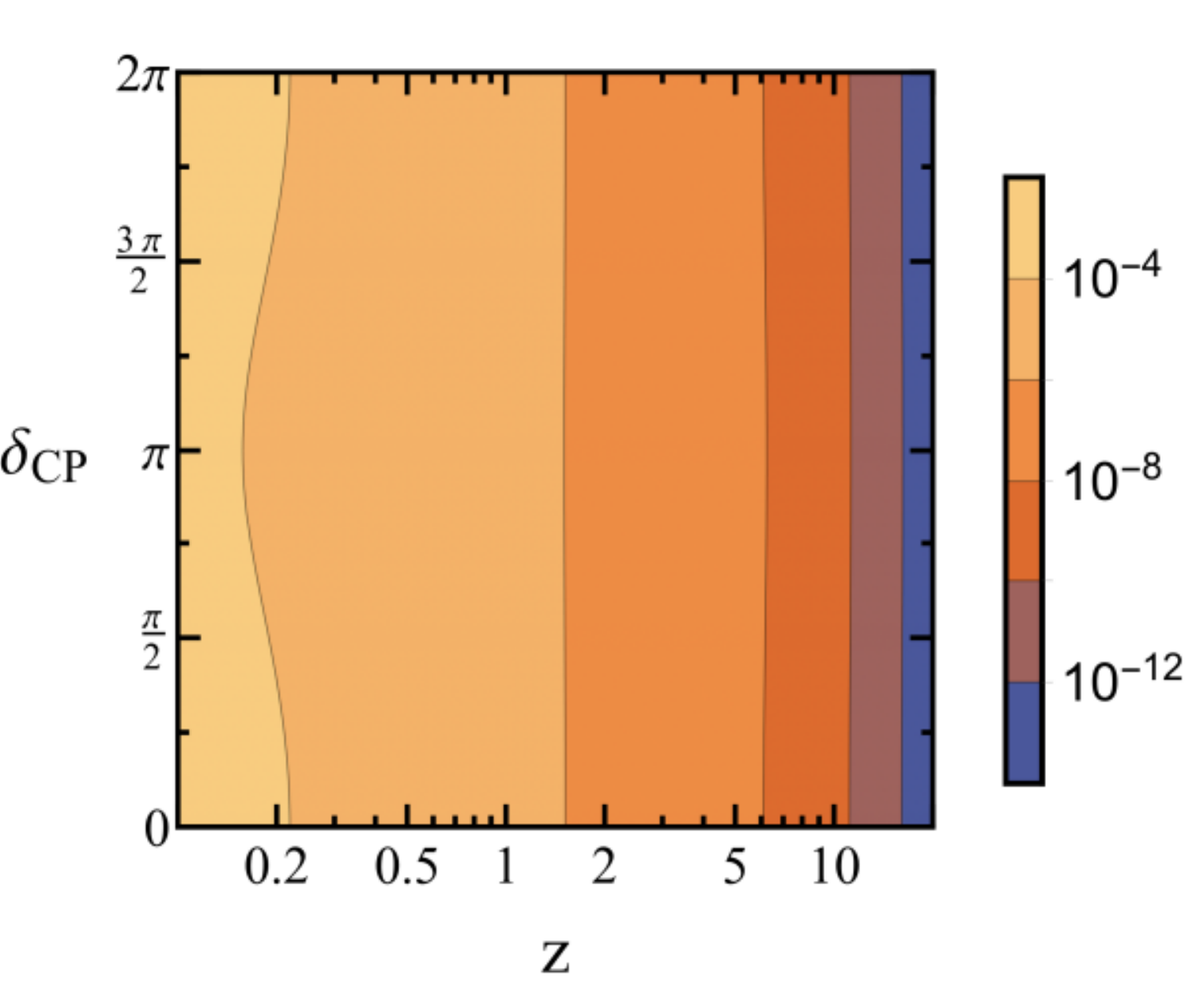}
  }
\caption{If the total scattering is maximised around $\delta_{CP} = 0,2\pi$, more asymmetry is washed out in this region and the baryon asymmetry will be maximised for values of $\delta_{CP}$ closer to $\pi$.}
%\label{deltaCP3}
%\end{figure}
%\begin{figure}[t]
\captionsetup[subfigure]{justification=centering,labelformat=empty}
\centering
\includegraphics[scale=0.6]{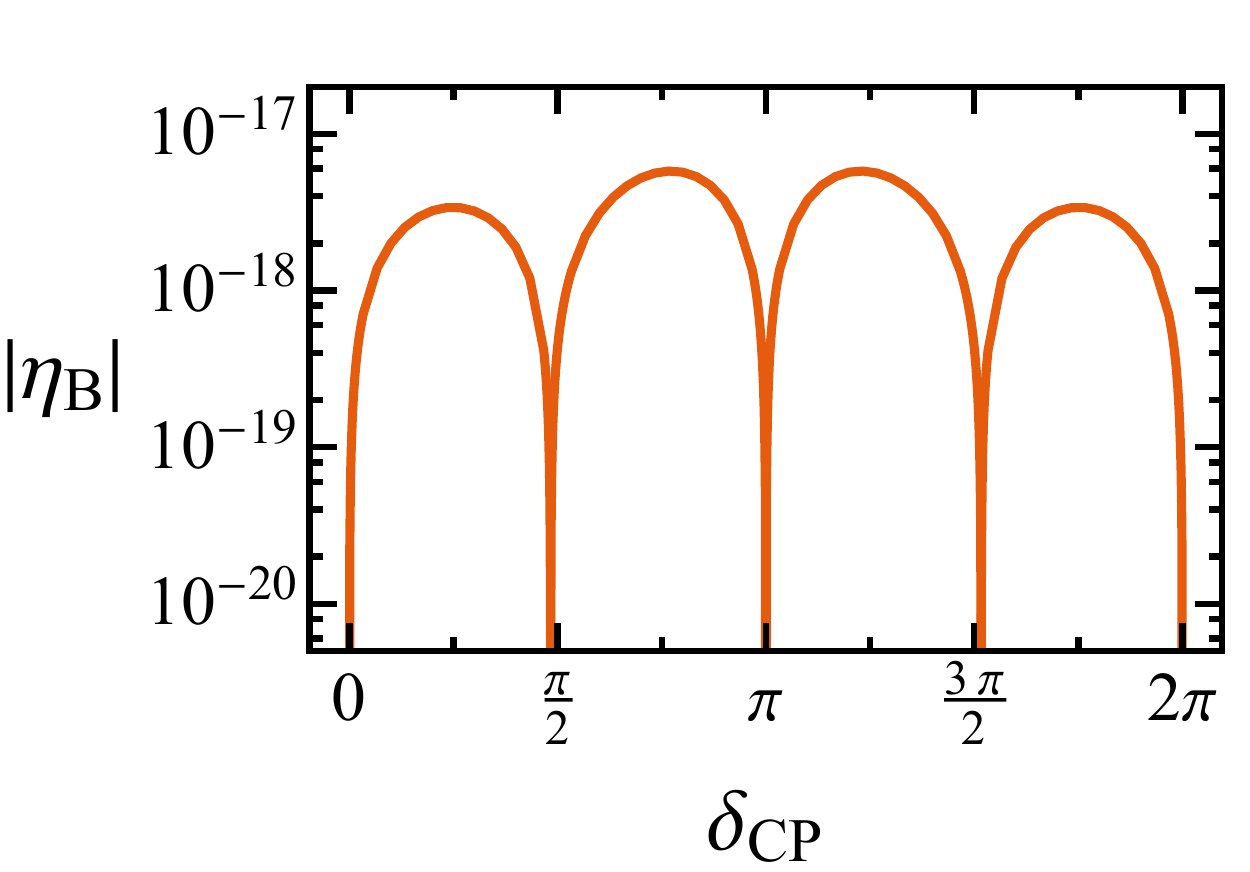}
\subfloat[\!\!\!\!\!\!\huge$\mathbf{\eta_e}$]{
  \includegraphics[width=0.3\textwidth]{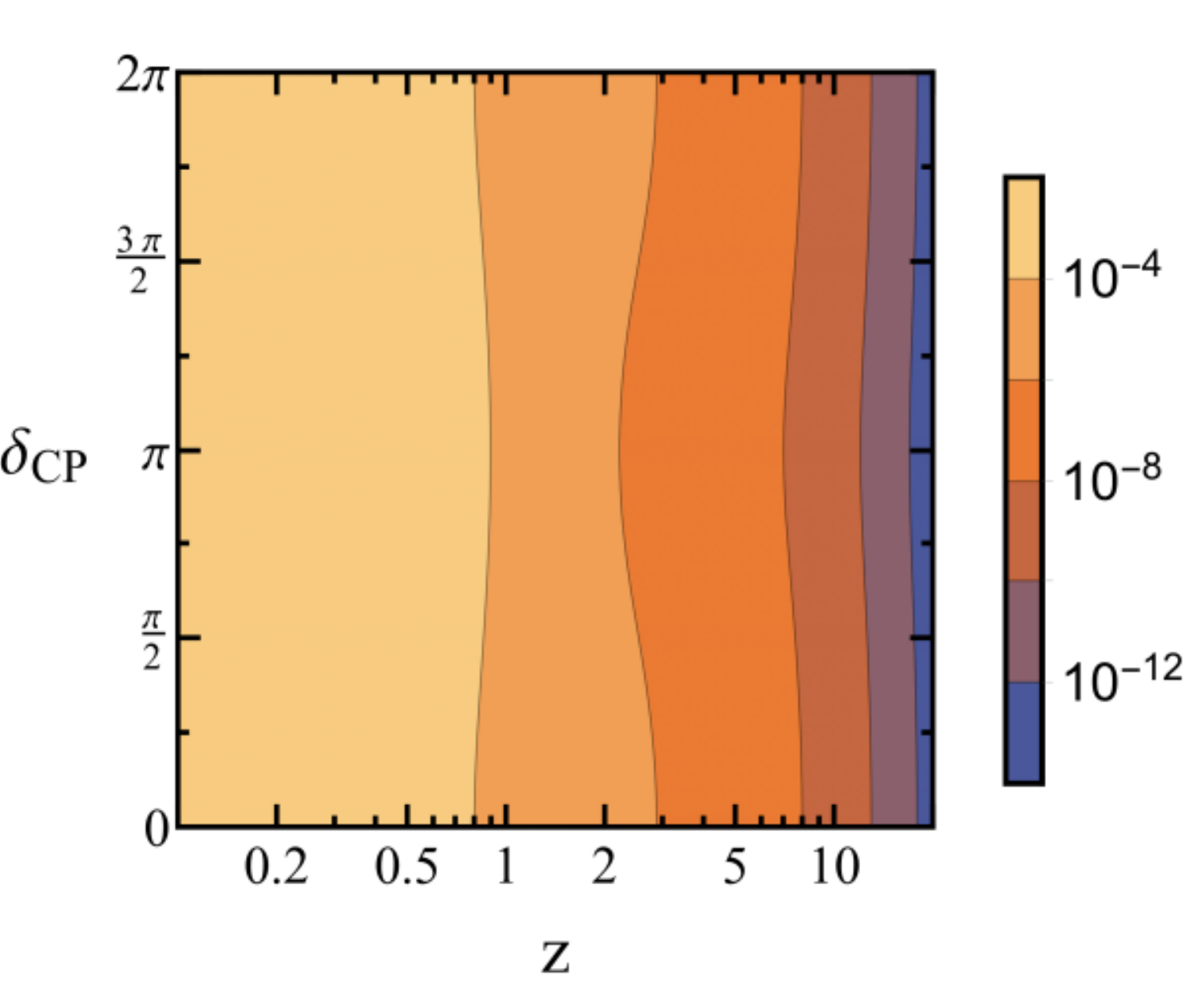}
  }
\subfloat[\!\!\!\!\!\!\huge$\mathbf{\eta_\mu}$]{
  \includegraphics[width=0.3\textwidth]{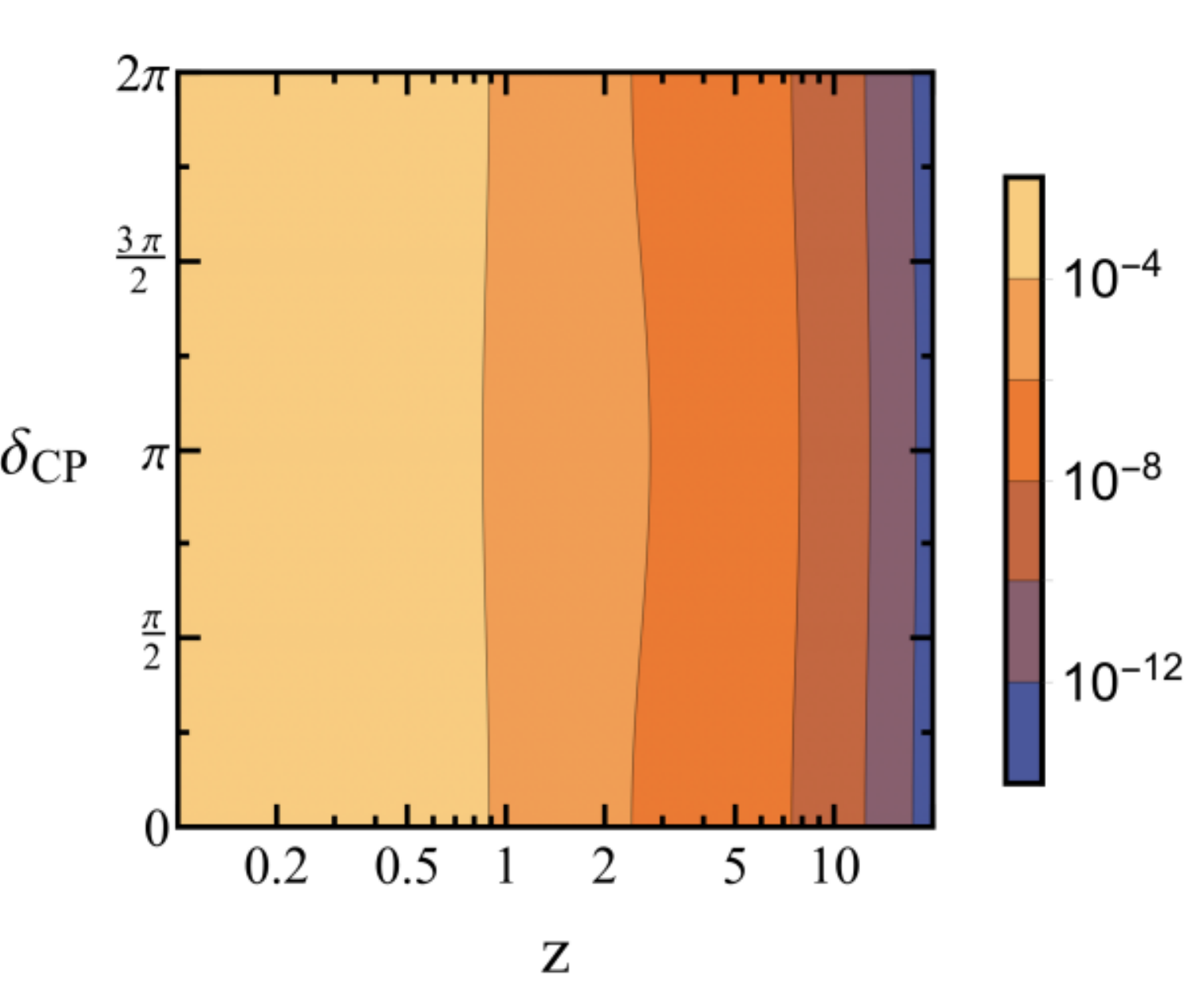}
  } 
\subfloat[\!\!\!\!\!\!\huge$\mathbf{\eta_\tau}$]{
  \includegraphics[width=0.3\textwidth]{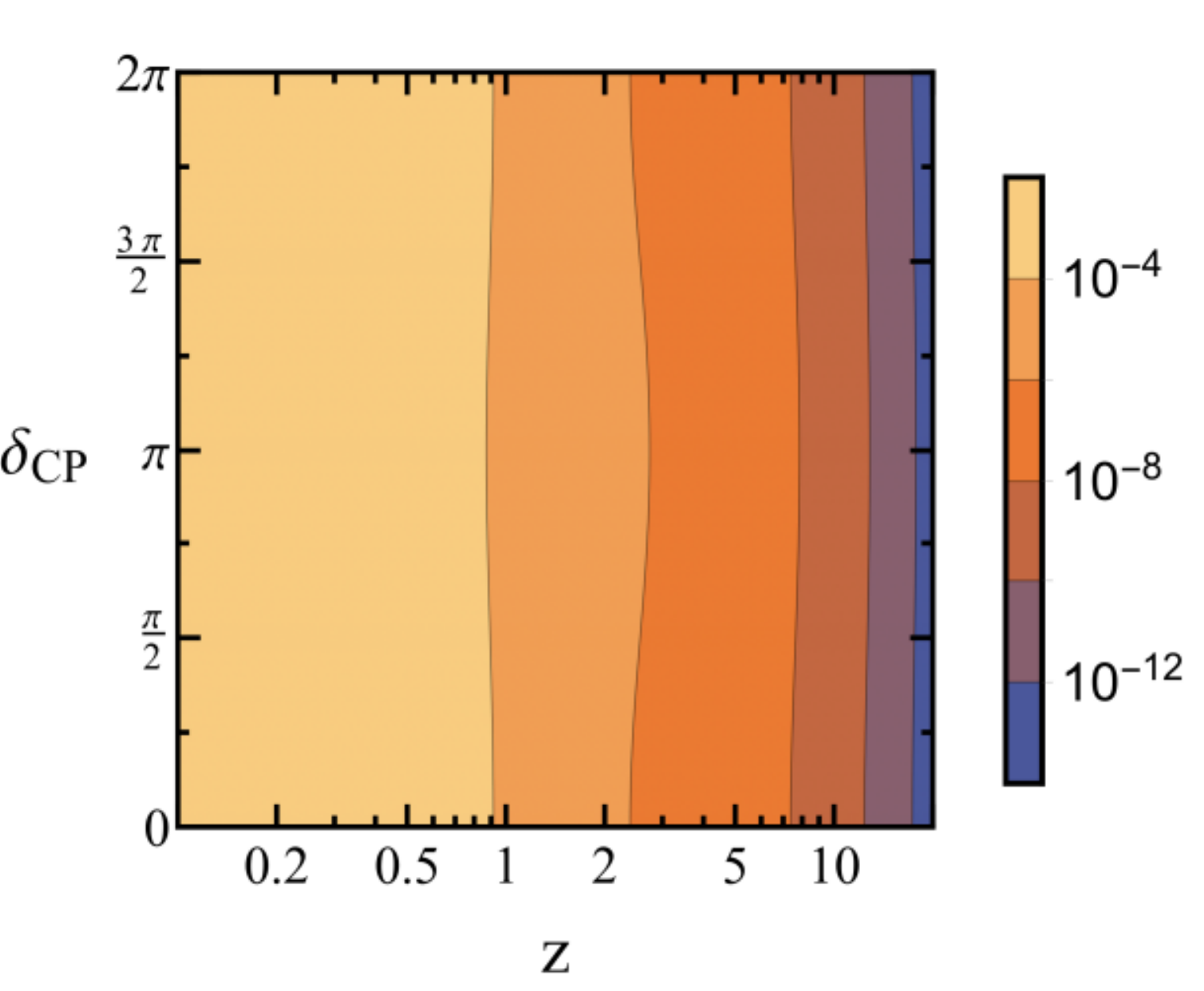}
  }
\caption{In the unique scenario that the scattering rates have competing efficiencies, e.g. the efficiency to $e$ is maximal for $\delta_{CP} \rightarrow 0,2\pi$ and the efficiency to $\mu$ and $\tau$ is maximal for $\delta_{CP} \rightarrow \pi$ as shown above, the net baryon asymmetry generated can actually be minimised for maximal CP violation.}
\label{deltaCP4}
\end{figure}

Due to the flavoured nature of the Boltzmann equations, an inherent $\delta_{CP}$ dependence exists on the strength of certain flavoured scattering terms within them. It is interesting to note the consequence that a maximally-violating phase, such as $\delta_{CP}$, actually does not necessarily result in a maximally generated net baryon asymmetry. Although maximal values of $\delta_{CP}$ (as suggested by current data) will lead to maximised generation of asymmetry per decay, if the efficiency of the washout is greater here compared to some other values of $\delta_{CP}$ the net surviving asymmetry may end up being smaller. It is even possible for the net asymmetry to be minimised when the CP violating phase is maximised. This phenomenon is not explored further in the present work and the CP-violating phase $\delta_{CP}$ is simply fixed to its suggestive value of $\delta_{CP} = 3\pi/2$. Below are example plots of the absolute value of the net baryon asymmetry\footnote{The absolute value is plotted due to the sign change between $0 < \delta_{CP} < \pi$ and $\pi < \delta_{CP} < 2\pi$.} $| \eta_B |$ as a function of $\delta_{CP}$ in the pure ISS scenario, chosen once again for computation convenience even though the asymmetry values are unrealistic. Alongside each plot is a contour plot of the sum of each $\Delta L = 0,1,2$ scattering term ($\sum_{X,i,\alpha} \Gamma^{X(i\alpha)}$) as a function of $\delta_{CP}$ and $z$ for a specific lepton flavour. As illustrated in Figs.~\ref{deltaCP1}-\ref{deltaCP4}, in scenarios where scattering terms without a $\delta_{CP}$ dependence dominate, the generated asymmetry behaves as na\"{i}vely expected. However, for some regions of parameter space the final asymmetry generated is instead maximised in regions of $\delta_{CP}$ which minimise the total washout.

\FloatBarrier

%\paragraph{Note added.} This is also a good position for notes added
%after the paper has been written.

\bibliographystyle{JHEP}
\bibliography{diracphase}

\end{document}